\documentclass[a4paper,11pt]{article}
\pdfoutput=1
\usepackage{geometry}
\usepackage{a4wide,slashbox}
\usepackage{graphicx}
\usepackage{epsf}
\usepackage{amsmath}
\usepackage[normalem]{ulem}
\usepackage{amssymb}

\usepackage{cite}
\usepackage{multirow,tabularx}
\usepackage{appendix}
\usepackage{tikz}
\usepackage{graphicx,amsmath,amsfonts,amssymb,amsthm,euscript,braket,xcolor}
\newcommand{\be}{\begin{equation}}
\newcommand{\ee}{\end{equation}}

\newcommand{\Rmnum}[1]{\expandafter\@slowromancap\romannumeral #1@}
\newcommand{\bea}{\begin{eqnarray}}
\newcommand{\eea}{\end{eqnarray}}

\numberwithin{equation}{section}

\newcommand*\circled[1]{\tikz[baseline=(char.base)]{
            \node[shape=circle,draw,inner sep=2pt] (char) {#1};}}

\begin{document}

\title{\bf Interplay between the holographic QCD phase diagram and mutual \& $n$-partite information }
\author{\textbf{Subhash Mahapatra}\thanks{mahapatrasub@nitrkl.ac.in}
 \\\\\
\textit{{\small Department of Physics and Astronomy, National Institute of Technology Rourkela, Rourkela - 769008, India}}}
\date{}

\maketitle
\abstract{In an earlier work, we studied holographic entanglement entropy in QCD phases using a dynamical Einstein-Maxwell-dilaton gravity model whose dual
boundary theory mimics essential features of QCD above and below deconfinement. The model although displays subtle differences compared to the standard QCD phases,
however, it introduces a notion of temperature in the phase below the deconfinement critical temperature and captures quite well the entanglement and thermodynamic properties of QCD
phases. Here we extend our analysis to study the mutual and $n$-partite information by
considering $n$ strips with equal lengths and equal separations, and investigate how these quantities leave their imprints in holographic QCD phases. We discover
a rich phase diagram with $n\geq2$ strips and the corresponding mutual and $n$-partite information shows rich structure, consistent with the thermodynamical
transitions, while again revealing some subtleties. Below the deconfinement critical temperature, we find no dependence of the mutual and $n$-partite information on temperature
and chemical potential.}

\section{Introduction}
Recent developments in string theory suggest that the idea of gauge/gravity duality \cite{Maldacena:1997re,Gubser:1998bc,Witten:1998qj} can shed new light on the intriguing connection between  quantum information notions, quantum field theories and spacetime geometries
\cite{Ryu:2006bv,Ryu:2006ef,Lewkowycz:2013nqa,Headrick:2010zt,Headrick:2014cta,VanRaamsdonk:2010pw,Balasubramanian:2013lsa,Hayden:2011ag}.
At the heart of these advancements is the seminal work of Ryu-Takayanagi \cite{Ryu:2006bv,Ryu:2006ef}, which gave a holographic framework for calculating entanglement entropy.  The Ryu-Takayanagi (RT) proposal relates the entanglement entropy of the boundary theory to the area of minimal surfaces, which are homologous to the boundary of the subsystem and extend into the bulk.  The Ryu-Takayanagi
entanglement entropy proposal is one of the most significant and useful suggestions that has emerged from the gauge/gravity duality, providing not
only a deep connection between quantum information and geometry but also opens a new way to calculate and understand other information theoretic quantities
such as the mutual or $n$-partite information
\cite{Hayden:2011ag,Headrick:2010zt,Allais:2011ys}.\\

One of the main and original motivations of the gauge/gravity duality was to understand quantum chromodynamics (QCD) at strong coupling.
Indeed, the lack of any non-perturbative theoretical tool as well as the large computational complexities and expenses involved in lattice simulations
make the gauge/gravity duality the only reliable tool at our disposal to investigate QCD at strong coupling. The main idea here is to construct a
gravity model whose dual boundary theory incorporates the essential features of QCD - such as confinement/deconfinement transition, running of the coupling
constant, temperature dependent Wilson and Polyakov loop expectation values, meson mass spectrum etc -  as accurately as possible. This research area,
sometimes called AdS/QCD or holographic QCD, has attracted a lot of interest lately and by now many holographic models, both string theory inspired
top-down as well as phenomenological bottom-up models, have been constructed which reproduce many QCD properties holographically \cite{Witten9803,
Polchinski0003,Sakai0412,Sakai0507,Kruczenski0311,Karch0205,Klebanov0007,Erlich,Gubser:2008yx,Gubser:2008ny,DeWolfe:2011ts,Gursoy,Gursoy:2010fj,Jarvinen:2015ofa,
Herzog0608,Karch0602,Callebaut:2011ab,Callebaut:2013ria,Dudal:2015wfn,Dudal:2014jfa,Dudal:2018rki,Fang:2015ytf,
Giataganas:2017koz,Panero:2009tv,Cai:2012xh,Noronha:2010hb,Paula,He:2013qq,Yang:2015aia,Arefeva:2018hyo,Knaute:2017opk}.  \\

Interactions in quantum field theories (QFT) via the entanglement in quantum states cause quantum information to be dispersed non-locally across space.
It is therefore of great interest to examine how this structure of shared information changes with the size of the subsystem (the length scale), as it might
provide important information about the confinement structure. With the exception of a few lattice related papers
 \cite{Buividovich:2008kq,Buividovich:2008gq,Itou:2015cyu,Rabenstein:2018bri},
the discussion of entanglement entropy in QCD like gauge theories has been rather limited. The conceptual as well as computational difficulties presented
in the definition of entanglement entropy for interacting field theories make it extremely difficult to get any reliable non-perturbative estimate of the
entanglement entropy relevant for QCD. On the other hand, the holographic RT proposal bypasses the technical difficulties presented in the computation
 of entanglement entropy of quantum field theories and therefore one can use this proposal to find the entanglement structure of QCD. This idea was first initiated
in \cite{Klebanov0709} in the top-down models of gauge/gravity, where a change in the entanglement entropy order was observed (from $\mathcal{O}(N^2)$ to $\mathcal{O}(N^0)$
or vice versa) as the size of the entangling region varied. In particular, a phase transition between connected and disconnected RT entangling surfaces
in the confining background was obtained, causing a non-analyticity in the structure of entanglement entropy. This transition was suggested as an
indication of (de)confinement in \cite{Klebanov0709}. Importantly, the non-analytic behaviour of entanglement entropy in the confining background has received
numerical confirmation from lattice papers as well \cite{Buividovich:2008kq,Buividovich:2008gq,Itou:2015cyu}. The idea of \cite{Klebanov0709} was then applied to
many other town-down confining as well as soft wall models of holographic QCD
\cite{Kola1403,Fujita0806,Lewkowycz,Kim,Ghodrati,Ali-Akbari:2017vtb,Knaute:2017lll,
Anber:2018ohz,Dudal:2016joz}. Only recently the entanglement entropy computations for the phenomenological bottom-up models,
which are somewhat more appropriate to model QCD holographically \cite{Dudal:2018ztm,Gursoy:2018ydr}, were performed
and the results were similar to those reported in  \cite{Klebanov0709}.\\

Most holographic discussion concerning the entanglement structure of QCD has been restricted to the entanglement entropy only (with one subsystem).
However, there are other information theoretic quantities such as the mutual and $n$-partite information that appear when two or more disjoint subsystems
are considered \cite{Calabrese:2009ez,Calabrese:2010he,Hayden:2011ag}.
These quantities do not suffer from the ambiguities associated with the entanglement entropy and can provide more information than the entanglement entropy alone.
For example, they are finite and do not suffer from the usual UV divergences. These quantities, therefore, provide a UV cutoff independent information as opposed
to the entanglement entropy which explicitly contains the UV cutoff. Likewise, the tripartite information ($n=3$), which quantifies the extensivity of the mutual
information ($n=2$), measures how much of the information
that were presented in one part of the system can only be retrieved when having access to both parts of a bipartite system.  \\

For two subsystems $A_1$ and $A_2$, the mutual information is defined as the amount of information that $A_1$ and $A_2$ can share. In terms of the
entanglement entropy it is written as
\begin{eqnarray}
I(A_1, A_2)=S(A_1) + S(A_2) - S(A_1 \cup A_2)
\label{IBdefi}
\end{eqnarray}
where $S(A_1)$, $S(A_2)$ and $S(A_1 \cup A_2)$ are as usual the entanglement entropies of $A_1$, $A_2$ and their union respectively. Form eq.~(\ref{IBdefi})
it is evident that $I(A_1, A_2)$ is zero for two uncorrelated subsystems. Moreover, the subadditivity property of the entanglement entropy
also ensures that $I(A_1, A_2)$ is non-negative \textit{i.e.} $I(A_1, A_2)$ provides an upper bound on the correlation functions between operators in
$A_1$ and $A_2$. For $n$ disjoint subsystems,
the above definition can be generalised to define $n$-partite information
\begin{eqnarray}
I^{[n]}(A_{\{i\}})=  \sum_{i=1}^{n}  S(A_{i}) -  \sum_{i<j}^{n} S(A_{i} \cup A_{j}) + \sum_{i<j<k}^{n} S(A_{i} \cup A_{j} \cup A_{k}) - \dots \nonumber \\  - (-1)^n S(A_1 \cup A_2 \cup ... \cup A_n)
\label{npartitedefi}
\end{eqnarray}
From the above definition, it is clear that $n$-partite information for $n\geq2$ is UV finite. Indeed, because of these desirable features,
the mutual and $n$-partite information have
been used, both from holography as well as from field theory point of view, to probe various interesting physics in a variety of systems, see for example
\cite{Balasubramanian:2011at,
Fischler:2012uv,Kundu:2016dyk,Casini:2015woa,Morrison:2012iz,Fonda:2014cca,Alishahiha:2014jxa,MolinaVilaplana:2011xt,Asplund:2013zba,
Agon:2015ftl,Mozaffar:2015xue,Tanhayi:2015cax,Hosseini:2015gua,Mirabi:2016elb,Hartnoll:2014ppa,Cardy:2013nua,Larkoski:2014pca}.
 \\

The discussion of mutual and $n$-partite information with two or more disjoint intervals in holographic QCD is however relatively new. A partial discussion
appeared in \cite{Ben-Ami:2014gsa}, where only the entanglement entropy phase diagram for $n$ disjoint intervals in a top-down gauge/gravity confining model
was discussed. However, as is
well known, the top-down holographic QCD models usually face several limitations in mimicking real QCD.
In particular, the boundary field theory of these top-down models generally contains additional Hilbert space sectors (arising from the KK modes of extra dimensions),
the non-running coupling constant, undesirable conformal symmetries etc, whose analogue in real QCD do not exist
\cite{Polchinski0003,Sakai0412,Sakai0507,Kruczenski0311,Karch0205,Klebanov0007}. On the other hand, the phenomenological bottom-up holographic QCD models,
 although often formulated in an ad-hoc manner to reproduce desirable features for the boundary QCD and lack solid gauge/gravity duality foundations, can overcome most
of the difficulties presented in the top-down models \cite{Gubser:2008yx,Gubser:2008ny,DeWolfe:2011ts,
Gursoy,Gursoy:2010fj,Jarvinen:2015ofa,
Callebaut:2011ab,Callebaut:2013ria,Dudal:2015wfn,Dudal:2014jfa,Dudal:2018rki,Fang:2015ytf,
Giataganas:2017koz,Panero:2009tv,Cai:2012xh,Noronha:2010hb,Paula,He:2013qq,Yang:2015aia,Arefeva:2018hyo,Knaute:2017opk}. Since, the amount of correlation
between two or more disjoint subsystems is actually characterised by the mutual or $n$-partite information, it is therefore of great interest to study
them in a consistent bottom-up holographic QCD model, in particular, to disclose further the additional information they can provide into the
QCD vacuum structure and confinement mechanism.\\

In this work, our main aim is to fill the above mentioned gap by studying the mutual and $n$-partite information in a self-consistent bottom-up holographic QCD
 model by considering two or more subsystems. By now many phenomenological holographic QCD models have been constructed, see for example
\cite{Gubser:2008yx,Gubser:2008ny,DeWolfe:2011ts,Gursoy,Gursoy:2010fj,Jarvinen:2015ofa,Cai:2012xh,Noronha:2010hb,He:2013qq,Yang:2015aia,Arefeva:2018hyo,Knaute:2017opk},
 with each having own merits and demerits. Here, we consider a particular phenomenological Einstein-Maxwell-dilaton (EMD) holographic QCD model constructed
 in \cite{Dudal:2017max}. An important advantage of
this model is that it can be solved exactly and the full gravity solution can be obtained analytically in terms of the gauge-dilaton coupling $f(\phi)$ and
scale factor $P(z)$ (see section 2 for more information). Moreover, by taking appropriate forms of these two functions, important real QCD properties like
vector meson mass spectrum, confinement/deconfinement  transition, Wilson loop area law etc can be realised holographically in this model as well. Importantly, by
taking suitable forms of $P(z)$ (see eq.(\ref{Aansatz1}) and (\ref{Aansatz2})), this model not only predicts the standard confined and deconfined phases but
also a novel specious-confined phase. This specious confined phase - although not exactly equivalent to the standard confined phase, however, shares many of its
properties - is dual to a small black hole phase in the gravity side, and therefore has a notion of temperature. This, in turn, allow us to investigate temperature
dependent profiles of various observables in the specious confined phase, which can be compared with
lattice QCD confined phase predictions. Indeed, it was found
that thermal behaviour of the quark-antiquark free energy, entropy and the speed of second sound in the specious-confined phase were qualitatively similar
to lattice QCD predictions \cite{Dudal:2017max}. Because of these analytic and interesting features, the model of
\cite{Dudal:2017max} has also been used in other holographic areas such as in holographic complexity \cite{Mahapatra:2018gig,Zhang:2018qnt},
and here we consider it to investigate the mutual and $n$-partite information in holographic QCD.
\\

In this work, following \cite{Dudal:2017max}, we take two different forms $P(z)$. Neither of these forms change the asymptotic structure
of the boundary, however cause non-trivial modifications in the bulk spacetime.  The first form $P(z)=P_1(z)$ gives thermal-AdS/black hole phase transition
 in the gravity side, which in the dual boundary
theory corresponds to standard-confined/deconfined phase transition.  We then examine the entanglement entropy, mutual and $n$-partite information in the obtained
confined/deconfined phases by considering one or more strip geometries of length $\ell$ as the subsystems. In the confined phase, with one strip, the
entanglement entropy again undergoes a connected to disconnected surface transition and exhibits non-analytic behaviour at the critical length $\ell_c$.
These features are the same as suggested by \cite{Klebanov0709}. However, with two or more strips, four minimal area  surfaces
$\{ S_A, S_B, S_C \ \text{and} \ S_D \}$ appear
(see Figure~\ref{ES2equalstrips} for more details) which compete with each other and lead to an interesting phase diagram in the parameter
space of $\ell$ and $x$ ($x$ being the separation length between the strips). In particular, two distinct tri-critical points appear where three minimal
area surfaces coexist. Interestingly, depending on the surfaces involved, the order of the mutual and $n$-partite information may or may not change
at the transition point. This is very different from the entanglement entropy behaviour, where the order always changes at the transition point.
On the other hand, in the deconfined phase, only two minimal area surfaces $( S_A \ \text{and} \ S_B)$ appear with two or more subsystems.
There is again a  phase transition between these two surfaces, and this phase transition is always accompanied by a change in the order
of mutual and $n$-partite information. We further find that these information theoretic results for the confined/deconfined phases qualitatively
remain the same even when chemical potential is considered. We then discuss the holographic QCD phase diagram by examining the mutual and $n$-partite
information in the temperature-chemical potential plane. We find that these quantities capture the signature of thermal-AdS/black hole
 phase transition (or dual confined/deconfined phase transition), suggesting that these non-local observables, like the entanglement entropy,
are also sensitive to the phase transition.

With the second form $P(z)=P_2(z)$, we instead find the small/large stable black hole phases, which on the dual boundary theory correspond to the
specious-confined/ deconfined phases. The entanglement entropy computations of the specious-confined phase reveal a novel connected to connected surface transition
(instead of a connected to disconnected transition), where the order of the entanglement entropy does not change. With two or more strips, unlike in the case
of standard confined phase, the $\ell-x$ phase diagram in the specious-confined phase contains only two phases  $(S_A \ \text{and} \ S_B)$ and the
transition between $S_A$ and $S_B$ is again accompanied by a change in the order
of mutual and $n$-partite information.  Moreover, the mutual information also behaves desirably in the specious-confined phase and satisfies non-negative property.
Interestingly, the mutual and $n$-partite information of the deconfined phase here are similar to
the deconfined phase mutual and $n$-partite information obtained using $P_1 (z)$.
Further, we investigate the dual specious-confined/deconfined QCD phase diagram by studying the
mutual and $n$-partite information in the temperature-chemical potential plane, and again find that these information theoretic quantities are
sensitive to the phase transition.\\

The paper is organised as follow. In the next section, we briefly review our EMD gravity solution and then derive the necessary entanglement entropy formulae.
In section 3, using the first form of $P(z)$, we first examine the thermodynamics of the gravity solution and then discuss the entanglement entropy,
mutual and $n$-partite information in the dual confined/deconfined phases. In section 4, we repeat the computations of section 3 with the second form of $P(z)$.
The last section is devoted to conclusions and an outlook to future research.

\section{Holographic set up}
In this section, we briefly describe the EMD gravity model as well as the holographic entanglement entropy and state only the useful expressions, which will be
important for our investigation in later sections. The holographic EMD gravity model at finite and zero temperature as well various expressions for the entanglement
  entropy have been discussed in great detail in \cite{Dudal:2017max,Dudal:2018ztm}, and we refer the reader to \cite{Dudal:2017max,Dudal:2018ztm} for more
  technical details. \\

The EMD action in five dimensions consists of Ricci scalar $R$, a field strength tensor $F_{MN}$ and a dilaton field $\phi$,
\begin{eqnarray}
&&S_{EM} =  -\frac{1}{16 \pi G_5} \int \mathrm{d^5}x \sqrt{-g} \ \ \left[R-\frac{f(\phi)}{4}F_{MN}F^{MN} -\frac{1}{2}\partial_{M}\phi \partial^{M}\phi -V(\phi)\right] \,.
\label{actionEF}
\end{eqnarray}
where the gauge kinetic function $f(\phi)$  represents the coupling between the gauge field $A_{M}$ and $\phi$, $G_5$ is the Newton constant in five dimensions
and $V(\phi)$ is the potential of the dilaton field. Interestingly, using the following Ans\"atze,
\begin{eqnarray}
& & ds^2=\frac{L^2 e^{2 P(z)}}{z^2}\biggl(-g(z)dt^2 + \frac{dz^2}{g(z)} + dy_{1}^2+dy_{3}^2+dy_{3}^2 \biggr)\,, \nonumber \\
& & A_{M}=A_{t}(z), \ \ \ \ \phi=\phi(z) \,.
\label{metric}
\end{eqnarray}
the equations of motion of the above EMD action can be explicitly solved analytically in terms of a scale function
$P(z)$ \cite{Dudal:2017max,Dudal:2018ztm,He:2013qq,Yang:2015aia},
{
\allowdisplaybreaks
\begin{eqnarray}
&&g(z)=1-\frac{1}{\int_{0}^{z_h} dx \ x^3 e^{-3P(x)}} \biggl[\int_{0}^{z} dx \ x^3 e^{-3P(x)} + \frac{2 c \mu^2}{(1-e^{-c z_{h}^2})^2} \det \mathcal{G}  \biggr],\nonumber \\
&&\phi'(z)=\sqrt{6(P'^2-P''-2 P'/z)}, \nonumber \\
&& A_{t}(z)=\mu \frac{e^{-c z^2}-e^{-c z_{h}^2}}{1-e^{-c z_{h}^2}}, \nonumber \\
&& f(z)=e^{c z^2 -P(z)}\,, \nonumber \\
&&V(z)=-3L^2z^2ge^{-2P}\left[P''+P' \bigl(3P'-\frac{6}{z}+\frac{3g'}{2g}\bigr)-\frac{1}{z}\bigl(-\frac{4}{z}+\frac{3g'}{2g}\bigr)+\frac{g''}{6g} \right] \,.
\label{metsolution1}
\end{eqnarray}}
where
\[
\det \mathcal{G} =
\begin{vmatrix}
\int_{0}^{z_h} dx \ x^3 e^{-3P(x)} & \int_{0}^{z_h} dx \ x^3 e^{-3P(x)- c x^2} \\
\int_{z_h}^{z} dx \ x^3 e^{-3P(x)} & \int_{z_h}^{z} dx \ x^3 e^{-3P(x)- c x^2}
\end{vmatrix}.
\]
The above (Einstein frame) gravity solution corresponds to a black hole with a horizon at $z=z_h$. This solution is obtained by using the boundary condition
that $\lim_{z \rightarrow 0}g(z)=1$ at the asymptotic boundary $z=0$, and that $g(z_h)=0$ at the horizon. Here $\mu$ is the chemical potential of
the boundary theory, which is obtained from the asymptotic expansion of the gauge field. The form of coupling function $f(z)$ in eq.~(\ref{metsolution1})
is also arbitrary and we chose $f(z)=e^{c z^2 - P(z)}$ so that the holographic meson mass spectrum of the boundary theory lies on a linear Regge trajectory,
as governed by QCD phenomenology. Similarly, the magnitude of the parameter $c=1.16 \ \text{GeV}^2$ is fixed by matching the holographic meson mass spectrum to
that of lowest lying (heavy) meson states. Let us also note the expressions of black hole temperature and entropy,
\begin{eqnarray}
&&\hspace{-5mm}T= \frac{z_{h}^3 e^{-3 P(z_h)}}{4 \pi \int_{0}^{z_h} dx \ x^3 e^{-3P(x)}} \biggl[ 1+\frac{2 c \mu^2 \bigl(e^{-c z_h^{2}}\int_{0}^{z_h} dx \ x^3 e^{-3P(x)}-\int_{0}^{z_h} dx \ x^3 e^{-3P(x)}e^{-c x^{2}} \bigr)}{(1-e^{-c z_h^{2}})^2} \biggr]\,,  \nonumber \\
&& \frac{S_{BH}}{V_3}= \frac{L^3 e^{3 P(z_h)}}{4 G_5 z_{h}^{3}}. 
\label{Htemp}
\end{eqnarray}
where $V_3$ is the volume of the three-dimensional plane.

Another solution of EMD action can be obtained by taking the limit $z_h \rightarrow \infty$, which implies $g(z)=1$.  This solution corresponds to
thermal-AdS (without horizon). The thermal-AdS solution has an asymptotically AdS structure at the boundary $z=0$, however it can have,
depending on $P(z)$, a non-trivial structure in the bulk. The non-trivial structure of thermal-AdS in the IR region is in fact the same reason
for having confinement behaviour in the dual boundary theory \cite{Erlich}.  \\
\\
Let us now briefly discuss the holographic entanglement entropy and its relevant expressions in EMD gravity model. We concentrate only on one entangling
surface, as the mutual and $n$-partite information can be obtained from it using eqs.~(\ref{IBdefi}) and (\ref{npartitedefi}).
According to the RT prescription, the entanglement entropy of the subsystem $A$ is given by the area of the minimal surface $\gamma_A$ which extends from the
AdS boundary into the bulk and shares the same boundary $\partial A$ as the subsystem $A$,
\begin{eqnarray}
S^{EE}=\frac{\text{Area}(\gamma_A)}{4 G_5}\,.
\label{RT}
\end{eqnarray}
Here we consider a strip of length $\ell$ as the subsystem, \textit{i.e.} the strip domain $-\ell/2\leq y_1 \leq \ell/2$, $0\leq y_2 \leq L_{y_2}$ and
$0\leq y_3 \leq L_{y_3}$  defines the entangling surface on the boundary. With the strip subsystem, there are two local minima surfaces of eq.~(\ref{RT}): a (U-shaped)
connected and a disconnected surface \cite{Dudal:2018ztm}.
The entanglement entropy of the connected surface is given by the following expression,
\begin{eqnarray}
S^{EE}_{con} (\ell)=\frac{L_{y_2} L_{y_3} L^3}{2 G_5} \int_{0}^{z_*} dz \ \frac{z_{*}^3}{z^3} \frac{e^{3 P(z)-3 P(z_*)}}{\sqrt{g(z)[z_{*}^6 e^{-6P(z_*)}-z^6 e^{-6P(z)}]}}
\label{SEEcon}
\end{eqnarray}
where $z_*$ is the turning point of the connected minimal area surface and is related to the strip length $\ell$ in the following way
\begin{eqnarray}
\ell=2\int_{0}^{z_*} dz \ \frac{z^3 e^{-3 P(z)}}{\sqrt{g(z)[z_{*}^6 e^{-6P(z_*)}-z^6 e^{-6P(z)}]}}\,.
\label{lengthSEEcon}
\end{eqnarray}
On the other hand, the entanglement entropy for the disconnected surface is given by
\begin{eqnarray}
S^{EE}_{discon}=\frac{L_{y_2} L_{y_3} L^3}{2 G_5} \biggl[ \frac{e^{3 P(z_d)}}{2 z_{d}^3} \ell + \int_{0}^{z_d} dz \ \frac{e^{3 P(z)}}{z^3\sqrt{g(z)}} \biggr]
\label{SEEdiscon}
\end{eqnarray}
where $z_d=\infty$ for thermal-AdS background and $z_d=z_h$ for AdS black hole background. The first term in eq.~(\ref{SEEdiscon}) comes from the entangling
surface along the horizon and therefore does not contribute for the thermal-AdS background. Hence, $S^{EE}_{discon}$ is actually independent of $\ell$ for the
thermal-AdS background. As found in \cite{Dudal:2018ztm}, this behaviour of disconnected entanglement entropy provided several interesting features in the
entanglement entropy phase diagram in the dual confined phase. As we will show shortly this behaviour provides even richer phase structure in the mutual
and $n$-partite information. For the AdS black hole background, however, the first term provides a finite contribution to the holographic entanglement entropy.

\section{Case I: the confined/deconfined phases}
As in \cite{Dudal:2017max,Dudal:2018ztm}, let us first consider the following simple form of the scale function $P(z)$,
\begin{eqnarray}
P(z)=P_{1}(z)=- \bar{a} z^2.
\label{Aansatz1}
\end{eqnarray}
It is easy to observe that $P_{1}(0)\rightarrow 0$, asserting that spacetime asymptotes to AdS. Also,
\begin{eqnarray}
& &V(z)|_{z\rightarrow 0}=-\frac{12}{L^2}+\frac{\Delta(\Delta-4)}{2}\phi^2(z)+\ldots, \nonumber \\
& & V(z)|_{z\rightarrow 0}=2 \Lambda + \frac{m^2 \phi^2}{2}+\ldots
\label{Vcase1exp}
\end{eqnarray}
where $m^2=\Delta(\Delta-4)$ with $\Delta=3$, satisfying the well known relation of the gauge/gravity duality. The parameter $\bar{a}= 0.145$ is fixed by
requiring the transition temperature $T_c$ of the thermal-AdS/black hole (or the dual confinement/deconfinement) phase transition to be
around $270 \ \text{MeV}$ at $\mu=0$ in the pure glue sector, as is observed in large $N$ lattice QCD \cite{Lucini:2003zr}.

\subsection{Black hole thermodynamics}
\begin{figure}[h!]
\begin{minipage}[b]{0.5\linewidth}
\centering
\includegraphics[width=2.8in,height=2.3in]{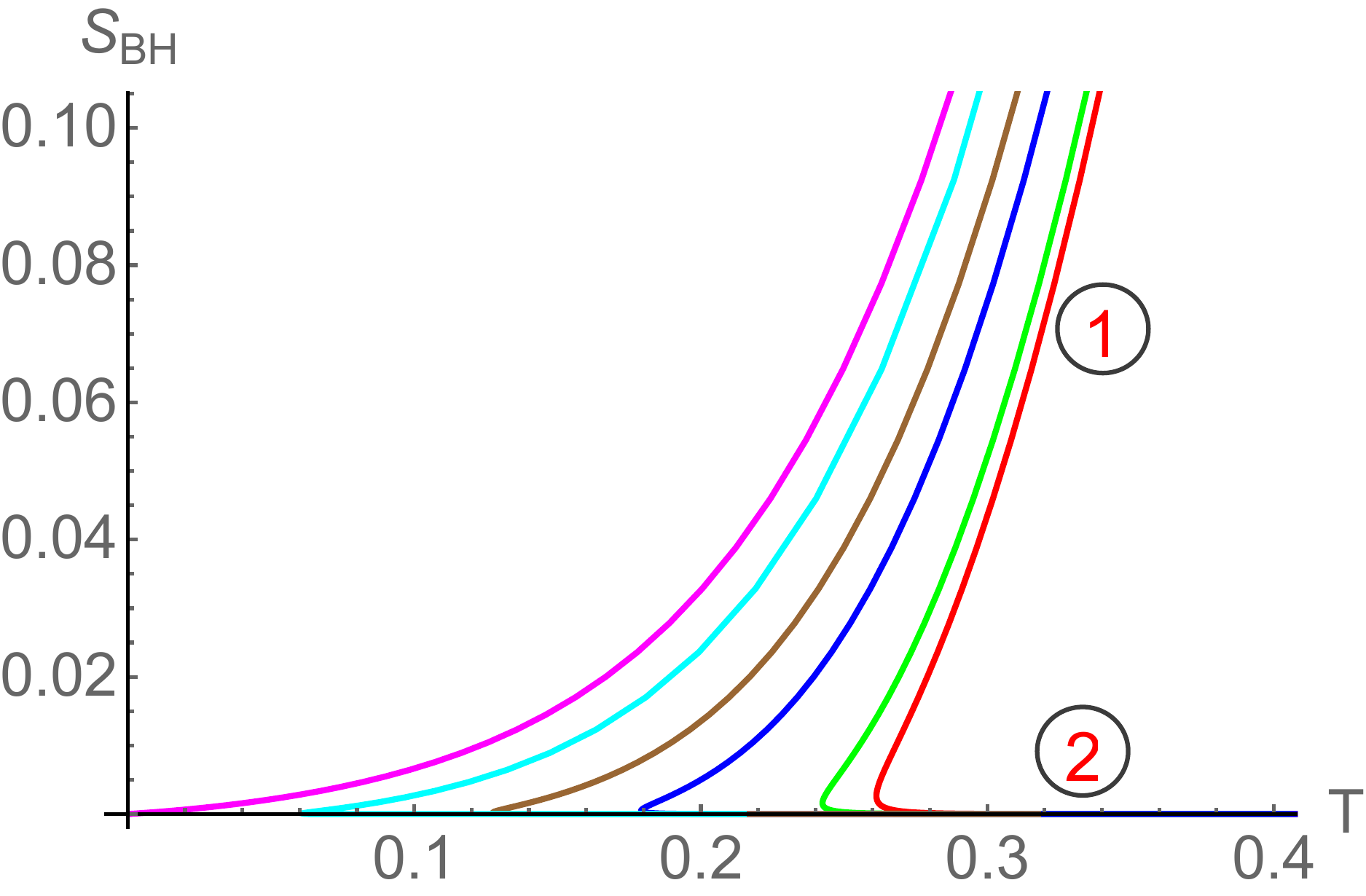}
\caption{ \small $S_{BH}$ as a function of $T$ for various values of $\mu$. Here red, green, blue, brown, cyan and magenta curves correspond
 to $\mu=0$, $0.2$, $0.4$, $0.5$, $0.6$ and $0.673$ respectively. In units \text{GeV}.}
\label{TvsSBHvsMucase1}
\end{minipage}
\hspace{0.4cm}
\begin{minipage}[b]{0.5\linewidth}
\centering
\includegraphics[width=2.8in,height=2.3in]{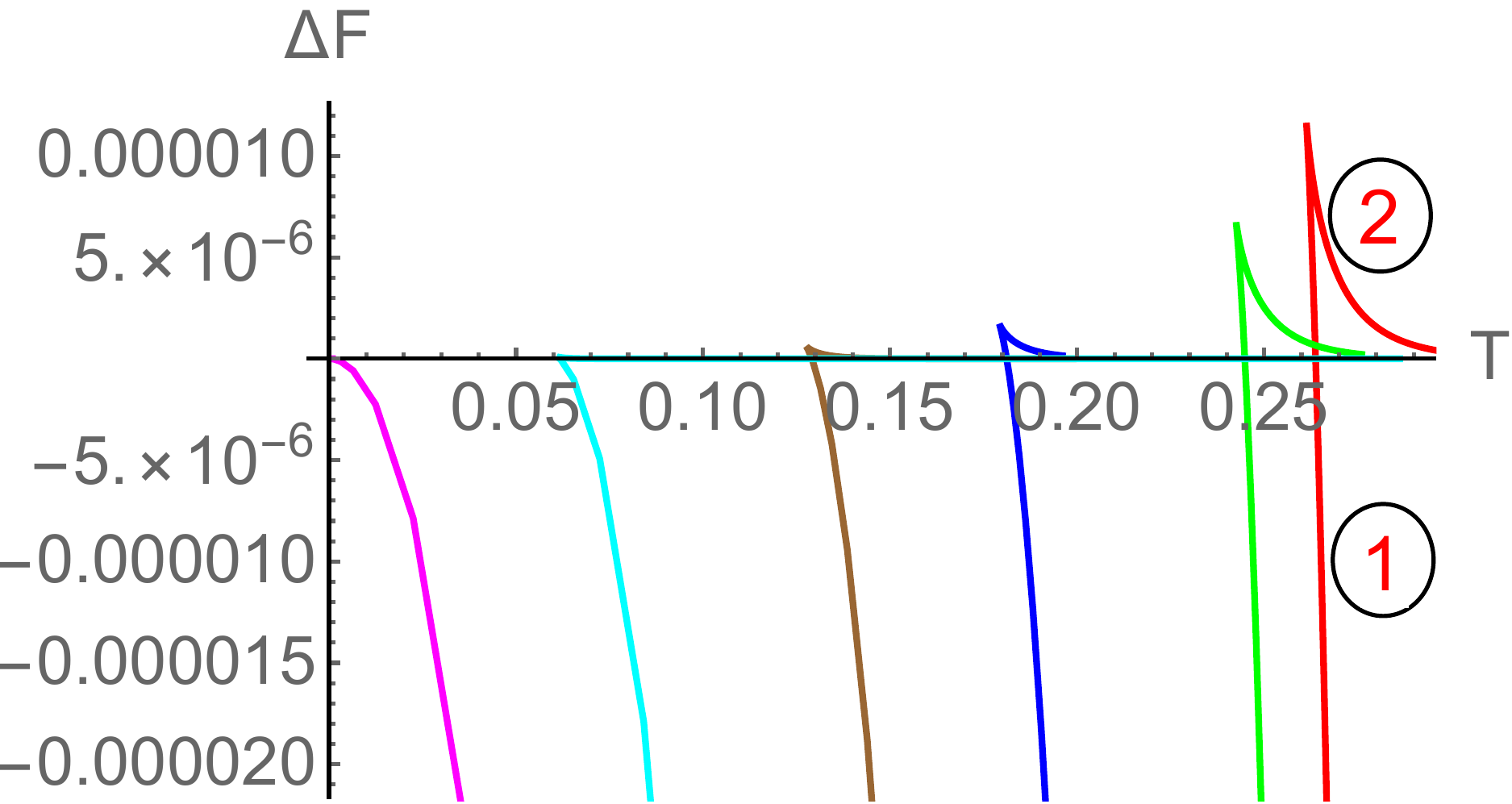}
\caption{\small $\triangle F = F_{\text{Black hole}}-F_{\text{Thermal-AdS}}$ as a function of $T$ for various values of  $\mu$. Here red, green, blue, brown, cyan and magenta curves correspond
to $\mu=0$, $0.2$, $0.4$, $0.5$, $0.6$ and $0.673$ respectively. In units \text{GeV}.}
\label{TvsFBHvsMucase1}
\end{minipage}
\end{figure}
The thermodynamics of the gravity solution with eq.~(\ref{Aansatz1}) has been discussed in \cite{Dudal:2017max} and here we just briefly highlight
its main features. The thermodynamic results are shown in Figures~\ref{TvsSBHvsMucase1} and \ref{TvsFBHvsMucase1}. For small values of $\mu$, there appears a
minimum temperature $T_{min}$ below which no black hole solution exists whereas above $T_{min}$
two black hole solutions -- a large stable black hole (marked by $\circled{1}$) and a small unstable black hole (marked by $\circled{2}$) -- exist at each temperature.
The entropy increases with temperature in the large black hole phase indicating its stable nature whereas the entropy decreases with temperature in the small
black hole phase indicating its unstable nature. The phase structure is shown in Figure~\ref{TvsFBHvsMucase1}, where a Hawking/Page type first order phase
transition between large black hole and thermal-AdS phases can be observed. The phase transition takes place at $T_{crit}>T_{min}$.

For higher values of $\mu$, the critical temperature of the Hawking/Page thermal-AdS/black hole phase transition however decreases and the phase transition
stops at a critical chemical potential $\mu_c$. In particular, at $\mu_c$ the unstable small black hole phase disappears and we have a single black hole phase which remains stable
at  all temperatures (indicated by a magenta line in Figures~\ref{TvsSBHvsMucase1} and \ref{TvsFBHvsMucase1}). For this model, we get $\mu_c=0.673 \ \text{GeV}$.
 Since finding an exact value of the QCD critical point in the QCD $(T,\mu)$ plane, if existing, is extremely hard
 \cite{Ratti:2018ksb,deForcrand:2002hgr,Brewer:2018abr}, a reasonable estimate is a few hundred MeV, so we observe our estimate for the critical point lies
 in the same range.
\\

In \cite{Dudal:2017max}, by calculating the free energy of the probe quark-antiquark pair, it was further shown that the above Hawking/Page phase
transition on the gravity side is dual to the confinement/deconfinement phase transition on the dual boundary side. In particular, the large black
hole phase was shown to be dual to the deconfined phase whereas the thermal-AdS phase was shown to be dual to the confined phase. Since the backreaction
of the dilaton field is included in a
self-consistent manner from the beginning in this model and $T_{crit}$ of the obtained
confinement/deconfinement phase transition decreases with $\mu$ as well, a result again in line with
lattice QCD, this model therefore provides a more realistic holographic QCD model compared to the soft and hard wall models.
It is therefore of great interest to investigate how the information theoretic quantities behave in
this more physical holographic QCD model.

\subsection{Entanglement in holographic QCD phases with multiple strips}
The aim of this subsection is to investigate the entanglement entropy, mutual and $n$-partite information in the above constructed
confined/deconfined phases holographically. Unfortunately, analytic results are difficult to obtain and therefore we provide only numerical results here.
We will first discuss the results in the thermal-AdS background and then discuss in the black hole background.

\subsubsection{With thermal-AdS background: one strip}
\begin{figure}[h!]
\begin{minipage}[b]{0.5\linewidth}
\centering
\includegraphics[width=2.8in,height=2.3in]{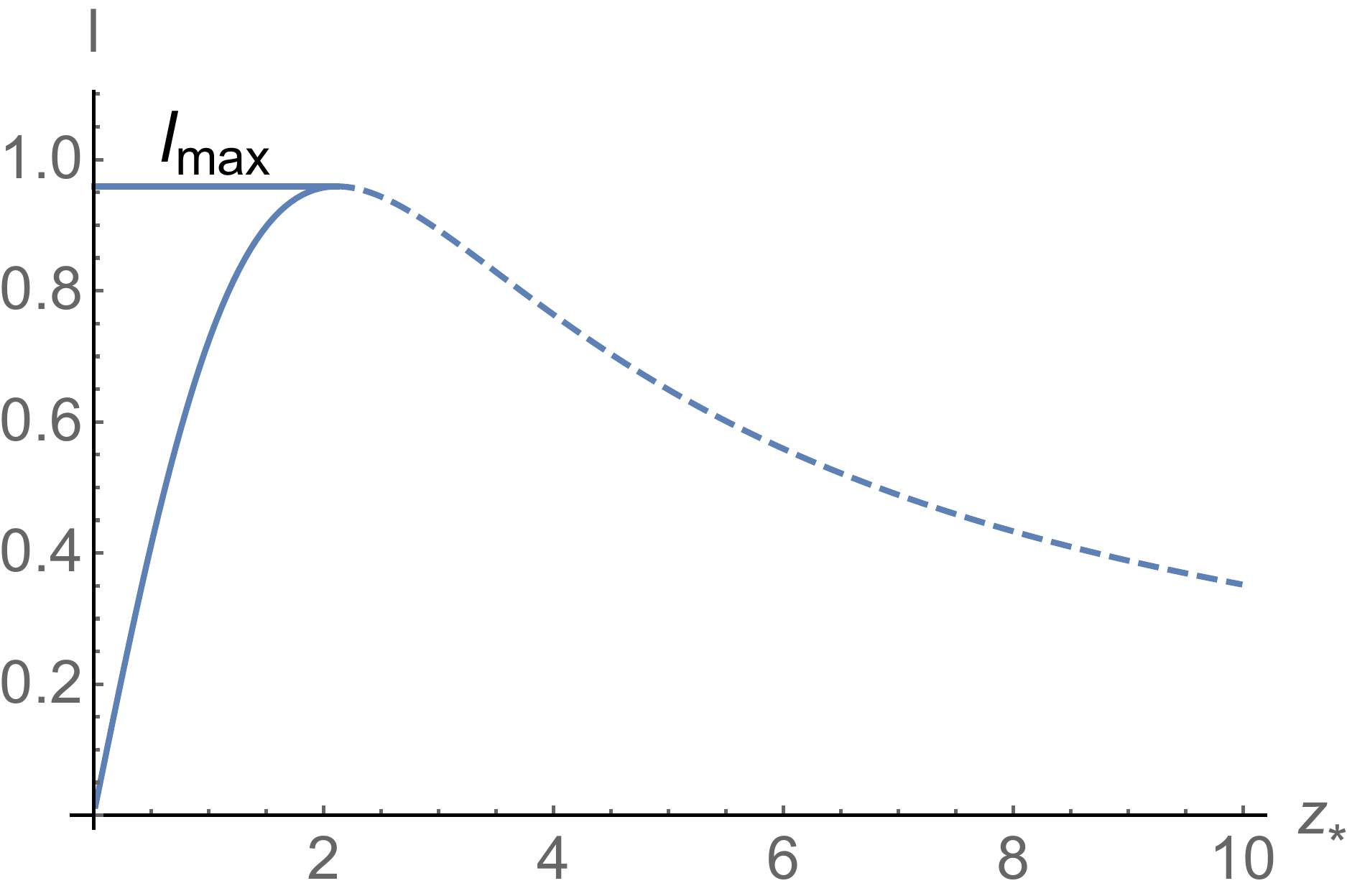}
\caption{ \small $\ell$ as a function of $z_*$ in the thermal-AdS background. In units \text{GeV}.}
\label{zsvslAdScase1}
\end{minipage}
\hspace{0.4cm}
\begin{minipage}[b]{0.5\linewidth}
\centering
\includegraphics[width=2.8in,height=2.3in]{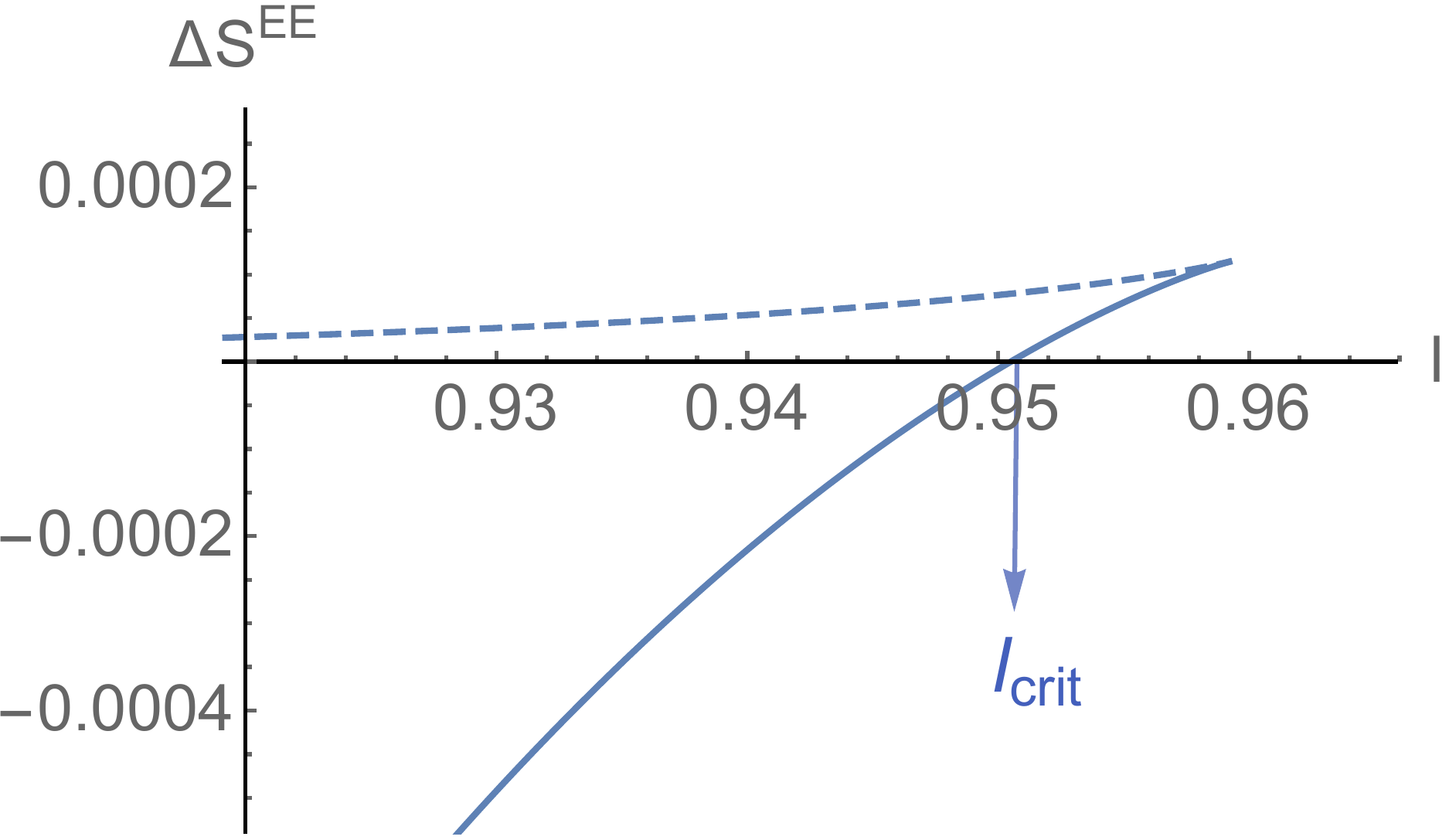}
\caption{\small $\Delta S^{EE}=S^{EE}_{con}-S^{EE}_{discon}$ as a function of $\ell$ in the thermal-AdS background. In units \text{GeV}.}
\label{lvsSEEAdScase1}
\end{minipage}
\end{figure}

Let us first discuss the results in the thermal-AdS background with one strip. This will set the notation and convention for the rest of the section. The results are shown in
Figures~\ref{zsvslAdScase1} and \ref{lvsSEEAdScase1}, where the variation of strip length $\ell$ with respect to the turning point $z_*$ of the
connected entangling surface and the difference between connected and disconnected entanglement entropies $\Delta S^{EE}=S^{EE}_{con}-S^{EE}_{discon}$
respectively are plotted. We find that there exist three RT minimal area surfaces: two connected and one disconnected.
The two connected surfaces, shown by solid and dashed lines in Figure~\ref{zsvslAdScase1}, only exist below a maximum length $\ell_{max}$,
and above this $\ell_{max}$ only the disconnected
entangling surface remains. Out of the two connected surfaces, the one that occurs at small $z_*$ (solid line) always has a lower the entanglement entropy
than the one which occurs at large $z_*$ (dashed line). This suggests that the small $z_*$ connected surface solution is the true minima of eq.~(\ref{SEEcon}) for small $\ell$.

\begin{figure}[h!]
\centering
\includegraphics[width=3.6in,height=1.5in]{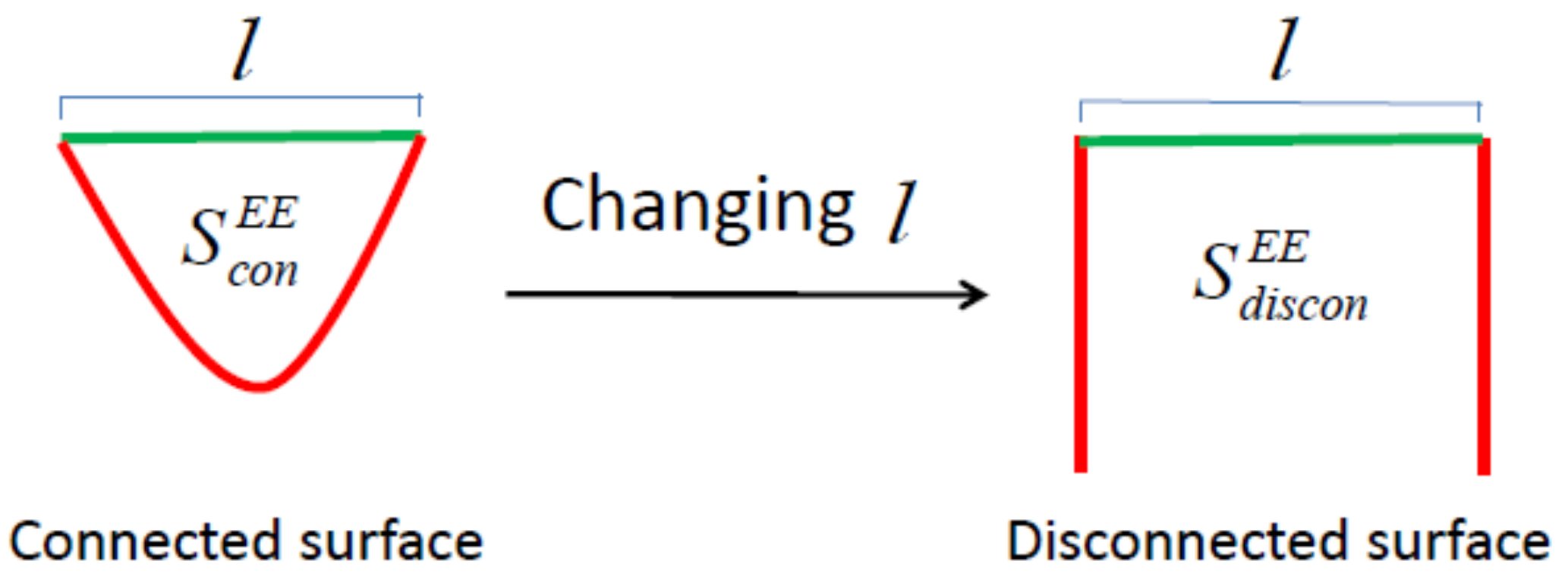}
\caption{ \small Illustration of two minimal area surfaces corresponding to one strip of length $\ell$ in the thermal-AdS background.
A phase transition from connected to disconnected surfaces takes place as the length of the strip increases.}
\label{ESonestripAdS}
\end{figure}

Interestingly, a connected to disconnected entanglement entropy phase transition takes place as we increase the strip size $\ell$. A pictorial illustration
of this is shown in  Figure~\ref{ESonestripAdS}. In particular, $\Delta S^{EE} = S^{EE}_{con}-S^{EE}_{discon}$ changes sign from a negative to positive value as
$\ell$ increased.
This implies that $S^{EE}_{discon}$ has a lower entanglement entropy for large $\ell$ whereas $S^{EE}_{con}$ has a lower entanglement entropy for small $\ell$.
The strip length at which this conncted/disconncted phase transition occur defines an $\ell_{crit}(<\ell_{max}$). For $c=1.16$,
we find $\ell_{crit}\simeq 0.96 \ \text{GeV}^{-1}$.

Since the area of the disconnected surface in the thermal-AdS background is actually independent of $\ell$, the corresponding entanglement entropy
becomes independent of $\ell$ as well. These results can be summarized as,
\begin{eqnarray}
\frac{\partial S^{EE}}{\partial \ell} &\propto &\frac{1}{G_N} = \mathcal{O}(N^2)\quad\text{for}\quad \ell < \ell_{crit}\,, \nonumber \\
&\propto& \frac{1}{G_{N}^{0}} = \mathcal{O}(N^0)\quad\text{for}\quad \ell > \ell_{crit}
\end{eqnarray}
where $N$ is the number of colors of the dual gauge group. \\

The above kind of connected/disconnected phase transition between two entangling surfaces, where the order of the entanglement entropy changes at
the critical point was first found in the top-down models of the gauge/gravity duality in \cite{Klebanov0709,Nishioka:2006gr}. The behavior that the
entanglement entropy scales as $N^0$ for large $\ell$ and as $N^2$ for small $\ell$, led the authors of \cite{Klebanov0709} to interpret the subsystem
length $\ell$ as the inverse temperature ``$T_{c}\propto \frac{1}{\ell_{c}}$''. Indeed, below the deconfinement critical temperature the color confined
degrees of freedom count as order $\mathcal{O}(N^0)$ whereas above the critical temperature the deconfined  gluon (colored) degrees of freedom count as
order $\mathcal{O}(N^2)$, the same counting as estimated by the holographic entanglement entropy. This further suggests that the entanglement entropy can
act as a tool to diagnose confinement.

In \cite{Klebanov0709}, the above type of connected/disconnected entanglement entropy phase transition was suggested to be a characteristic
feature of confining theories. Our analysis further validates this claim as we found identical results for the holographic entanglement entropy, however
now in a genuine self-consistent bottom-up confining model where the running of the coupling constant is incorporated from the beginning. Moreover, a similar type of
non-analyticity in the entanglement entropy has also been observed in SU(2) gauge theory using lattice simulations \cite{Buividovich:2008kq}.
Therefore, such non-analyticity in the structure of entanglement entropy seems to be a universal feature of all confining theories. Interestingly, our
gauge/gravity duality estimate for the length scale at which non-analyticity in the entanglement entropy appears ($\ell_{crit} \sim 0.2 \ \text{fm}$)
is roughly of the same order as estimated by lattice simulations ($\ell_crit \sim 0.5 \ \text{fm}$) \cite{Buividovich:2008kq,Itou:2015cyu}, lending
further support to the idea that the
gauge/gravity duality can yield compelling predictions for QCD-like theories. As we will see shortly, this non-analytic behavior persists even when two or more
subsystems are considered, albeit in those cases various other types of non-analyticity also emerge.

\subsubsection{With thermal-AdS background: two strips}
\begin{figure}[h!]
\begin{minipage}[b]{0.5\linewidth}
\centering
\includegraphics[width=2.8in,height=2.3in]{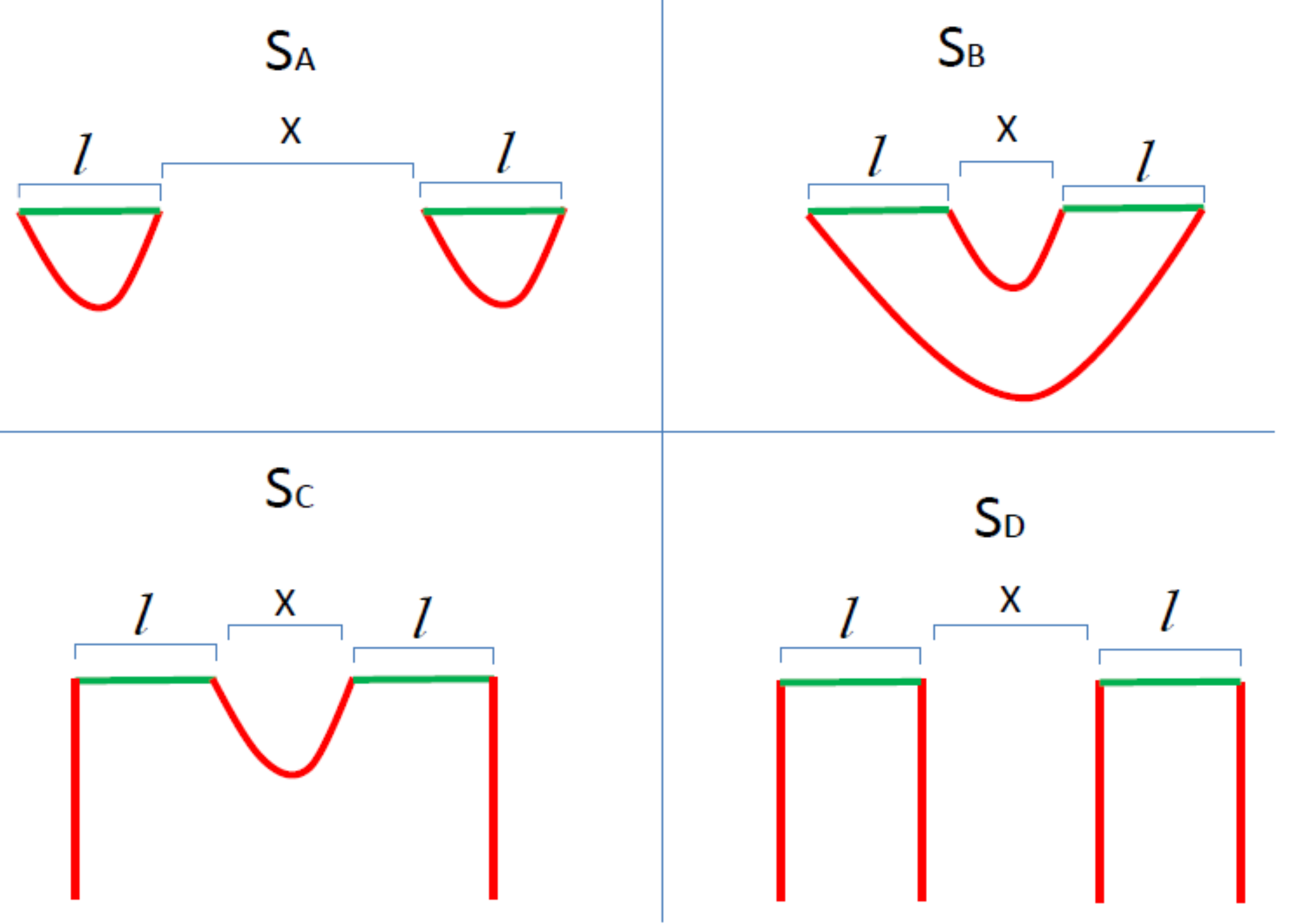}
\caption{ \small Illustration of the four different minimal area surfaces for two strips of equal length $\ell$ separated by a distance
 $x$ in the thermal-AdS background. $S_A$ corresponds to minimal area surface for small $\ell$ and large $x$, $S_B$ corresponds
to minimal area surface for small $\ell$ and small $x$, $S_c$ corresponds to minimal area surface for large $\ell$ and small $x$ and $S_D$
corresponds to minimal area surface for large $\ell$ and large $x$.}
\label{ES2equalstrips}
\end{minipage}
\hspace{0.4cm}
\begin{minipage}[b]{0.5\linewidth}
\centering
\includegraphics[width=2.8in,height=2.3in]{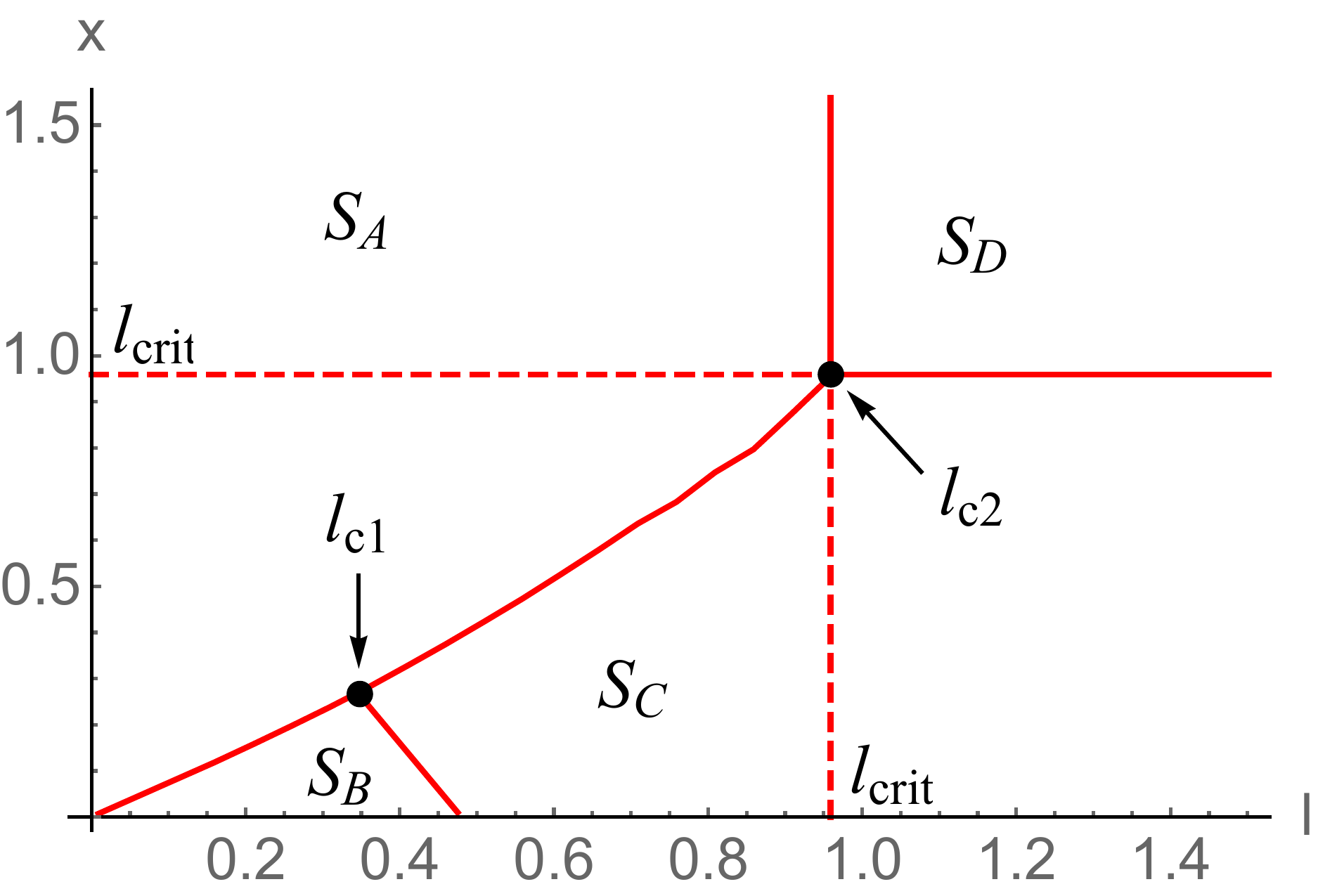}
\caption{\small The phase diagram of various entangling surfaces for the case of two strips of equal length $\ell$ separated by a
distance $x$ in the thermal-AdS background. The four different phases correspond to the four bulk surfaces of Figure~\ref{ES2equalstrips}.
$\ell_{c1}$ and $\ell_{c2}$ indicate two tri-critical points. The vertical and horizontal dashed lines indicate the critical length $\ell_{crit}$.
In units \text{GeV}.}
\label{phaseDiag2stripsAdS}
\end{minipage}
\end{figure}

Having discussed the holographic entanglement entropy with one strip, we now move on to discuss it with two strips. For convenience, we concentrate only
on equal size strips, where $\ell_1=\ell_2=\ell$. Interestingly, depending on the size $\ell$ and
separation $x$ between the two strips, there can now be four minimal area surfaces. These surfaces are shown in Figure~\ref{ES2equalstrips}.
We see that with two strips there can be connected ($S_A$ and $S_B$), disconnected ($S_D$) as well as a combination of connected and disconnected ($S_C$) surfaces.

The entanglement entropy expressions of these four minimal surfaces can be written down as,
\begin{eqnarray}
 S_A  (\ell,x) &=& 2 S^{EE}_{con} (\ell), \hspace{2.0cm} S_B(\ell,x)= S^{EE}_{con} (x) + S^{EE}_{con} (2\ell+x)  \,, \nonumber \\
  S_C  (\ell,x) &=& S^{EE}_{con} (x) + S^{EE}_{discon} , \hspace{0.7cm} S_D(\ell,x)= 2 S^{EE}_{discon} \,.
 \label{eq2strips}
\end{eqnarray}
where as usual $S^{EE}_{con}$ and $S^{EE}_{discon}$ are the entanglement entropies of the connected and disconnected entangling surfaces with one strip.

With two strips, there can be various kinds of phase transitions between different entangling surfaces. The complete phase diagram in the phase space
of $(\ell,x)$ is shown in Figure~\ref{phaseDiag2stripsAdS}. We find that for small $x, \ell \ll \ell_{crit}$ it is the $S_A$ phase which has
the lowest entanglement entropy. The $S_B$ phase, however, becomes more favorable as $\ell$ increases. For a further increase in $\ell$, keeping
$x (\ll \ell_{crit})$ small, a phase transition from $S_B$ to $S_C$ takes place. For $x=0$, this phase transition from $S_B$ to $S_C$
occurs at $\ell=\ell_{crit}/2$, as can also be observed from eq.~(\ref{eq2strips}). In general, the phase transition line between $S_B$ and $S_C$ is
given by $2\ell+x=\ell_{crit}$. With a further increase in both $x, \ell \gg \ell_{crit}$ the $S_D$ configuration ultimately becomes more favorable. In
the near origin region, where both $x, \ell \ll \ell_{crit}$ are small, the results are similar to the CFT expectation. This is interesting considering that
QCD is expected to become conformal at extremely high temperature \cite{Panero:2009tv}. This further provides support to the analogy of strip length
as the inverse temperature.

Interestingly two tri-critical points appear in the two strips phase space, which has no analogue in the one strip case. These two tri-critical
points are shown by black dots in Figures~\ref{phaseDiag2stripsAdS} and are denoted by $\ell_{c1}$ and $\ell_{c2}$ respectively. The first tri-critical
point, where ($S_A$, $S_B$ and $S_C$) phases coexist,  occurs at $(\ell=0.35 \ GeV^{-1}, x = 0.27  \ GeV^{-1})$ whereas the second tri-critical point, where
($S_B$, $S_C$ and $S_D$) phases coexist, occurs at $(\ell=0.96 \ GeV^{-1}, x=0.96  \ GeV^{-1})$. These two tri-critical points again suggest non-analyticity
in the structure of entanglement entropy. Importantly, the order of the entanglement entropy does not change as we go from one phase to another via
the first tri-critical point, however, the order may or may not change as we go from one phase to another via the second tri-critical point.
In particular, if one of the phases involved in the transition is $S_D$ then only the order changes (from $N^2$ to $N^0$ or visa versa),
otherwise, it does not change.

\begin{figure}[h!]
\begin{minipage}[b]{0.5\linewidth}
\centering
\includegraphics[width=2.8in,height=2.3in]{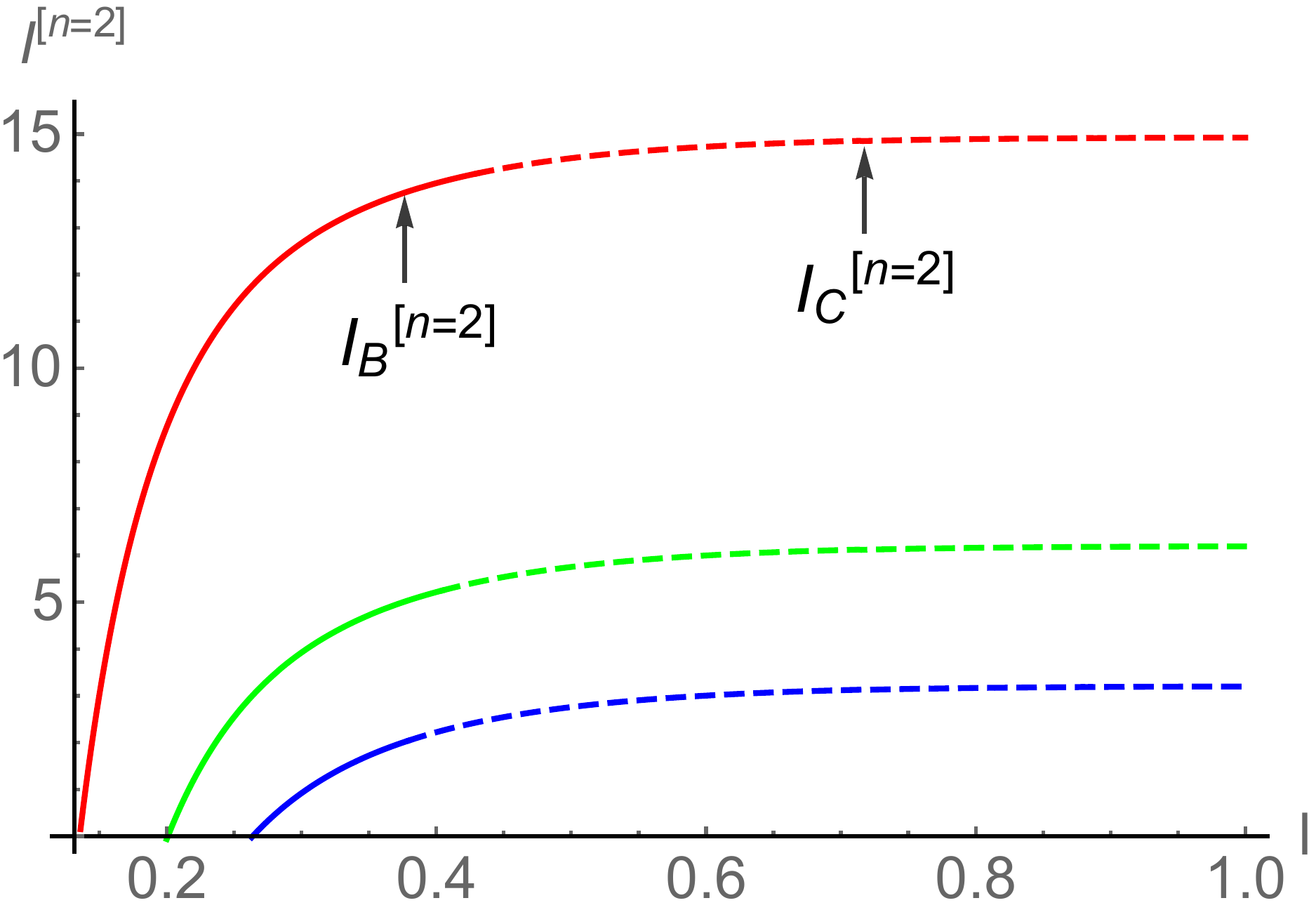}
\caption{ \small Mutual Information of $S_B$ and $S_C$ phases as a function of strip length $\ell$. The solid lines correspond to
$I_{B}^{[n=2]}$ whereas the dashed lines corresponds to $I_{C}^{[n=2]}$.  The red, green and blue lines correspond to separation length $x=0.10$, $0.15$ and $0.20$ respectively.
 In units \text{GeV}. }
\label{Mutualinfo2stripXpt2}
\end{minipage}
\hspace{0.4cm}
\begin{minipage}[b]{0.5\linewidth}
\centering
\includegraphics[width=2.8in,height=2.3in]{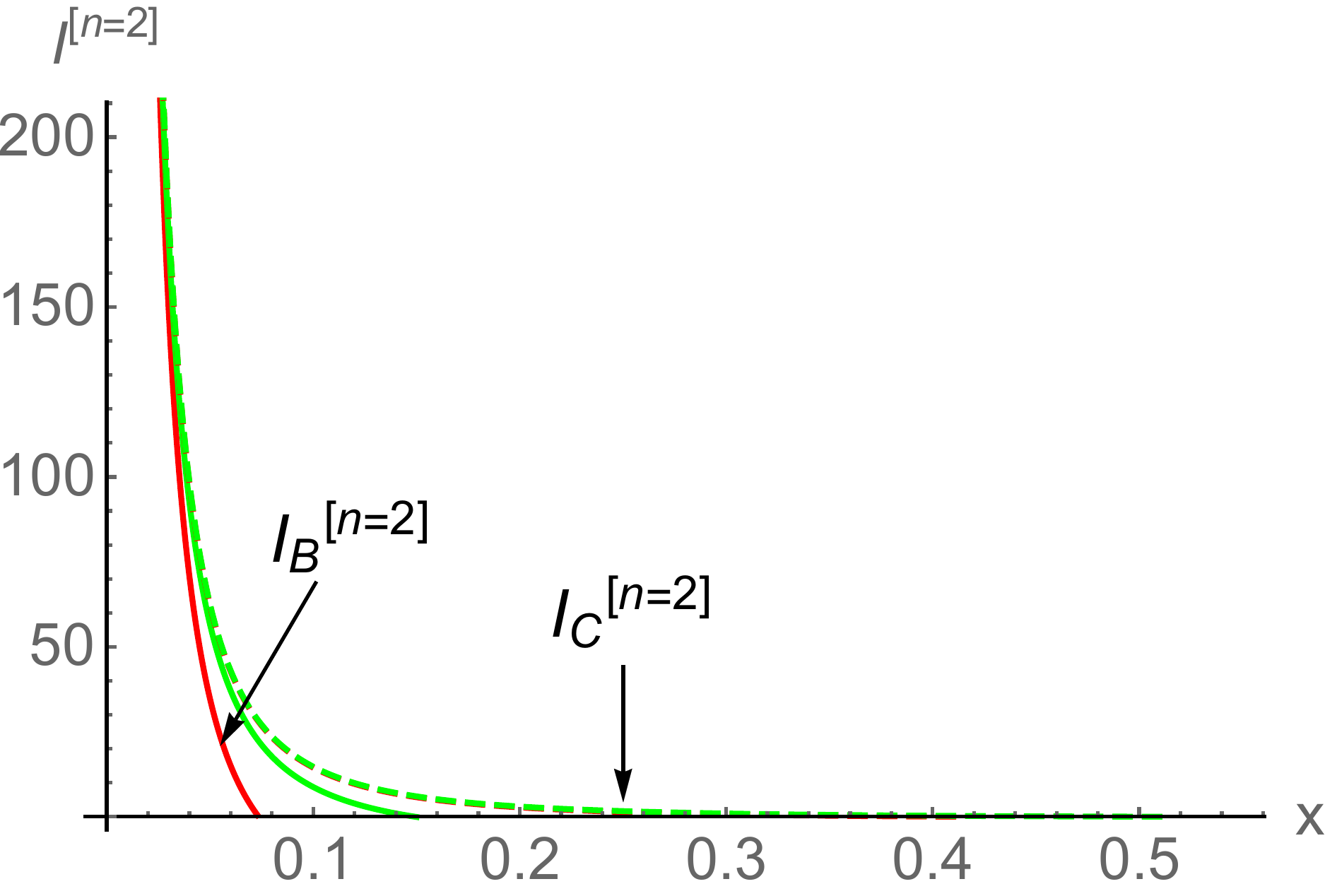}
\caption{\small Mutual Information of $S_B$ and $S_C$ phases as a function of separation length $x$. The solid and dashed lines correspond to $I_{B}^{[n=2]}$ and $I_{C}^{[n=2]}$ respectively.
The solid-red, solid-green, dashed-red and dashed-green lines correspond to strip length $\ell=0.1$, $0.2$, $0.5$ and $0.6$ respectively. In units \text{GeV}. }
\label{Mutualinfo2stripvsx}
\end{minipage}
\end{figure}

It is also interesting to analyze the structure of mutual information in the above phases. For this, let us first note the expressions of mutual
information $I^{[n=2]}=S_1+S_2 - S_1 \cup S_2$ in these four phases
\begin{eqnarray}
 I_A^{[n=2]}  (\ell,x) &=& S^{EE}_{con}(\ell) + S^{EE}_{con}(\ell) - 2 S^{EE}_{con}(\ell) = 0 \, \nonumber \\
 I_B^{[n=2]} (\ell,x)  &=&  S^{EE}_{con}(\ell) + S^{EE}_{con}(\ell)-S^{EE}_{con} (x) - S^{EE}_{con} (2\ell+x) \geq 0  \, \nonumber \\
 I_C^{[n=2]}  (\ell,x) &=& S^{EE}_{con}(\ell) + S^{EE}_{con}(\ell) - S^{EE}_{con} (x) - S^{EE}_{discon} \geq 0 \, \nonumber \\
 I_D^{[n=2]} (\ell,x)  &=& S^{EE}_{discon} + S^{EE}_{discon} - 2 S^{EE}_{discon} =0 \,.
 \label{mutual2strips}
\end{eqnarray}
which also implies
\begin{eqnarray}
\frac{\partial I_A^{[n=2]}}{\partial \ell} \propto \frac{1}{G_{N}^0} = \mathcal{O}(N^0), \ \ \ \ \ \ \frac{\partial I_B^{[n=2]}}{\partial \ell} \propto \frac{1}{G_N} = \mathcal{O}(N^2)  \nonumber \\
\frac{\partial I_C^{[n=2]}}{\partial \ell} \propto \frac{1}{G_N} = \mathcal{O}(N^2), \ \ \ \ \ \ \frac{\partial I_D^{[n=2]}}{\partial \ell} \propto \frac{1}{G_{N}^{0}} = \mathcal{O}(N^0) \,.
\end{eqnarray}

We see that depending on the critical point the order of the mutual information may or may not change as we go from one phase to another.
For example, going from $S_B$ to $S_A$ phase (by increasing $x$) causes a change in the mutual information order (from $\mathcal{O}(N^2)$
to $\mathcal{O}(N^0)$) whereas going from $S_B$ to $S_C$ phase (by increasing $\ell$) does not cause such change in order. If we presume that
the mutual information, like the entanglement entropy, also carries the information about the degrees of freedom of the system then our analysis
suggests that the mutual information can also be used to extract useful information of QCD phases. Moreover, the mutual information also varies
smoothly as we pass from one phase to another. For example, the mutual information connects smoothly between $S_B$ and $S_C$ phases as we pass
through the $S_B-S_C$ critical line. This result is shown in Figure~\ref{Mutualinfo2stripXpt2}, where the variation of $I_B^{[n=2]}$ and $I_C^{[n=2]}$ as a
function of $\ell$ is shown. Similarly, the mutual information also smoothly goes to zero as we approach $S_A$ (or $S_D$) phase  from $S_B$ (or $S_C$) phase.
This is shown in Figure~\ref{Mutualinfo2stripvsx}. These interesting results from holographic might have analogous realization in real QCD. Unfortunately,
unlike the entanglement entropy, we do not have any lattice results for the QCD mutual information yet.
In this regard, these results for the mutual information can be considered as a prediction from holography.

\subsubsection{With thermal-AdS background: $n$ strips}
\begin{figure}[h!]
\begin{minipage}[b]{0.5\linewidth}
\centering
\includegraphics[width=2.8in,height=2.3in]{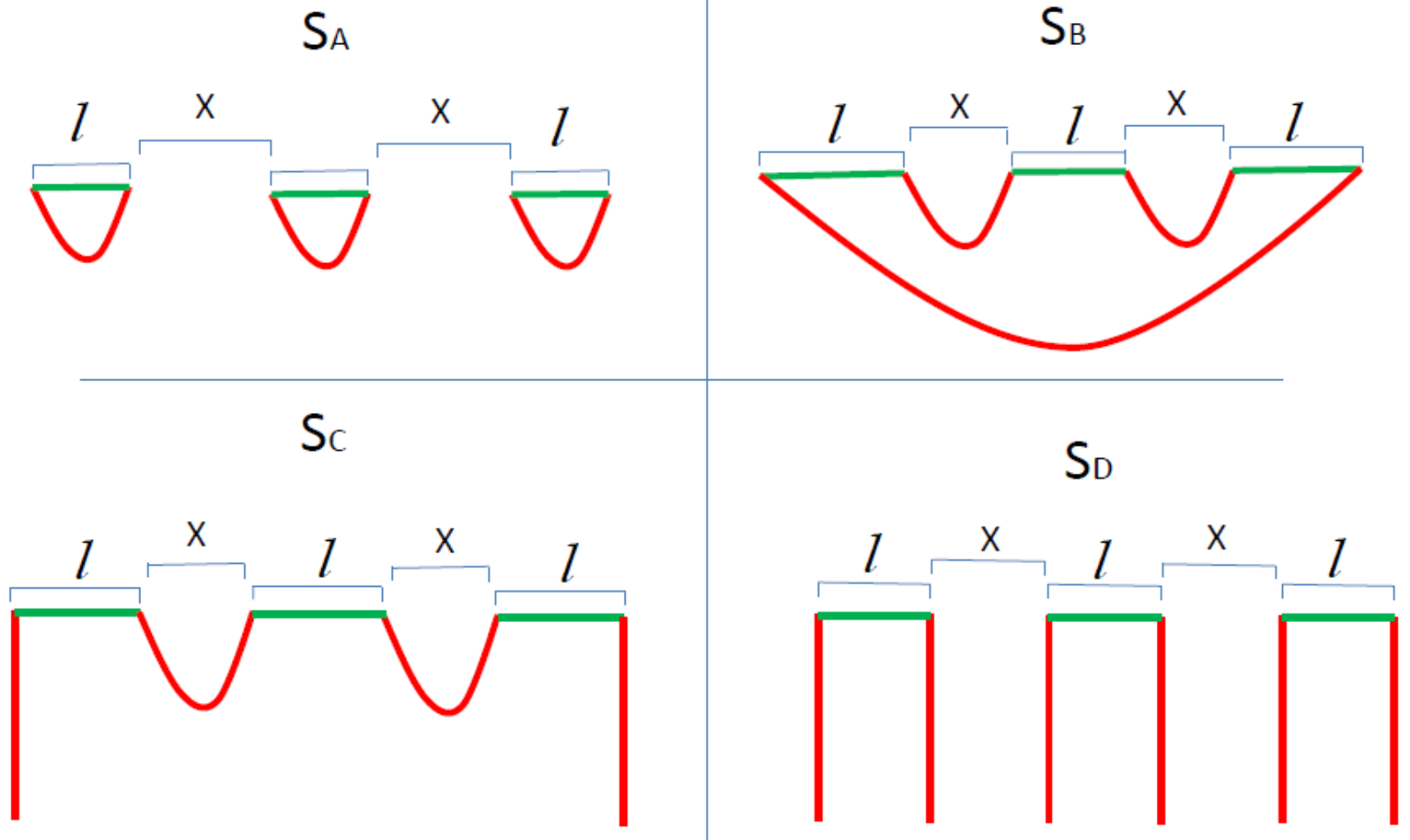}
\caption{\small Illustration of the four different minimal area surfaces for three strips of equal lengths $\ell$ and
separation $x$ in the thermal-AdS background. Similar minimal area surfaces appear for higher $n$ as well.}
\label{ES3equalstrips}
\end{minipage}
\hspace{0.4cm}
\begin{minipage}[b]{0.5\linewidth}
\centering
\includegraphics[width=2.8in,height=2.3in]{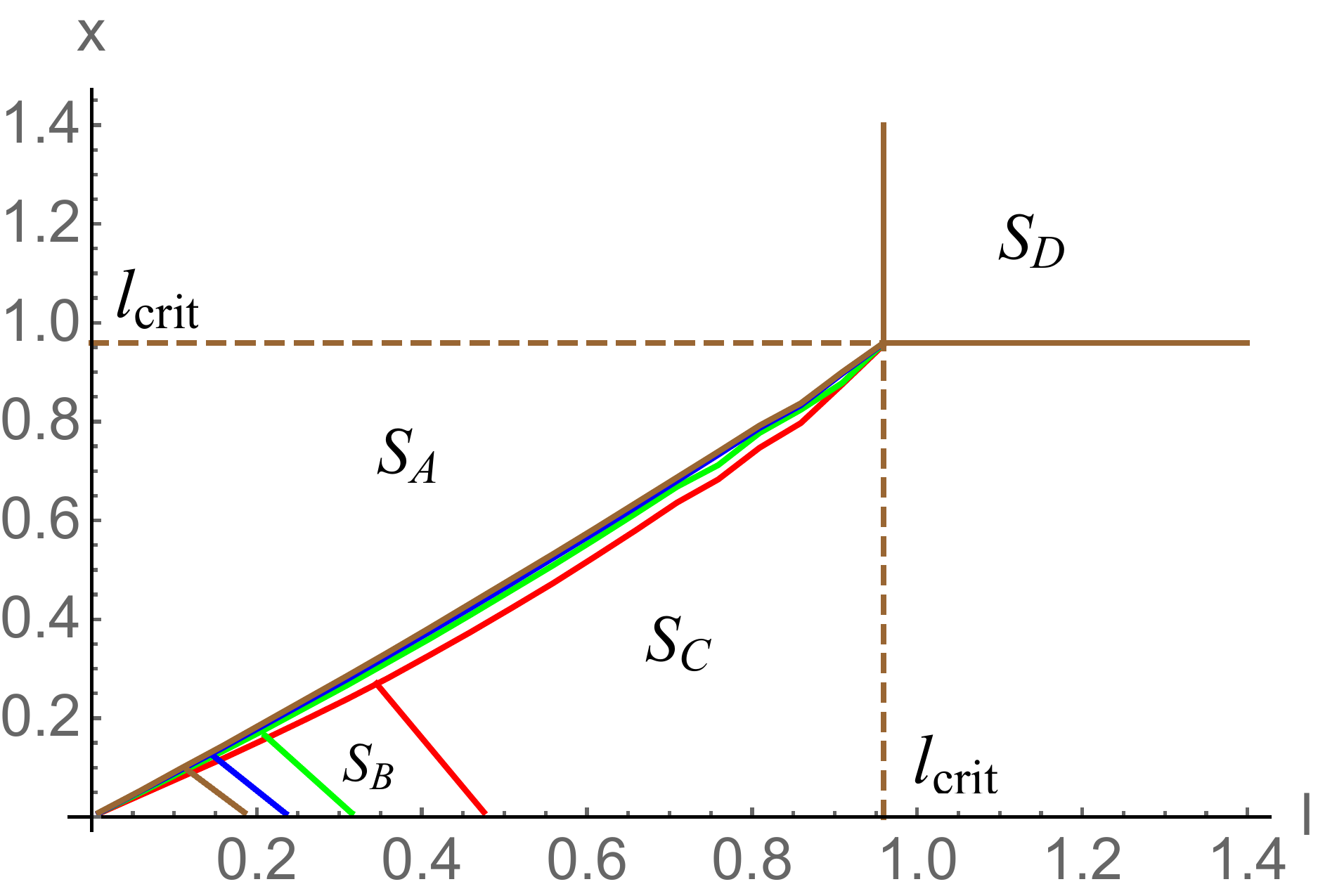}
\caption{\small The $n$ strip phase diagram of various entangling surfaces in the thermal-AdS background. The red, green, blue and brown curves
correspond to $n=2$, $3$, $4$ and $n=5$ respectively. In units \text{GeV}.}
\label{phaseDiagmultistripsAdS}
\end{minipage}
\end{figure}
As in the case of $n=2$ strips, there will again be four minimal area surfaces for $n>2$ strips. In fact, these can be the only minimal area surfaces have
been proved in \cite{Ben-Ami:2014gsa} as well. These surfaces for $n=3$ are shown in Figure~\ref{ES3equalstrips} and can be easily generalized to higher $n$ as well. The
entanglement entropy expressions of these four minimal surfaces are now given by,
\begin{eqnarray}
 S_A  (\ell,x) &=& n S^{EE}_{con} (\ell), \hspace{1.0cm} S_B(\ell,x)= (n-1) S^{EE}_{con} (x) + S^{EE}_{con} (n \ell+(n-1)x)  \,, \nonumber \\
  S_C  (\ell,x) &=& (n-1) S^{EE}_{con} (x) + S^{EE}_{discon} , \hspace{0.7cm} S_D(\ell,x)= n S^{EE}_{discon} \,.
 \label{eqnstrips}
\end{eqnarray}

Our results for different $n$ are shown in Figure~\ref{phaseDiagmultistripsAdS}, where it can be seen that the phase diagram is quite close to $n=2$ case.
In particular, there are again two tri-critical points where three different entangling surfaces coexist. The main difference from the $n=2$ case arises in
the size of region $S_B$. As can be seen, the region in the parameter space where phase $S_B$ is most stable decreases as the number of strips
increases. Although the location of the second tri-critical point ($\ell_{c2}$) does not change with different $n$, however the first tri-critical point
($\ell_{c1}$) moves more and more towards the origin. Therefore, the region $S_B$ gets smaller and smaller and eventually will disappear for $n \rightarrow \infty$.
 The behaviour that $S_B$ shrinks to zero as $n \rightarrow \infty$ can be seen analytically as well. Notice from eq.~(\ref{eqnstrips}) that the transition line between $S_B$ and $S_C$
 satisfies $n \ell +(n-1)x=\ell_{crit}$, which implies $x, \ell \rightarrow 0$ as $n \rightarrow \infty$. Therefore, the size of $S_B$ goes to zero
 as $n \rightarrow \infty$.

\begin{figure}[h!]
\begin{minipage}[b]{0.5\linewidth}
\centering
\includegraphics[width=2.8in,height=2.3in]{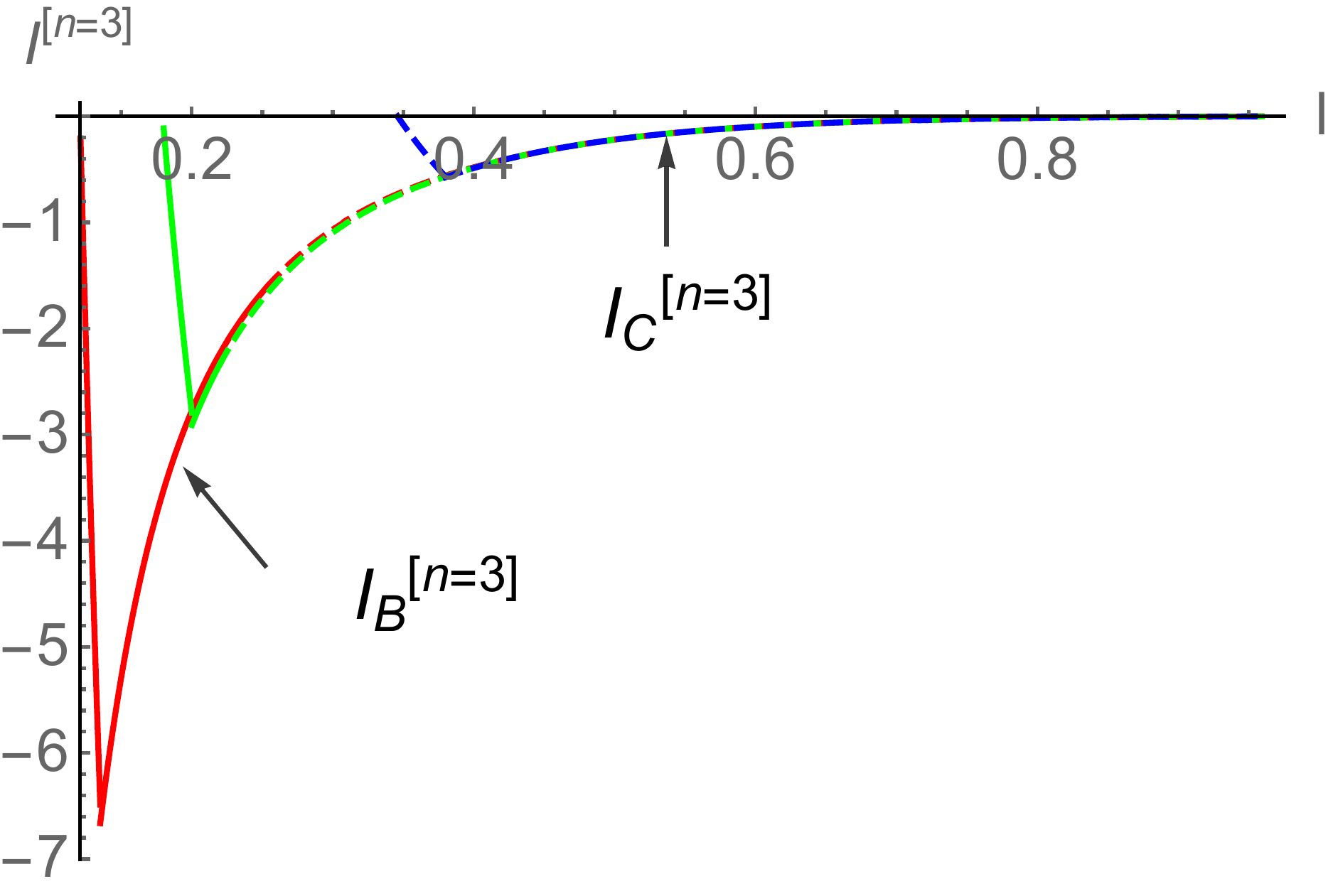}
\caption{ \small $I^{[n=3]}$ for $S_B$ and $S_C$ phases as a function of strip length $\ell$. The solid lines correspond to
$I_B^{[n=3]}$ whereas the dashed lines corresponds to $I_C^{[n=3]}$.  The red, green and blue lines correspond to separation length $x=0.10$, $0.15$ and $0.30$
respectively. In units \text{GeV}. }
\label{I3vsLAdScase1}
\end{minipage}
\hspace{0.4cm}
\begin{minipage}[b]{0.5\linewidth}
\centering
\includegraphics[width=2.8in,height=2.3in]{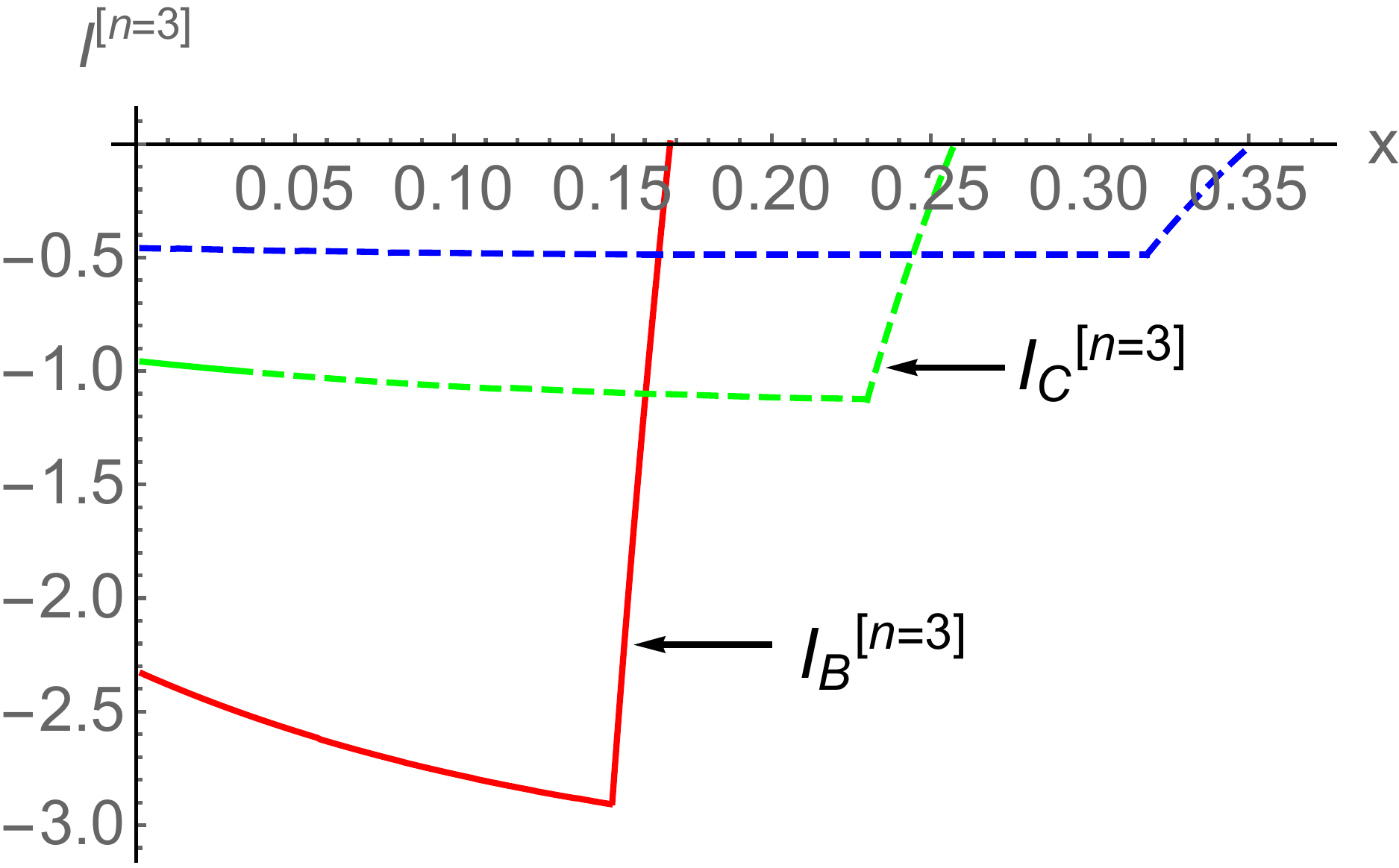}
\caption{\small $I^{[n=3]}$ for $S_B$ and $S_C$ phases as a function of $x$. The solid lines correspond to
$I_B^{[n=3]}$ whereas the dashed lines corresponds to $I_C^{[n=3]}$.  The red, green and blue lines correspond to strip length $\ell=0.2$, $0.3$ and $0.4$
respectively. In units \text{GeV}.}
\label{I3vsXAdScase1}
\end{minipage}
\end{figure}

We now calculate the corresponding $n$-partite information. It is not hard to see that the $n$-partite information again vanishes in
$S_A$ and $S_D$ phases. However, its computation in $S_B$ and $S_C$ phases is now more non-trivial as we need to be careful in evaluating the contributions
of $(n-1, n-2,..., 1)$ intervals entanglement entropy to the $n$-partite information. Therefore, depending on the values of $\ell$ and $x$, the
explicit expression of $n$-partite information will vary. For example, we have
the following expression for $3$-partite information at $x=0.1~GeV^{-1}$,
\[
    I^{[n=3]}=
\begin{cases}
    0, &  0 \leq \ell \leq 0.12 \\
    S^{EE}_{con}(3\ell+2x)-3S^{EE}_{con}(\ell)+2S^{EE}_{con}(x), & 0.12 \leq \ell \leq 0.135 \\
    S^{EE}_{con}(3\ell+2x)-2S^{EE}_{con}(2\ell+x)+S^{EE}_{con}(\ell), & 0.135 \leq \ell \leq 0.254 \\
    S^{EE}_{con}(\ell)-2S^{EE}_{con}(2\ell+x)+S^{EE}_{discon}, & 0.254 \leq \ell \leq 0.430 \\
    S^{EE}_{con}(\ell)-S^{EE}_{discon}, & 0.430 \leq \ell \leq \ell_{crit} \\
    0, & \ell \geq \ell_{crit} \\
\end{cases}
\]

The $3$-partite information as a function of $\ell$ and $x$ is shown in Figures~\ref{I3vsLAdScase1} and \ref{I3vsXAdScase1}.
Interestingly, unlike the mutual information,
the $3$-partite information is always non-positive. This indicates the monogamy of mutual information. At this point, it is instructive to point out that in
quantum theories without gravity duals the $3$-partite information can be negative, positive or zero. However,
in the context of field theories with gravitational dual the 3-partite information is always negative, which points to the monogamous nature of the mutual
information in holographic theories \cite{Hayden:2011ag}. Here, we find a similar result of the $3$-partite information in holographic QCD theories. Moreover,
since the entanglement entropy phase diagram and the transition lines for various entangling surfaces for $n=2$ and $n=3$ are different, they generate various non-analyticities
in the structure of $3$-partite information. This should be
contrasted with the mutual information where no such non-analytic behaviour was seen. Furthermore, we find that the length $\ell$ at which
non-analyticity in the $3$-partite information appears, increases with $x$.

Similar results can be obtained for other values of $n$ as well. We find that the $4$-partite information, on the other hand, is always non-negative.
This is also in line with the holographic suggestion that the $n$-partite information is
positive (negative) for even (odd) $n$ \cite{Alishahiha:2014jxa}. The $4$-partite information also exhibits non-analytic behaviour at various places. Interestingly,
the number of points where non-analyticity appears, increases with $n$. This is an interesting new result, and it would be interesting to find
an analogous realization in real QCD using lattice simulations.

\subsubsection{With black hole background: $1$ strip}
\begin{figure}[h!]
\begin{minipage}[b]{0.5\linewidth}
\centering
\includegraphics[width=2.8in,height=2.3in]{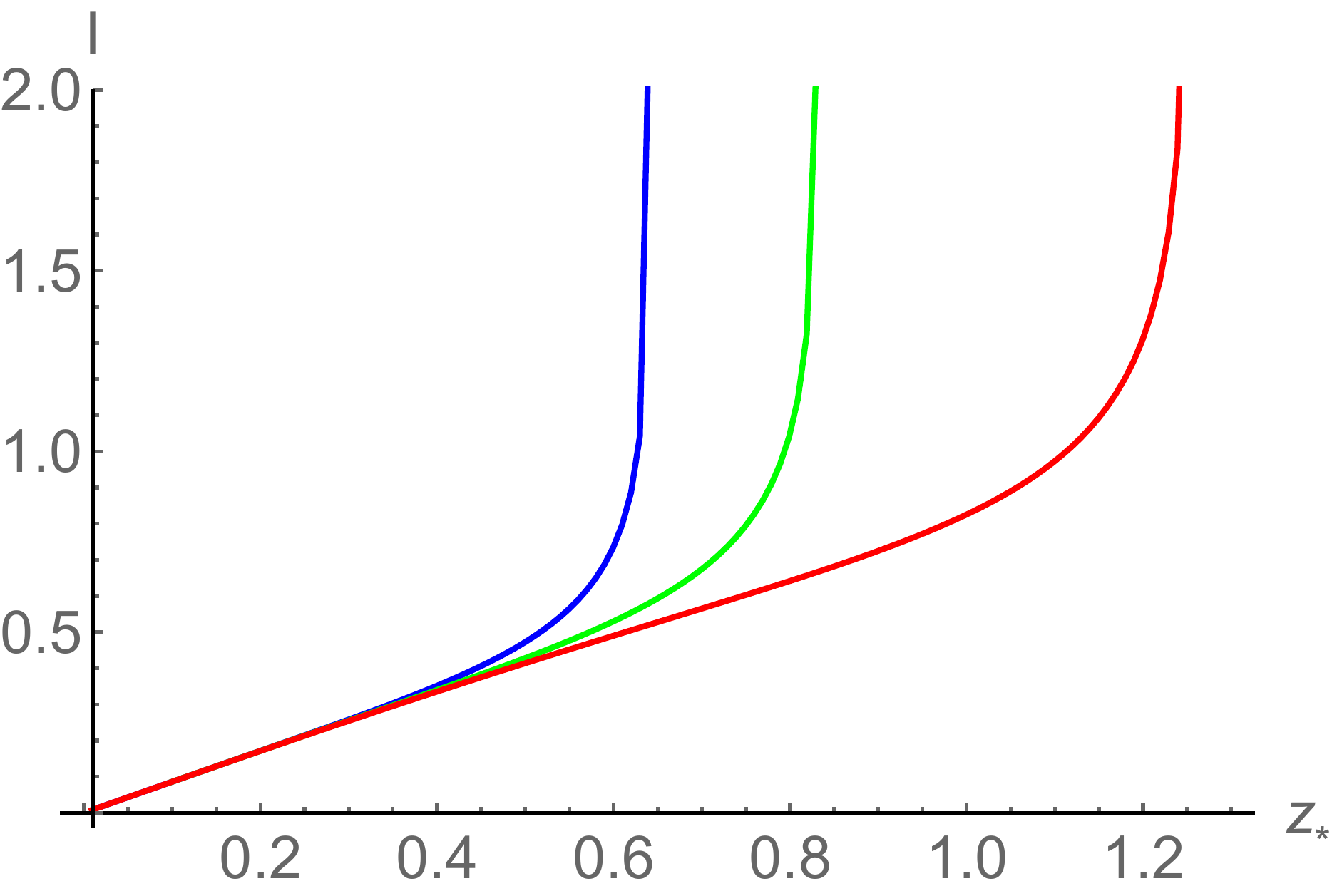}
\caption{ \small $\ell$ as a function of $z_*$ in the deconfinement phase. Here $\mu=0$ and red, green and blue curves correspond to
$T/T_{c}=1.2$, $1.6$ and $2.0$ respectively. In units \text{GeV}.}
\label{zsvslAdSBHMu0case1}
\end{minipage}
\hspace{0.4cm}
\begin{minipage}[b]{0.5\linewidth}
\centering
\includegraphics[width=2.8in,height=2.3in]{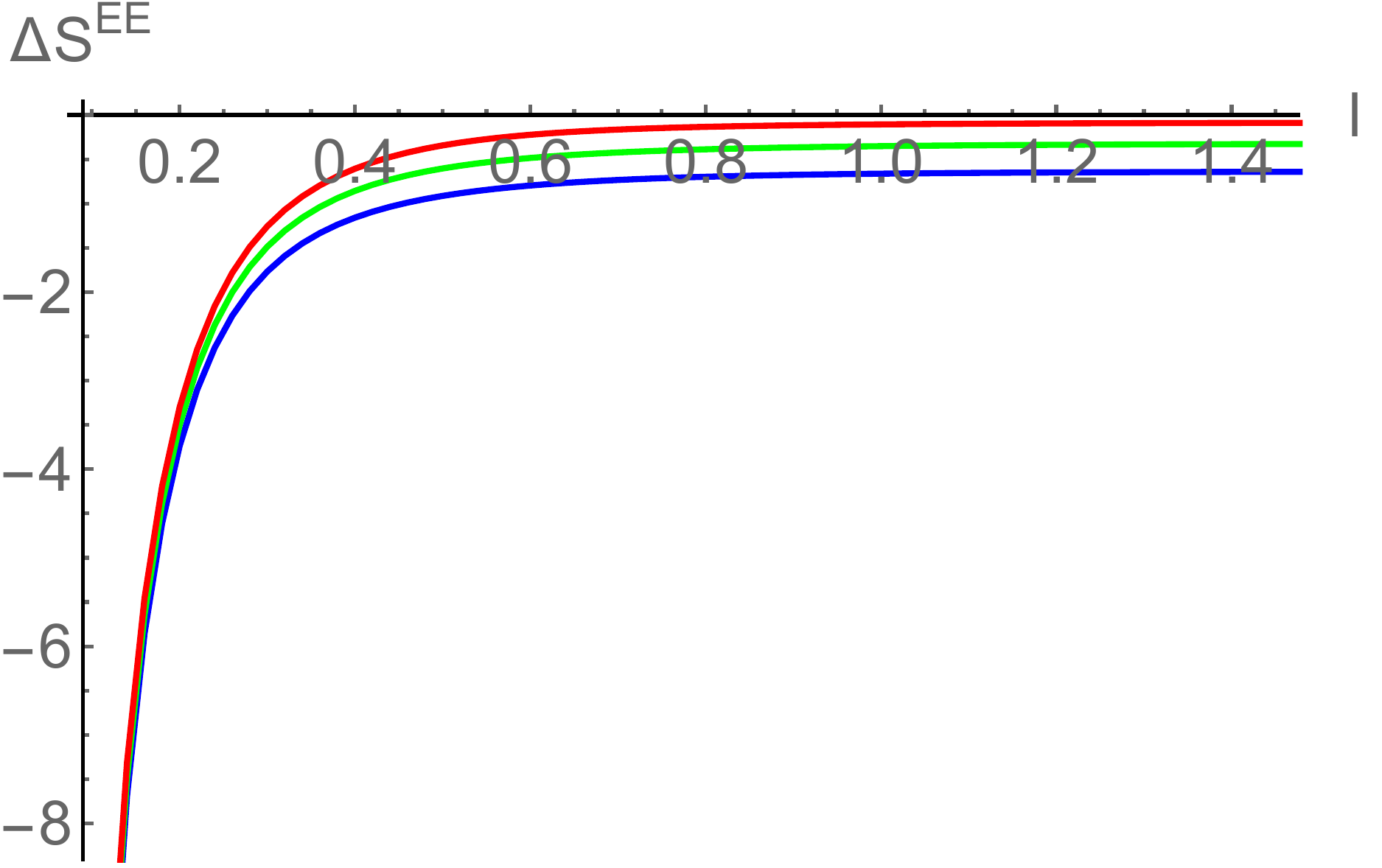}
\caption{\small $\Delta S^{EE}=S^{EE}_{con}-S^{EE}_{discon}$ as a function of $\ell$ in the deconfinement phase. Here $\mu=0$ and red, green
and blue curves correspond to $T/T_{c}=1.2$, $1.6$ and $2.0$ respectively. In units \text{GeV}.}
\label{lvsSEEAdSBhMu0case1}
\end{minipage}
\end{figure}

Having discussed the holographic entanglement entropy in the dual confined phase, we now move on to discuss it in the dual deconfined phase.
Let us consider $n=1$ case first. The results are shown in Figures~\ref{zsvslAdSBHMu0case1} and \ref{lvsSEEAdSBhMu0case1}.\\

In the deconfined phase, the entanglement entropy behavior is quite different. In particular, no maximum length ($\ell_{max}$) appears and the
connected entangling surface solution of eq.~(\ref{lengthSEEcon}) persists for all $\ell$ \textit{i.e} now a one-to-one relation between  $\ell$ and $z_*$ appears.
This is shown in Figure~\ref{zsvslAdSBHMu0case1}. This also indicates that the turning point ($z_*$) of the connected surface shifts more towards the horizon
$z_h$ as the subsystem size increases. Moreover, no phase transition between disconnected/connected surfaces appears as well. In particular, $\Delta S^{EE}$ is
always greater than zero, indicating $S^{EE}_{con}\leq S^{EE}_{discon}$. It is the first term of eq.~(\ref{SEEdiscon}) that makes $S^{EE}_{con} \leq S^{EE}_{discon}$.
The equality sign here is realized only when the size of the subsystem $A$ approaches the full system size, i.e.~when $\ell \rightarrow \infty$.
Not surprisingly, in the limit $\ell \rightarrow \infty$,  the entanglement entropy reduces to the Bekenstein--Hawking black hole entropy,
\begin{eqnarray}
S^{EE}_{con} =S^{EE}_{discon} = \frac{L_{y_2} L_{y_3} L^3}{4 G_5} \frac{e^{3 A(z_h)}}{ z_{h}^3} \ell = S_{BH}
\end{eqnarray}
which is expected from the general property of the entanglement entropy that it reduces to the thermal entropy at finite temperature when the size of
the subsystem $A$ approaches its full system size. Moreover, for the AdS black hole background we always have,
\begin{eqnarray}
\frac{\partial S^{EE}}{\partial \ell} \propto \frac{1}{G_N} = \mathcal{O}(N^2)\,.
\end{eqnarray}

Similar results for the entanglement entropy appear for finite values of $\mu$ as well. In particular, we again have $S^{EE}_{con}\leq S^{EE}_{discon}$,
indicating no phase transition between disconnected/connected surfaces as $\ell$ varies.

\subsubsection{With black hole background: $2$ strips}
\begin{figure}[h!]
\begin{minipage}[b]{0.5\linewidth}
\centering
\includegraphics[width=2.8in,height=2.3in]{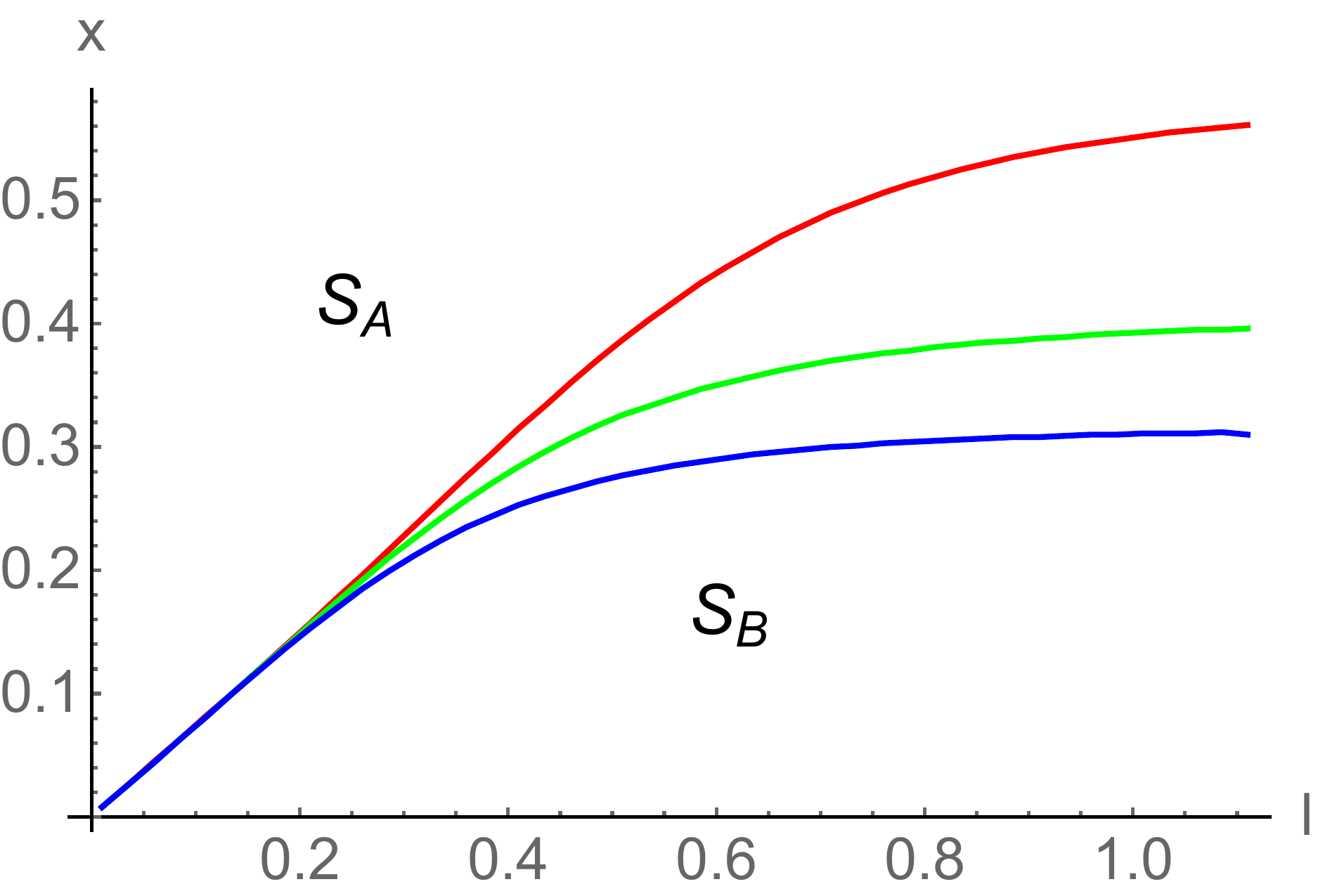}
\caption{ \small  Two strip phase diagram in the deconfined phase for various values of temperature. Here $\mu=0$ is fixed and red, green and blue
curves correspond to $T/T_c=1.2$, $1.6$ and $2.0$. In units \text{GeV}.}
\label{phaseDiag2stripsAdSBHvsTMu0}
\end{minipage}
\hspace{0.4cm}
\begin{minipage}[b]{0.5\linewidth}
\centering
\includegraphics[width=2.8in,height=2.3in]{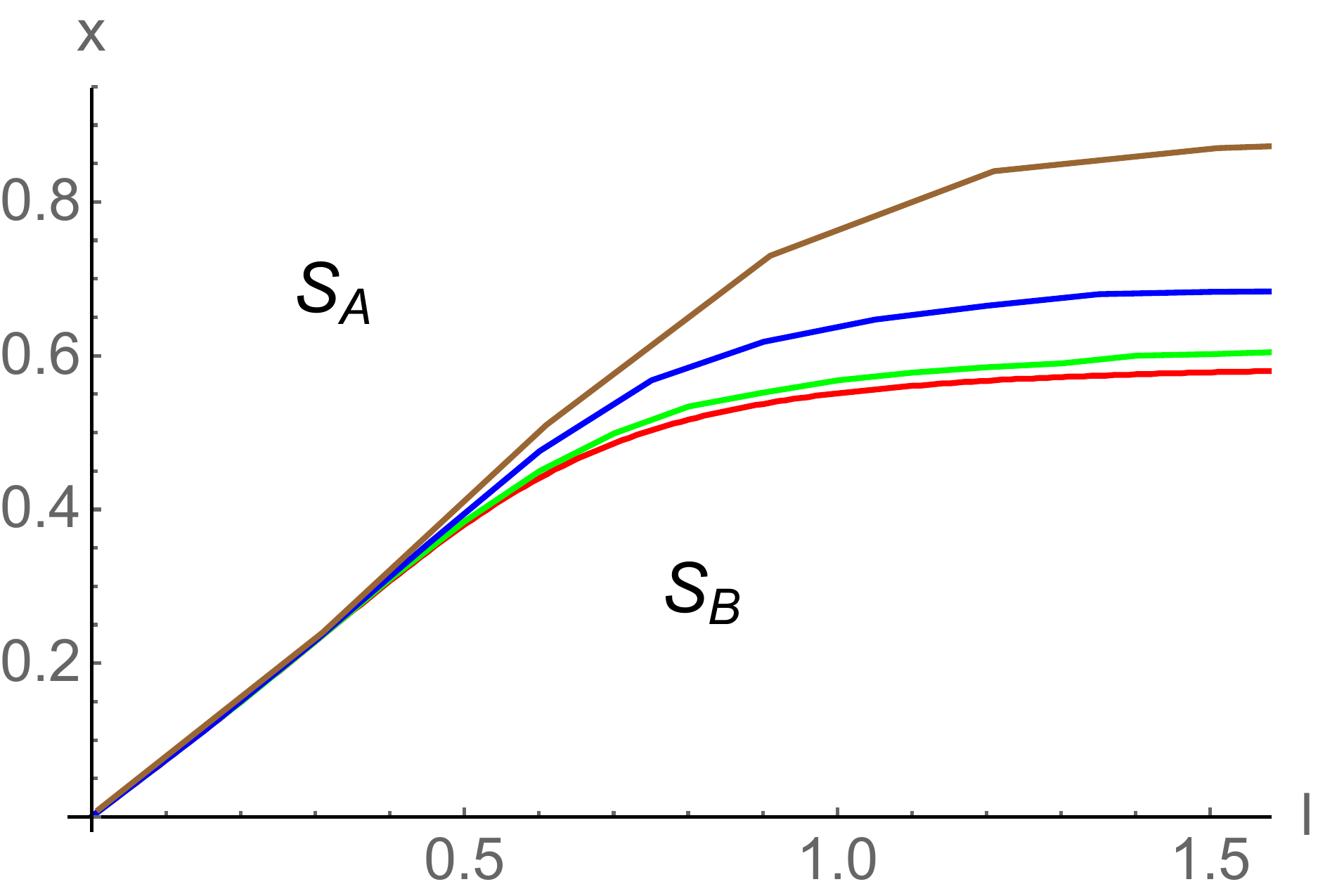}
\caption{\small Two strip phase diagram in the deconfined phase for various values of chemical potential. Here $T=1.2 \ T_c$ is fixed and red,
 green,  blue and brown curves correspond to $\mu=0$, $0.15$, $0.30$ and $0.45$. In units \text{GeV}.}
\label{phaseDiag2stripsAdSBHvsMuT1Pt2}
\end{minipage}
\end{figure}
Since the entanglement entropy of the connected entangling surface is always favored in the black hole background, $S_A$ and $S_B$ are now the only phases
which appear in the deconfinement phase with two strips. Correspondingly, the phase transition appears only between $S_A$ and $S_B$ connected phases as opposed to the
confined phase where the phase transition to disconnected phases ($S_C$ and $S_D$) also occurred. The two strip phase diagram of $S_A$ and $S_B$
at $\mu=0$ for various temperatures is shown in Figure~\ref{phaseDiag2stripsAdSBHvsTMu0}. Again, as in the confined phase, $S_A$ is more favorable at larger
$x$ whereas $S_B$ is more favorable at smaller $x$. Moreover, we find that $S_A$ becomes relativity more favorable than $S_B$ as we increase
the temperature. This can be observed by comparing red and blue lines of Figure~\ref{phaseDiag2stripsAdSBHvsTMu0}, where $T=1.2 \ T_c$ and $T=2.0 \ T_c$
respectively are used.

The above two strips phase diagram persists for a finite $\mu$ as well. This is shown in Figure~\ref{phaseDiag2stripsAdSBHvsMuT1Pt2}, where different
values of $\mu$ at constant temperature $T=1.2 \ T_c$ are considered. Interestingly, higher values of $\mu$ instead try to make $S_B$ more favorable.

\begin{figure}[h!]
\begin{minipage}[b]{0.5\linewidth}
\centering
\includegraphics[width=2.8in,height=2.3in]{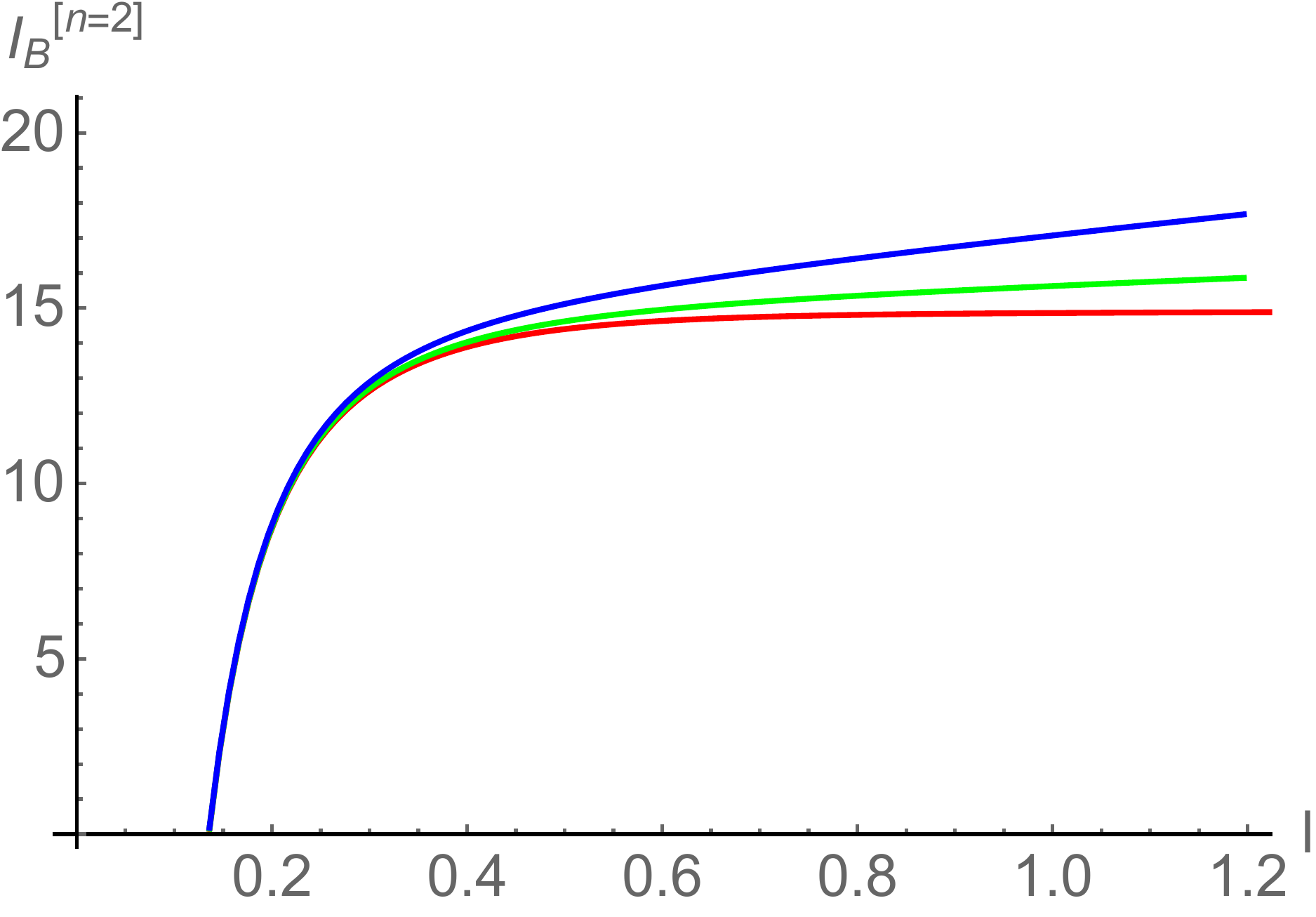}
\caption{ \small Mutual information $I_B^{[n=2]}$ as a function of $\ell$ for various values of temperature in the deconfined background. Here $\mu=0$ and
$x=0.1$ are fixed and red, green and blue curves correspond to $T/T_c=1.2$, $1.6$ and $2.0$ respectively. In units \text{GeV}.}
\label{MutualinfoAdSBh2stripXpt1vsLvsT}
\end{minipage}
\hspace{0.4cm}
\begin{minipage}[b]{0.5\linewidth}
\centering
\includegraphics[width=2.8in,height=2.3in]{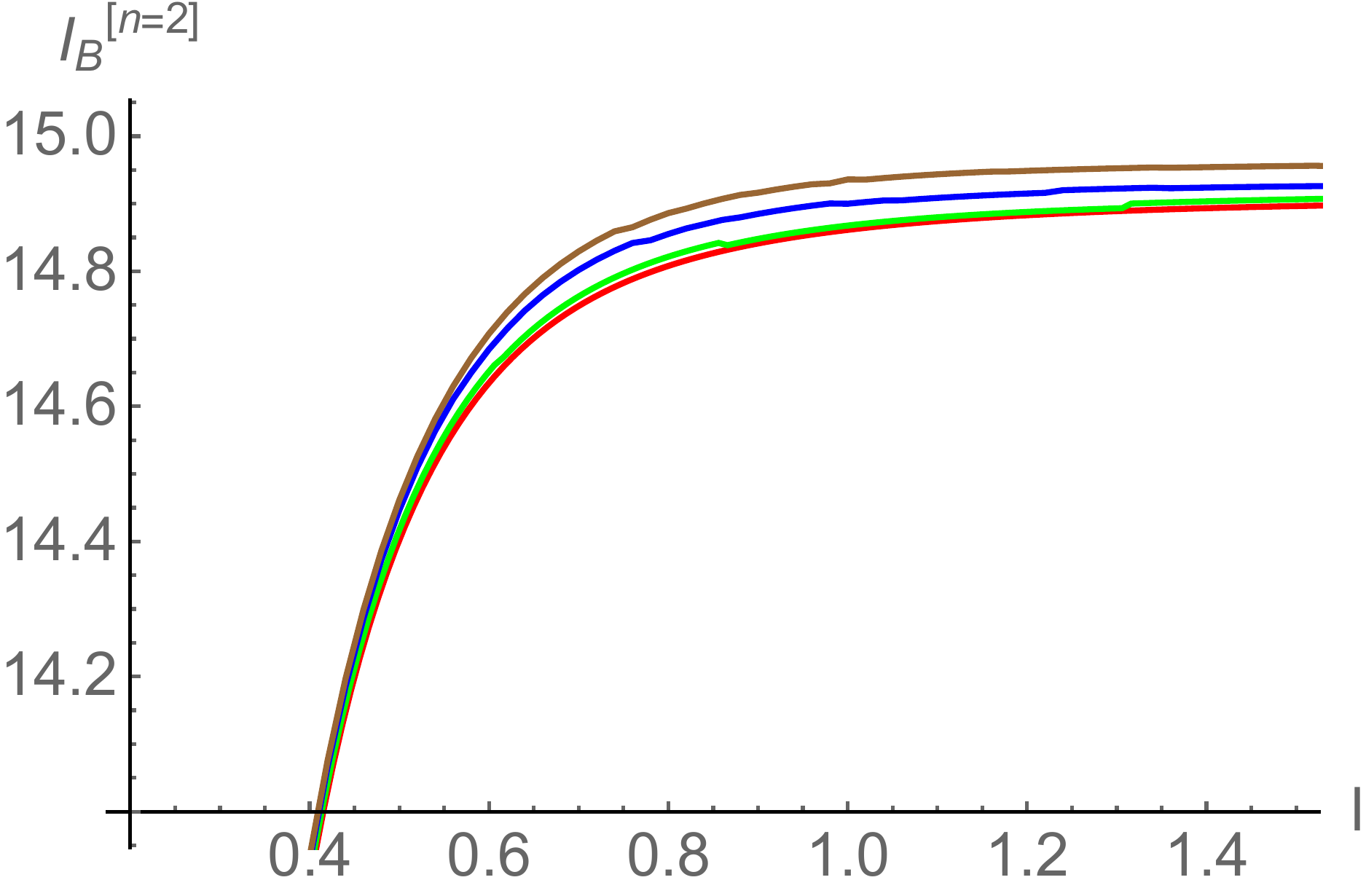}
\caption{\small Mutual information $I_B^{[n=2]}$ as a function of $\ell$ for various values of chemical potential in the deconfined background.
Here $T=1.2 \ T_c$ and $x=0.1$
are fixed and red, green, blue and brown curves correspond to $\mu=0$, $0.15$, $0.30$ and $0.45$ respectively. In units \text{GeV}.}
\label{MutualinfoAdSBH2stripXpt1T1Pt2vsMu}
\end{minipage}
\end{figure}

Next we analyse the behavior of mutual information in $S_A$ and $S_B$. By definition $I_A^{[n=2]}=0$ again (see eq.~(\ref{mutual2strips})). On the other hand,
$I_B^{[n=2]} \geq 0$ and approaches a temperature dependent constant value for large $\ell$. In particular, the large $\ell$ asymptotic value of
$I_B^{[n=2]}$ gets higher for higher
temperatures. This is shown in Figure~\ref{MutualinfoAdSBh2stripXpt1vsLvsT}. Importantly, since $I_A^{[n=2]} \propto \mathcal{O}(N^0)$ and
$I_B^{[n=2]} \propto \mathcal{O}(N^2)$, the order of the mutual information changes as we go from $S_A$ to $S_B$ and visa versa.
The phase transition in the deconfined background
is therefore always accompanied by a change in the order of mutual information as opposed to the confined background where
the order of mutual information may or may not change at the transition line. Although not shown here for brevity, the mutual information also
varies smoothly as we pass from
$S_B$ to $S_A$ by changing $x$.

The mutual information also behaves smoothly when finite chemical potential is considered, and most of the above mentioned results remain
true with chemical potential as well. The main difference appears in the large $\ell$ asymptotic value of $I_B^{[n=2]}$, which gets enhanced with $\mu$. The results
are shown in Figure~\ref{MutualinfoAdSBH2stripXpt1T1Pt2vsMu}. Although we have presented results only for $T=1.2 \ T_c$, however, similar results occur
for other values of $T$ as well.

\subsubsection{With black hole background: $n$ strips}
\begin{figure}[h!]
\begin{minipage}[b]{0.5\linewidth}
\centering
\includegraphics[width=2.8in,height=2.3in]{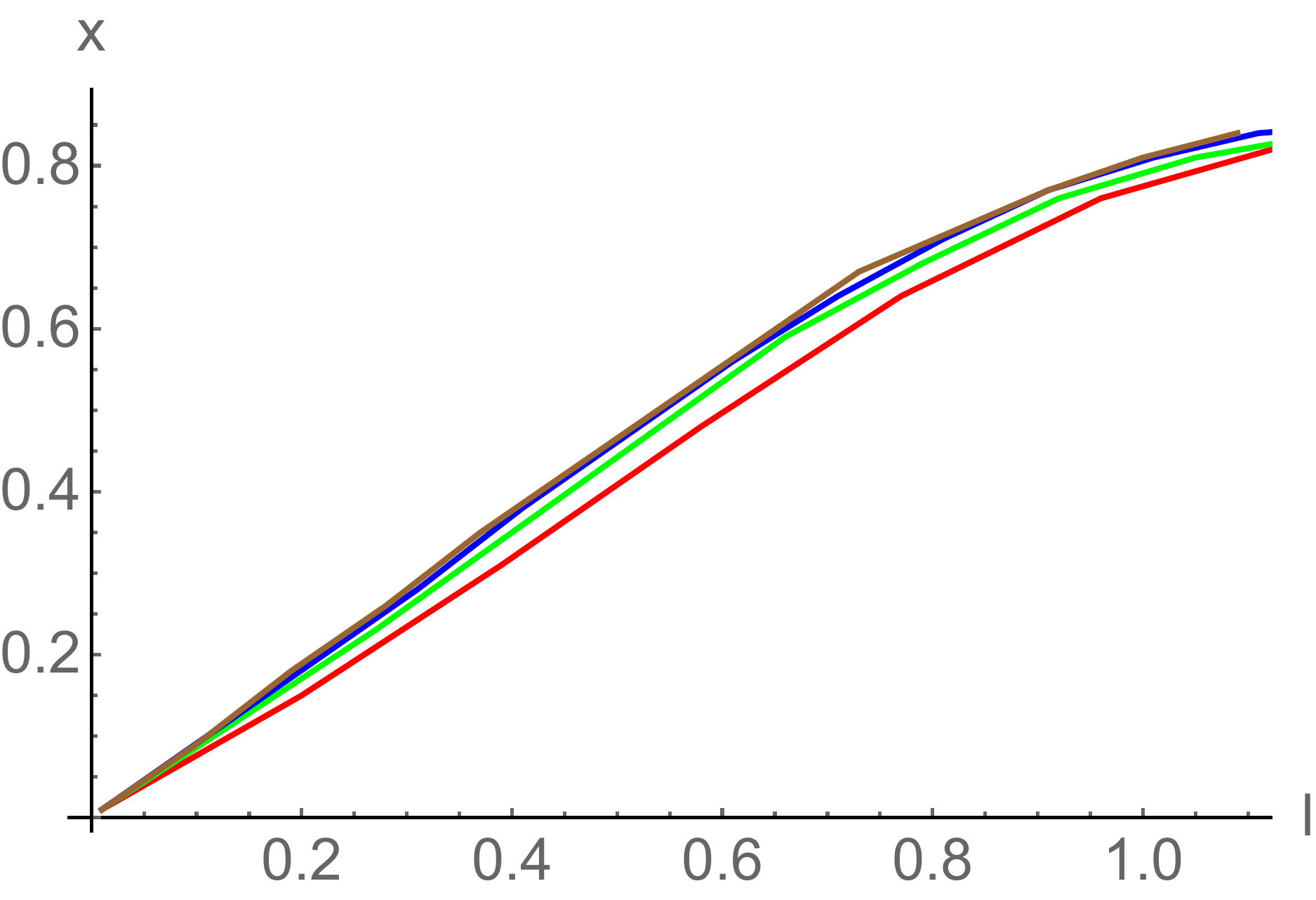}
\caption{ \small \small $n$ strip phase diagram in the deconfined phase. Here $T=1.2 \ T_c$ and $\mu=0.45$ are used and red, green,  blue and brown curves
correspond to $n=2$, $3$, $4$ and $5$ respectively. In units \text{GeV}.}
\label{phaseDiagNstripsAdSBHvsLMu0T1Pt2}
\end{minipage}
\hspace{0.4cm}
\begin{minipage}[b]{0.5\linewidth}
\centering
\includegraphics[width=2.8in,height=2.3in]{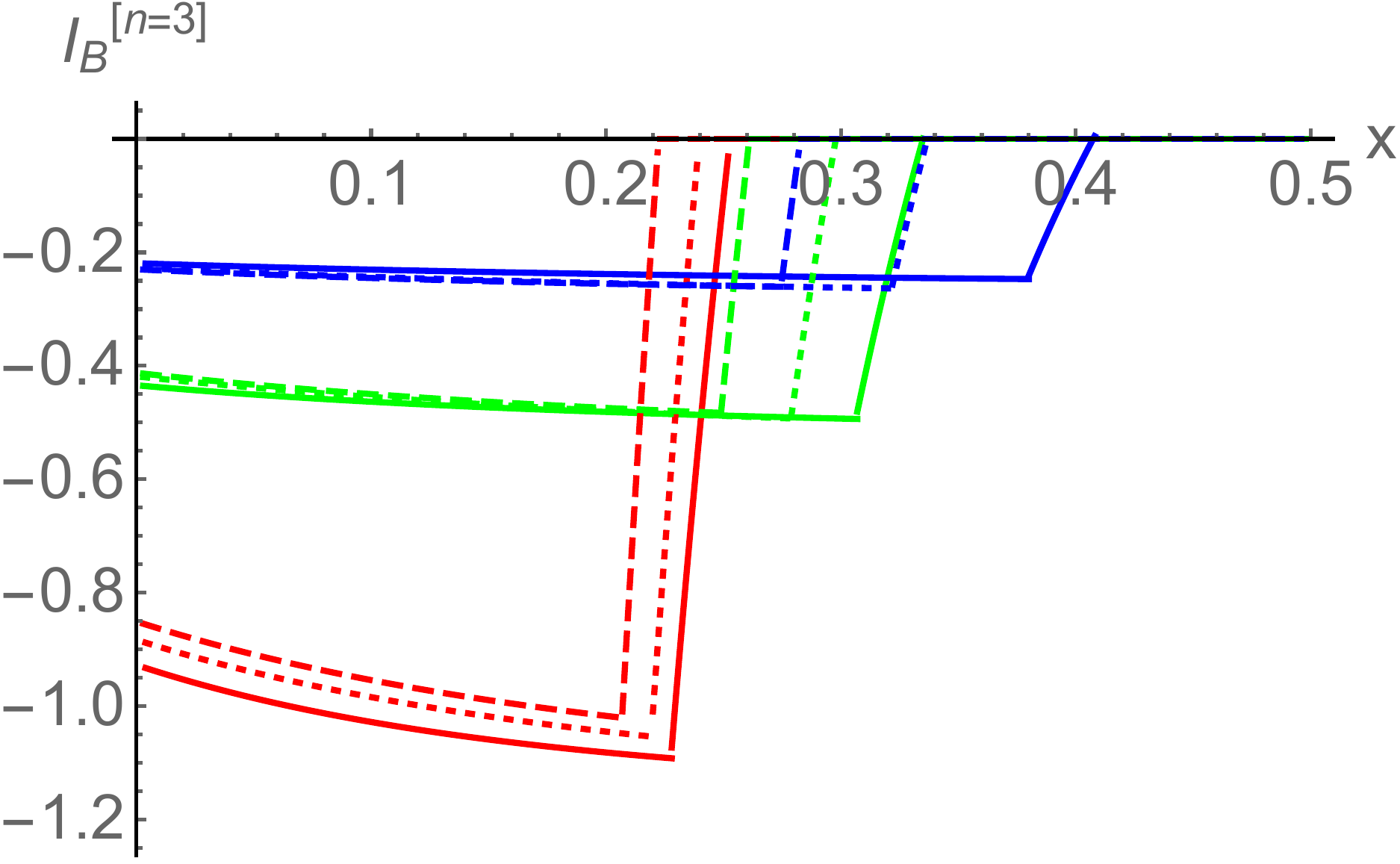}
\caption{\small Tri-partite information $I_B^{[n=3]}$ as a function of $x$ for various values of $\ell$ and $T$ at $\mu=0$. Here red, green and blue curves correspond to
$\ell$ = $0.3$, $0.4$ and $0.5$ respectively. The solid, dotted and dashed curves correspond to  $T/T_c=1.2$, $1.6$ and $2.0$ respectively. In units \text{GeV}.}
\label{I3vsXvsLAdSBHMu0T1Pt2case1}
\end{minipage}
\end{figure}

Let us now briefly discuss the phase diagram with $n > 2$ number of strips. The results for different $n$ are shown in
Figure~\ref{phaseDiagNstripsAdSBHvsLMu0T1Pt2}, where it can be observed that the phase diagram is quite similar to $n=2$ case. In particular,
again a phase transition between $S_A$ and $S_B$ occurs as the separation between the strips varies. Due to numerical limitations, it is
difficult to exactly establish the phase diagram for large $\ell$, however, the numerical trend suggests that the critical separation length $x_{crit}$
approaches a constant value \textit{i.e.} independent of $n$, for large $\ell$.\\

The behavior of tri-partite information ($I^{[n=3]}$) as a function of $x$ for various values of temperature and strip length is shown in
Figure~\ref{I3vsXvsLAdSBHMu0T1Pt2case1}. Again $I_{A}^{[n=3]}=0$ in the $S_A$ phase whereas it is $I_{B}^{[n=3]}\leq 0$ and have a non-trivial structure in
the $S_B$ phase. The non-positive profile of $I_{B}^{[n=3]}$ again indicates the monogamous nature of mutual information, however now in the deconfined phase.
We further find that these results hold
for finite chemical potential as well. Moreover, $I^{[n=3]}$ also exhibits non-analytic behaviour. The separation length $x$ at which non-analyticity appears
decreases with temperature whereas it increases (only slightly) when higher values of chemical potential are considered.    \\

Similarly, the $4$-partite information also behaves desirably in the deconfined phase. We find that it is always non-negative and exhibits non-analyticities at
various places. We further investigate how the $4$-partite information varies with temperature and chemical potential. Again, we find that, like the
$3$-partite information, the separation length
at which non-analyticity in $4$-partite information appears decreases with temperature whereas it enhances with chemical potential.
\\

At this point, it is instructive to point out that the above results for $n$-strip phase diagram and $n$-partite information in the dual deconfined
phase of our gravity model are qualitatively similar to what one gets in the dual deconfined phase of the AdS-Schwarzschild black hole. This suggests that,
as far as the entanglement structure is concerned, the excited profile of the dilaton field does not lead to a significant effect in the deconfined phase.
As we will show shortly, the above mentioned entanglement features of the dual deconfined phase will remain true even when other scale factors are considered. This
suggests some type of universality in the entanglement structure of holographic deconfined phases.

\subsubsection{Thermal-AdS/black hole phase transition and mutual information}
In recent years the holographic entanglement entropy has been used to probe and investigate black hole phase transitions. The main idea here is that
since the entangling surface propagates from asymptotic boundary into bulk it therefore might be able to probe a change in the spacetime geometry
which occurs during the phase transition. This idea has become a fruitful arena of research lately and has been applied in many different contexts, let us just mention
a few \cite{Johnson:2013dka,Dey:2015ytd,Zeng:2015wtt,Caceres:2015vsa}. In \cite{Dudal:2018ztm}, we performed a similar analysis for the EMD gravity model and
found that the holographic entanglement entropy does
indeed capture the essence of thermal-AdS/black hole phase transition (discussed earlier in this section).  However, one might also wonder whether other
information quantities like mutual or $n$-partite information can similarly be used as a diagnostic tool to probe black hole phase transition. Here we
take this analysis for the EMD model under consideration and found the answer in affirmative.

\begin{figure}[h!]
\centering
\includegraphics[width=2.8in,height=2.3in]{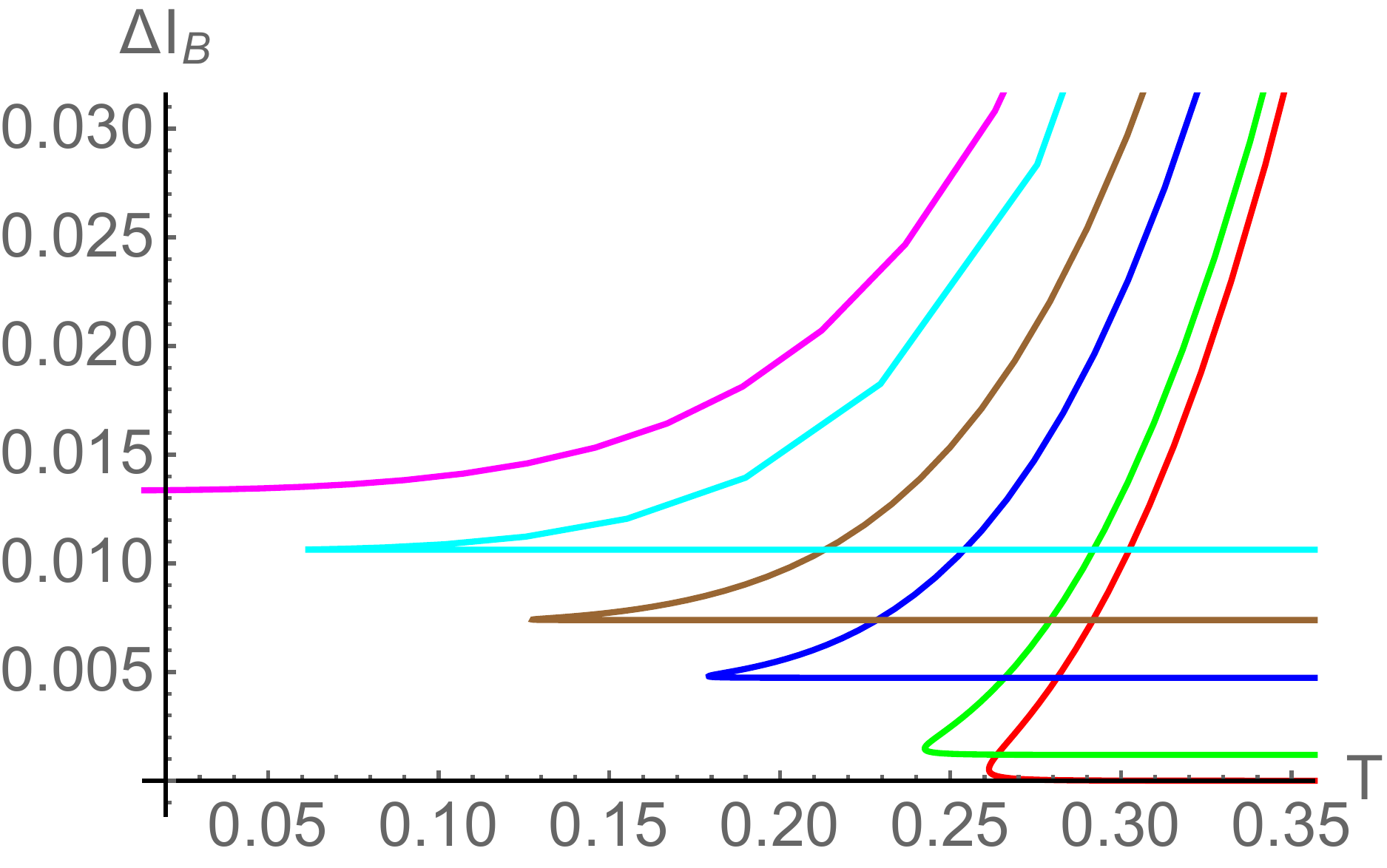}
\caption{ \small $\Delta I_B$ as a function of $T$ for various values $\mu$. Here $\ell=0.2$ and $x=0.1$ are used and red, green, blue, brown,
cyan and magenta curves correspond to $\mu=0$, $0.2$, $0.4$, $0.5$, $0.6$ and $0.673$ respectively. In units \text{GeV}.}
\label{IBvsTvsMu2stripsBHthermodynamicscase1}
\end{figure}

Our results are shown in Figure~\ref{IBvsTvsMu2stripsBHthermodynamicscase1}, where $\Delta I_B = I_{B}^{Thermal-AdS}-I_{B}^{Black hole}$ as a function of
temperature for various values of chemical potential is shown. $I_{B}^{Thermal-AdS}$ is the mutual information of the thermal-AdS space, which is
independent of temperature and chemical potential and is constant for a fixed $\ell$ and $x$. Here, we have used fixed
$\ell=0.2 \ GeV^{-1}$ and $x=0.1  \ GeV^{-1}$ so that $S_B$ is the most stable phase. We find that, just like the entanglement entropy,
the structure of mutual information also displays a striking similarity with the Bekenstein-Hawking thermal entropy. In particular, for small $\mu$,
there are again two branches in $\Delta I_B$ and these two branches exist only above $T>T_{min}$. The negative slope branch in
Figure~\ref{IBvsTvsMu2stripsBHthermodynamicscase1} corresponds to the unstable solution whereas the positive slope branch corresponds to the stable solution.

We see that, like the Bekenstein-Hawking entropy, $\Delta I_B$ also displays double valuedness for $\mu<\mu_c$ - an indication of black hole phase transition -
whereas this double valuedness disappears for $\mu>\mu_c$. An analogous similarity between Bekenstein-Hawking entropy and entanglement
entropy has been used
in recent years to advocate that the entanglement entropy can be used as a diagnostic tool to probe black hole phase transition \cite{Johnson:2013dka,Dey:2015ytd,Zeng:2015wtt,
Caceres:2015vsa}. We find that the similarity with the
Bekenstein-Hawking entropy goes beyond the entanglement entropy and even the mutual information exhibits similar features in the $T-\mu$ plane.
In Figure~\ref{IBvsTvsMu2stripsBHthermodynamicscase1}, we have used $\ell=0.2 \ GeV^{-1}$ and $x=0.1  \ GeV^{-1}$ however we have checked that
physical quantities like $T_{min}$ and $\mu_c$ do not change even when other values of $\ell$ and $x$ are considered. Moreover, although not
shown here for brevity, we find that similar results hold for $3$ and $4$-partite information as well. Our analysis therefore not only confirms the suggestions
of \cite{Johnson:2013dka,Dey:2015ytd,Zeng:2015wtt,Caceres:2015vsa} in a more advanced holographic bottom-up model but also put further weight on the expectation that other information theoretic quantities can
also be used to investigate phase transitions.

\section{Case II: the specious-confined/deconfined phases}
In \cite{Dudal:2017max}, by taking the following second form for the scale factor,
\begin{eqnarray}
P(z)=P_{2}(z)=-\frac{3}{4}\ln{(a z^2+1)}+\frac{1}{2}\ln{(b z^3+1)}-\frac{3}{4}\ln{(a z^4+1)}
\label{Aansatz2}
\end{eqnarray}
a novel specious-confined phase on the dual boundary side was revealed. This specious-confined phase did not strictly correspond to the
standard confined phase, but exhibited many properties which resembled quite well with the standard QCD confined phase. The novelty of the
specious-confined phase lies in the fact that it is dual to a non-extremal small black hole phase in the gravity side. It therefore has the notion
of temperature, which in turn allows to investigate temperature dependent properties of the dual specious-confined phase.  Indeed, it was shown
in \cite{Dudal:2017max} that thermal behaviour of the quark-antiquark free energy and entropy, as well as the speed of second sound in the specious-confined phase, were qualitatively
similar to lattice QCD results. It is important to mention that for both choices of scale factor $P_{1,2}(z)$, the potential is bounded from above by its UV
boundary value \textit{i.e.} $V(z)\leq V(0)$. Therefore, for both choices $P_{1,2}(z)$ the EMD model under consideration satisfies the Gubser
criterion to have a well-defined dual boundary theory \cite{Gubser:2000nd}.

With the above scale factor, the metric again asymptotes to AdS at the boundary $z\rightarrow 0$. However, it causes non-trivial modifications in
the bulk IR region which in turn greatly modifies thermodynamic properties of the system. As in the case of first scale factor $P_{1}(z)$, the
parameters $a=c/9$ and $b=5c/16$ in $P_{2}(z)$ are again fixed by demanding the specious-confined/deconfined phase transition to be
around $270 \ MeV$ at zero chemical potential, as is observed in lattice QCD for pure glue sector.

\subsection{Black hole thermodynamics}
\begin{figure}[h!]
\begin{minipage}[b]{0.5\linewidth}
\centering
\includegraphics[width=2.8in,height=2.3in]{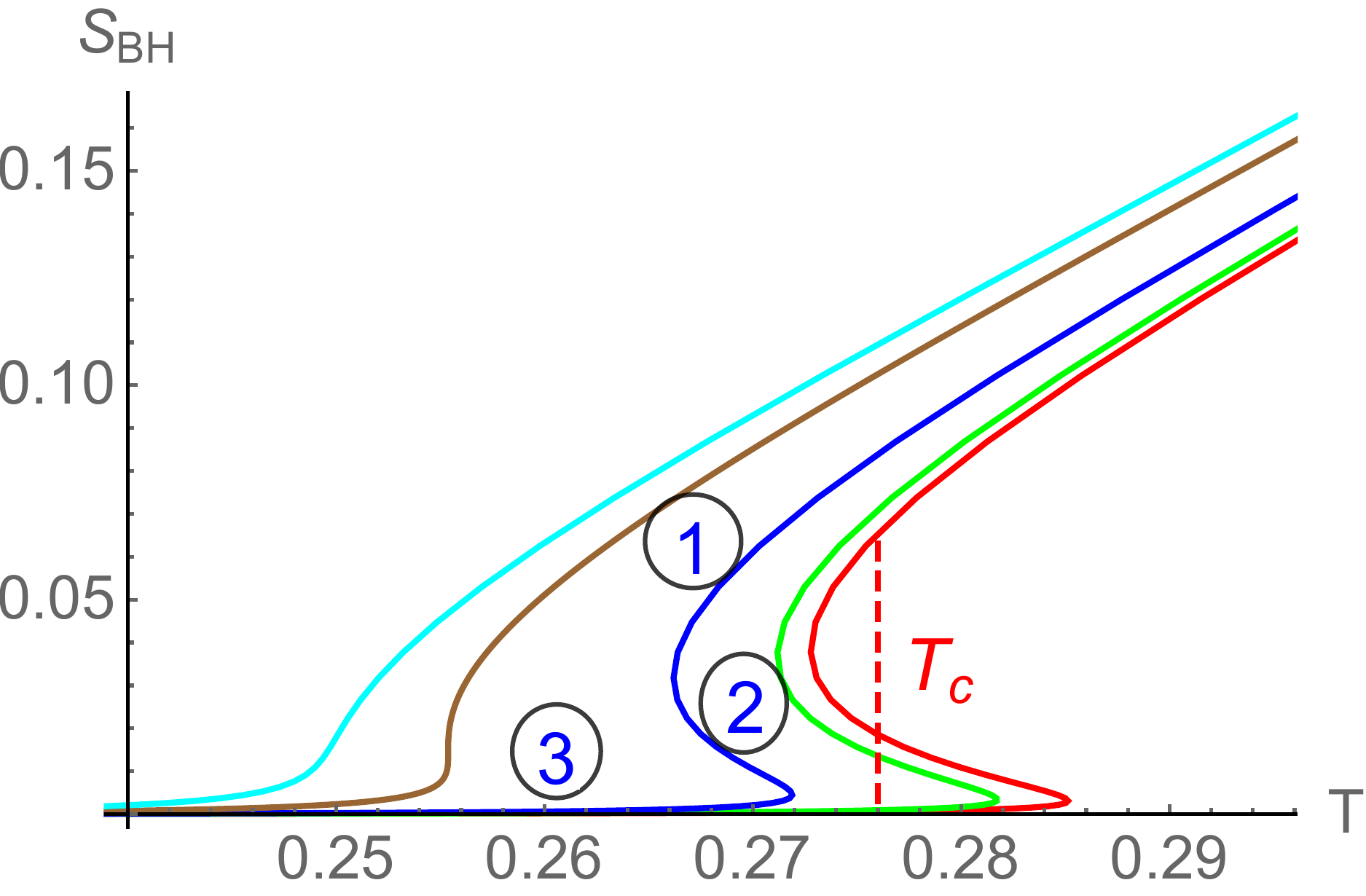}
\caption{ \small Bekenstein-Hawking entropy as a function of $T$ for various values of $\mu$. Here red, green, blue, brown and cyan curves
correspond to $\mu=0$, $0.1$, $0.2$, $0.312$ and $0.35$ respectively. In units \text{GeV}.}
\label{TvsSBHvsMucase2}
\end{minipage}
\hspace{0.4cm}
\begin{minipage}[b]{0.5\linewidth}
\centering
\includegraphics[width=2.8in,height=2.3in]{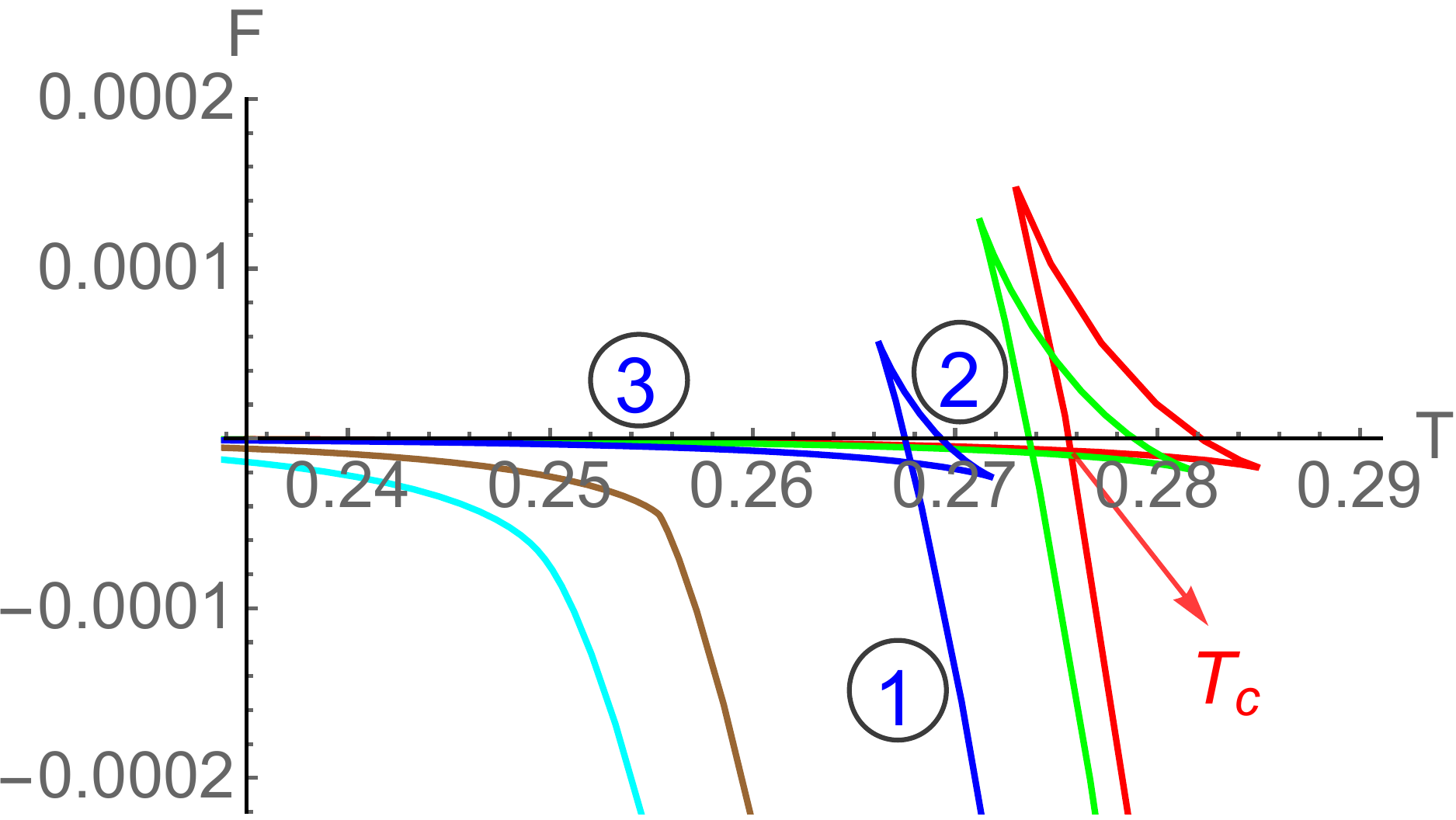}
\caption{\small Free energy as a function of $T$ for various values of $\mu$. Here red, green, blue, brown and cyan curves correspond to
$\mu=0$, $0.1$, $0.2$, $0.312$ and $0.35$ respectively. In units \text{GeV}.}
\label{TvsFBHvsMucase2}
\end{minipage}
\end{figure}
As we can see from Figures~\ref{TvsSBHvsMucase2} and \ref{TvsFBHvsMucase2}, the thermodynamic behaviour of EMD gravity model gets greatly
modified with $P_2(z)$. In particular, on the top of a large stable black hole phase (marked by $\circled{1}$) and an unstable black hole phase (marked by $\circled{2}$), now a new stable phase appears at low temperatures. This new stable phase is marked by $\circled{3}$ in
Figure~\ref{TvsSBHvsMucase2} and corresponds to a small black hole phase (large $z_h$). Importantly, with $P_2(z)$ at least one black hole
phase always exists at all temperature. Apart from these three black hole phases, there also exists a thermal AdS phase. However, we find
that the free energy of this thermal-AdS phase is always greater than the stable black hole phases, indicating that it is thermodynamically
unfavourable at all temperature. The normalised free energy behaviour, plotted in Figure~\ref{TvsFBHvsMucase2}, further suggests a first
order phase transition between small and large hole phases as the Hawking temperature varied. For $\mu=0$, the small/large phase transition
occurs at $T=0.276 \ GeV$. Therefore, the large black hole phase is thermodynamically favoured at $T>T_c$ whereas the small black hole phase is
favoured at $T<T_c$.

The small/large black hole phase transition persists for finite chemical potential as well. The complete phase diagram and the dependence of $T_c$
on $\mu$ can be found in \cite{Dudal:2017max}, where it was shown that $T_c$ decreases
with $\mu$ for $\mu<\mu_c=0.312 \ GeV$, mimicking yet another important feature of lattice QCD. At $\mu_c$, the small/large black hole phase
transition ceases to exist and we have a single stable black hole phase which exists at all temperatures (shown by cyan curve in Figure~\ref{TvsFBHvsMucase2}).
Overall, this gravity model shows a Van der Waals type black hole phase transition, however now with a planar horizon instead of a spherical horizon \cite{Chamblin:1999tk,Chamblin:1999hg,Mahapatra:2016dae,Caldarelli:1999xj}.

In \cite{He:2013qq,Yang:2015aia}, the above small/large black hole phases were suggested to be dual to confined/deconfined phases in the dual
boundary theory. In particular, the small black hole phase was suggested to be dual to the confined phase whereas the large black hole phase was
suggested to be dual to the deconfined phase. However, as was pointed out in \cite{Dudal:2017max}, the small black hole phase does not strictly
correspond to the confined phase, as the Polyakov and Wilson loop expectation value do not strictly exhibit the standard behaviour.
Interestingly, the dual boundary theory does however exhibit properties, such as the quark-antiquark free energy and entropy etc, which are
qualitatively similar to the standard QCD confined phase. For this reason, the dual boundary theory of the small black hole
phase was named as specious-confined phase.

Since the specious-confined phase has the notion of temperature (thereby allowing us to study temperature dependent properties of many
important observables) and shares many interesting lattice QCD properties, it becomes important to investigate mutual and $n$-partite information
in this phase as well.

\subsubsection{With small black hole background: one strip}
\begin{figure}[h!]
\begin{minipage}[b]{0.5\linewidth}
\centering
\includegraphics[width=2.8in,height=2.3in]{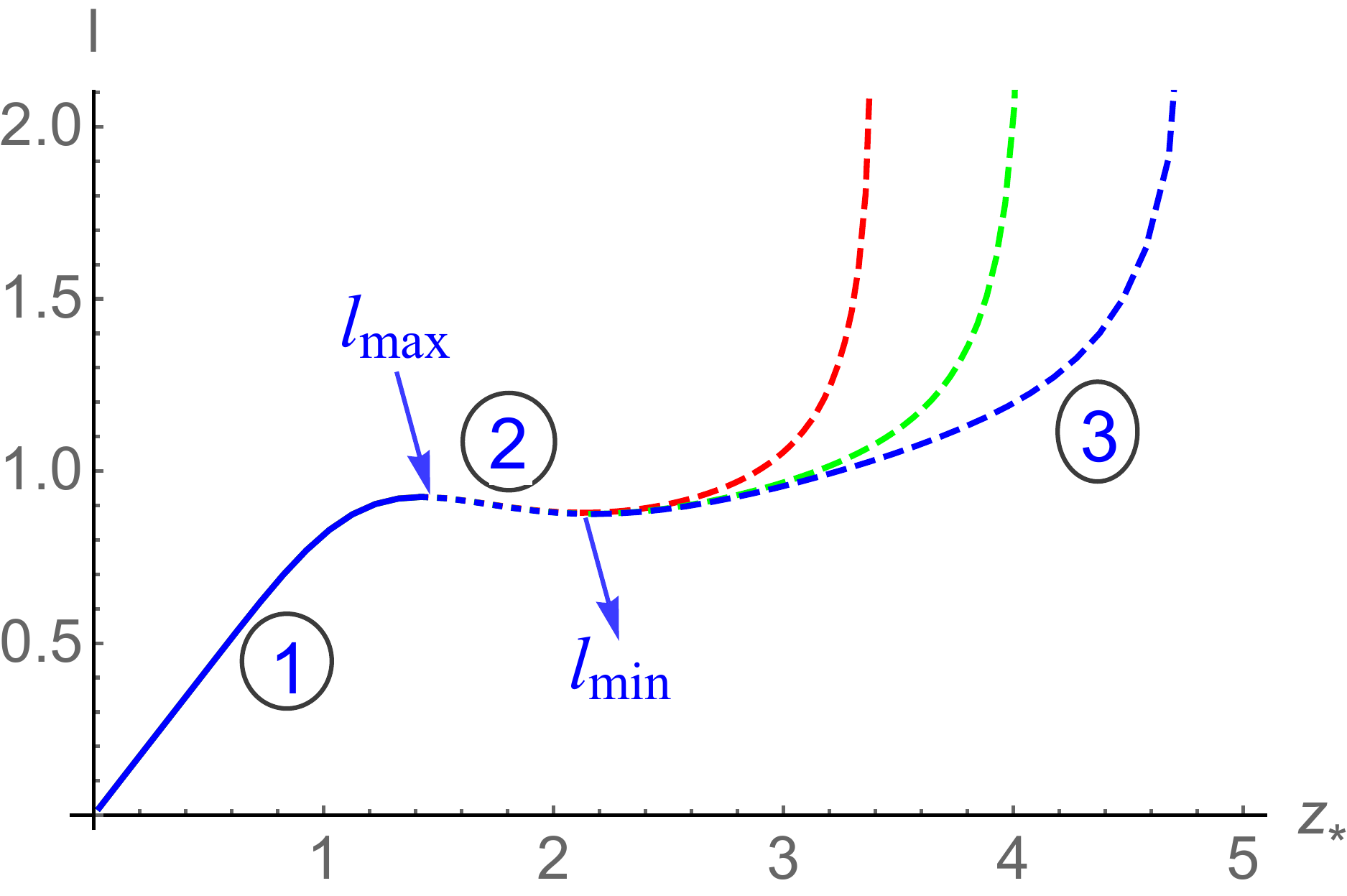}
\caption{ \small $\ell$ vs $z_*$ behaviour in the small black hole background. Here $\mu=0$ and red, green and blue curves correspond
 to $T/T_{c}=0.9$, $0.8$ and $0.7$ respectively. In units \text{GeV}.}
\label{zsvslAdSBHMu0smallcase2}
\end{minipage}
\hspace{0.4cm}
\begin{minipage}[b]{0.5\linewidth}
\centering
\includegraphics[width=2.8in,height=2.3in]{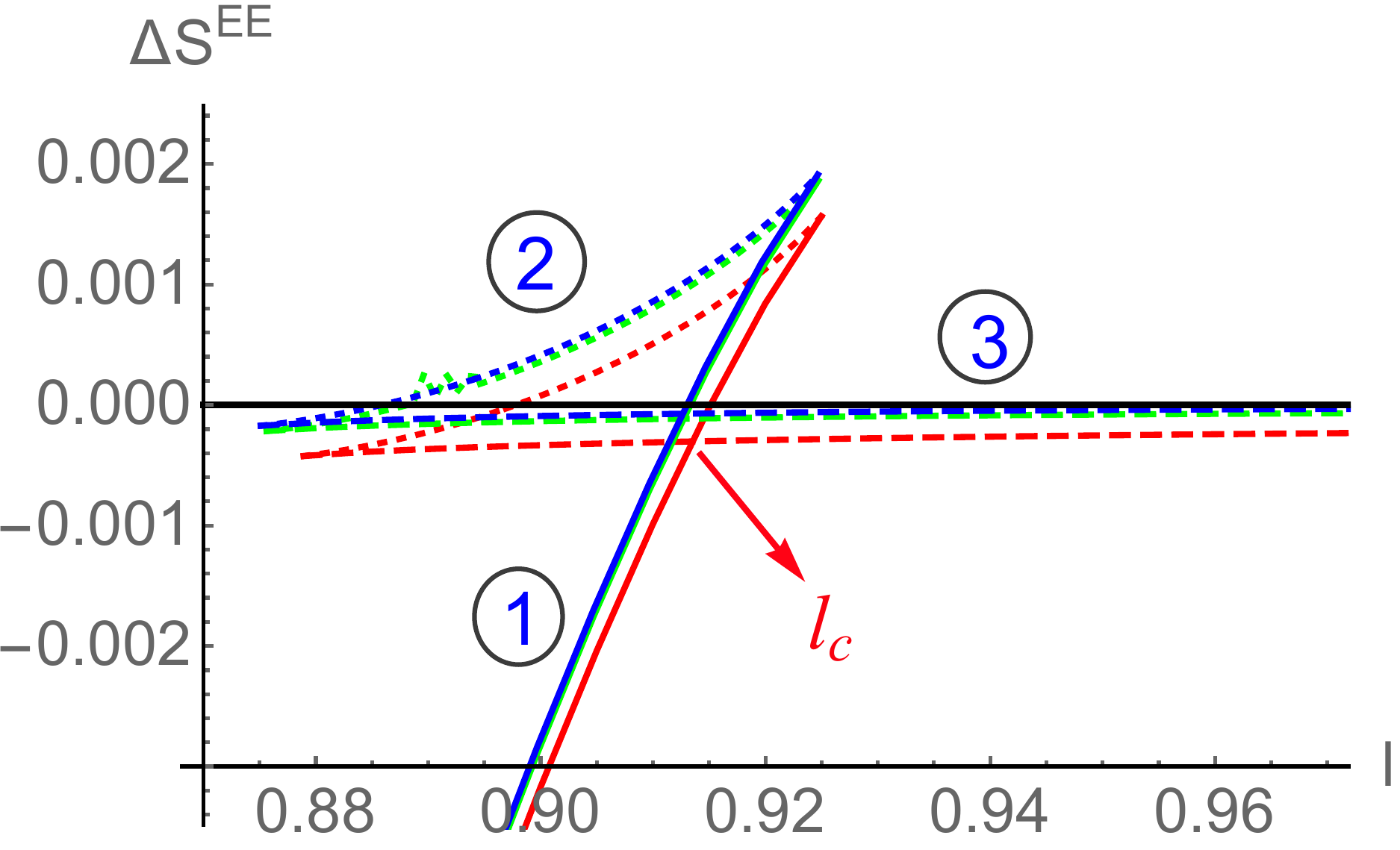}
\caption{\small $\Delta S^{EE}=S^{EE}_{con}-S^{EE}_{discon}$ vs $\ell$ behaviour in the small black hole background. Here $\mu=0$ and red,
green and blue curves correspond to $T/T_{c}=0.9$, $0.8$ and $0.7$ respectively. In units \text{GeV}.}
\label{lvsSEEAdSBhMu0smallcase2}
\end{minipage}
\end{figure}

Let us first briefly discuss the results for the entanglement entropy (one strip) in the specious-confined phase. The results,
shown in Figures~\ref{zsvslAdSBHMu0smallcase2} and \ref{lvsSEEAdSBhMu0smallcase2} for three different temperatures at zero chemical
potential, suggest a significant departure from the standard confined phase. In particular, the connected surface now exists
for all $\ell$. Further, the $\ell$ vs $z_*$ behaviour is now divided into three regions (instead of two as in the case of standard confined phase): $\ell$ first
increases with $z_*$ then decreases and finally increases again. These three regions are marked by $\circled{1}$, $\circled{2}$ and $\circled{3}$ respectively
 in Figure~\ref{zsvslAdSBHMu0smallcase2}. Importantly, between $\ell_{min}$ and $\ell_{max}$, these three solutions for the connected entangling surface
coexist for a given $\ell$. This makes the transition between different entangling surfaces more non-trivial in the specious-confined phase.

The difference between connected and disconnected entanglement entropy is shown in Figure~\ref{lvsSEEAdSBhMu0smallcase2}. It turns out that for all $\ell$,
the area of the connected surface is always smaller than the disconnected surface. It implies that connected surface is always more favourable,
and hence no connected/disconnected phase transition takes place in the specious confined phase. Subsequently, the entanglement
entropy is always of order $\mathcal{O}(N^2)$ at any subsystem size. This is one of the biggest differences between standard confined and specious confined
phases. Interestingly, however, now a new type of connected/connected surface phase transition appears in the specious confined phase. This connected/connected
surface phase transition is shown in Figure~\ref{lvsSEEAdSBhMu0smallcase2}, where one can clearly observe a transition between connected surfaces
$\circled{1}$ and $\circled{3}$. The critical strip size at which this phase transition occurs is indicated by $\ell_c$. It is
important to emphasise that this connected/connected surface transition is very different from the connected/disconnected surface transition observed
in the standard confined phase. In particular, the order of the entanglement entropy does not change at the connected/connected critical point,
\begin{eqnarray}
\frac{\partial S^{EE}}{\partial \ell} &\propto& \frac{1}{G_N} = \mathcal{O}(N^2)\quad\text{for both}\quad  \ell < \ell_{c}\quad \text{and}\quad   \ell > \ell_{c}\,.
\end{eqnarray}
This important result further confirms that a non-trivial interpretation of small black hole phase as the gravity dual of standard confined
phase is not entirely correct, as otherwise mentioned in \cite{He:2013qq,Yang:2015aia}. In order to further highlight the subtle
relation between standard confined and specious-confined phases, we also like to emphasize that although $\frac{\partial S^{EE}}{\partial \ell}$ in the
specious confined phase is not strictly zero, however it is very small. For example, for larger $\ell$, the entanglement entropy depends only
mildly on $\ell$ (as can be seen from Figure~\ref{lvsSEEAdSBhMu0smallcase2}) and is practically independent of it. When going from let's say
$\ell$ to $\ell/2$, the change in the magnitude of entanglement entropy occurs only at the fifth decimal place. This feature of entanglement
entropy in the specious-confined phase again resemble approximately -- however, not exactly -- to the standard confined phase for which
$\frac{\partial S^{EE}}{\partial \ell}=0$ for large $\ell$, highlighting once more the non-trivial similarities as well as differences
between standard confined and specious-confined phases as was first pointed out in \cite{Dudal:2017max}.

\begin{figure}[h!]
\begin{minipage}[b]{0.5\linewidth}
\centering
\includegraphics[width=2.8in,height=2.3in]{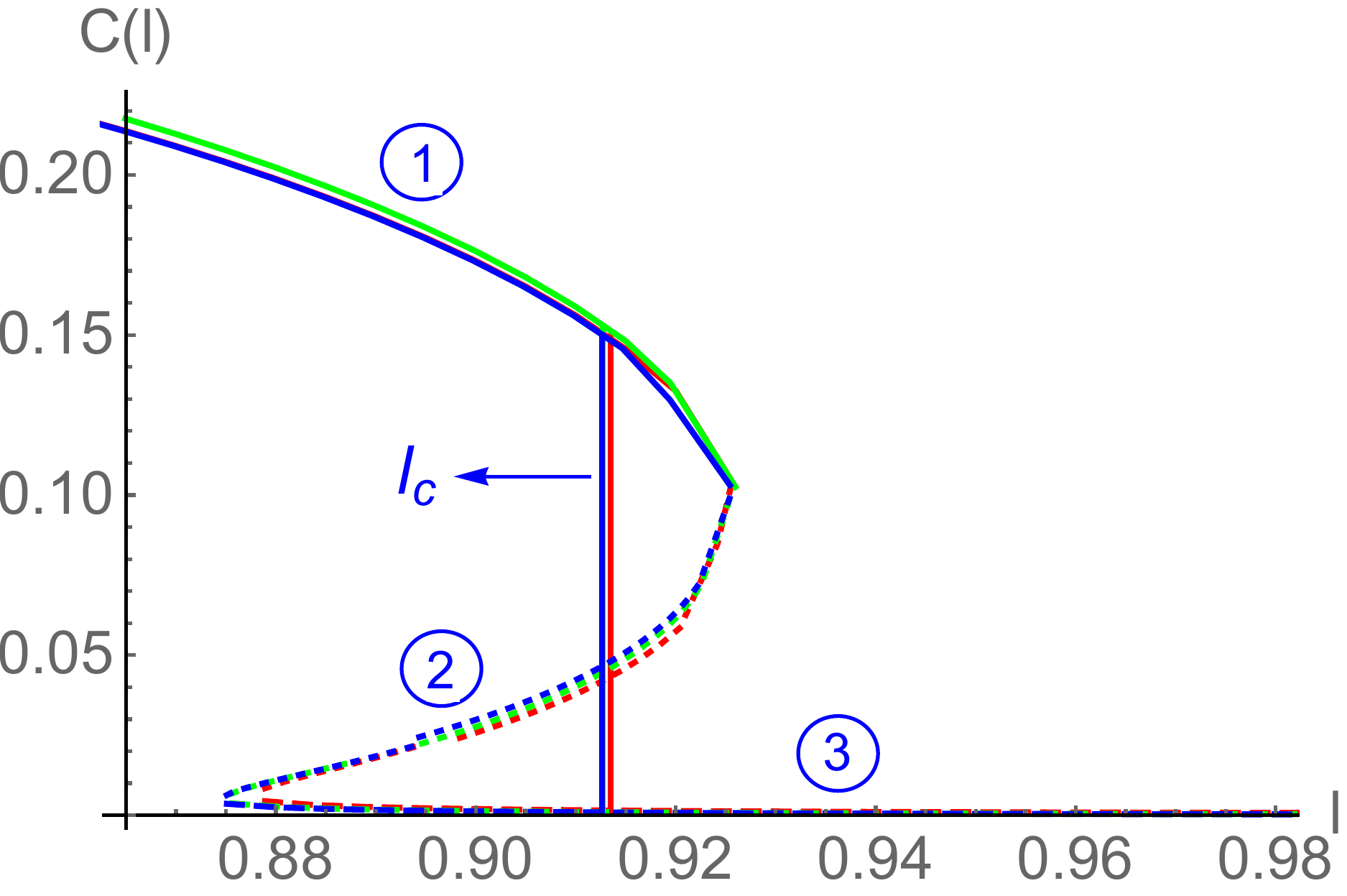}
\caption{ \small The entropic $\mathcal{C}$-function as a function of $\ell$ in the specious-confined phase. Here $\mu=0$ and red, green and blue
curves correspond to $T/T_{c}=0.9$, $0.8$ and $0.7$ respectively. In units \text{GeV}.}
\label{CvslsmallBHcase2}
\end{minipage}
\hspace{0.4cm}
\begin{minipage}[b]{0.5\linewidth}
\centering
\includegraphics[width=2.8in,height=2.3in]{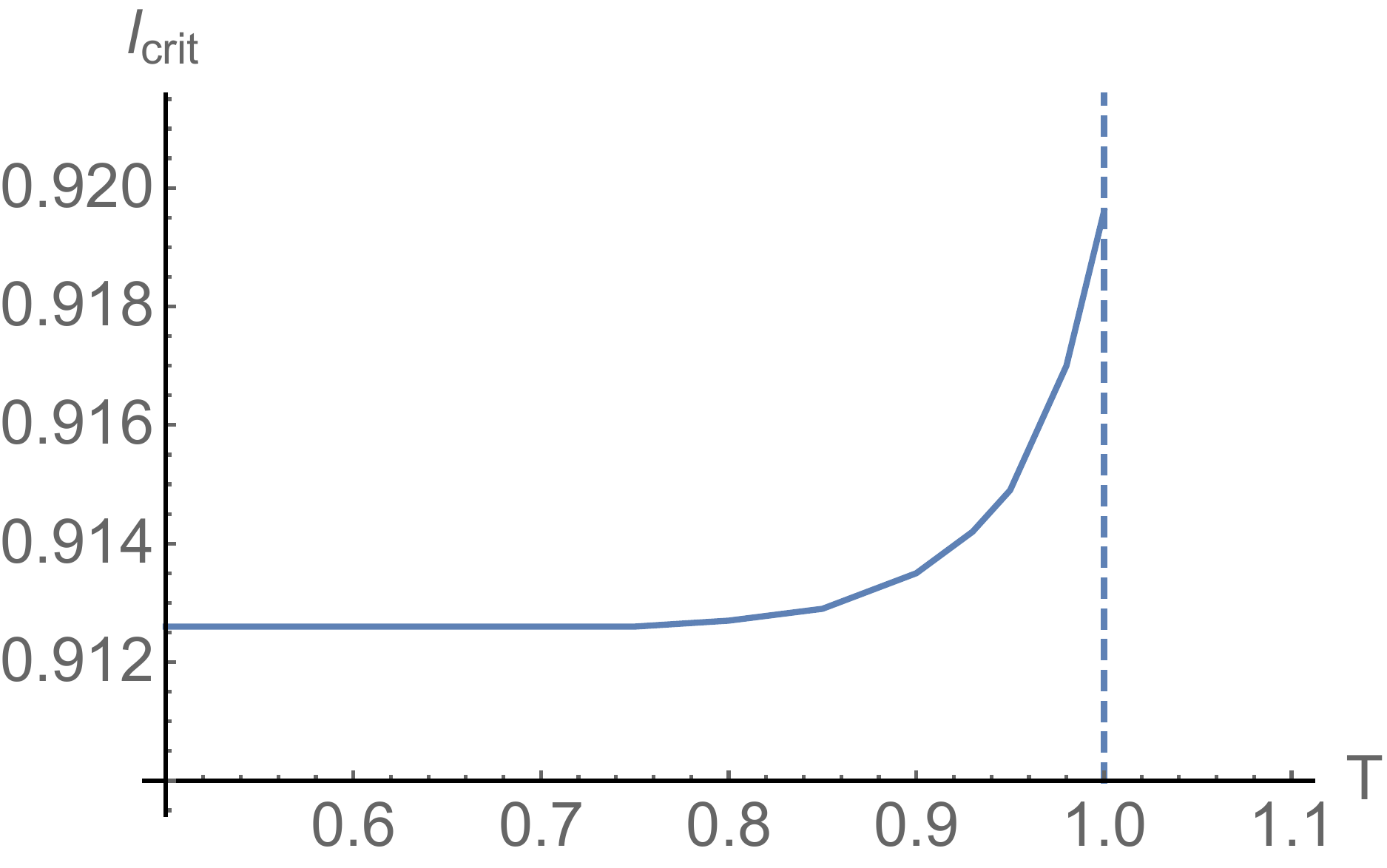}
\caption{\small $\ell_{c}$ vs $T$ in the specious-confined phase at $\mu=0$. This $(T, \ell_{c})$ holographic phase diagram can be compared with
the SU(2) lattice gauge theory conjecture of \cite[Figure 8]{Buividovich:2008kq}. In units \text{GeV}.}
\label{lcritvsTMu0smallBHcase1}
\end{minipage}
\end{figure}

Interestingly the entropic $\mathcal{C}$-function, which quantifies the number of degrees of freedom in a system at length scale $\ell$ (or the energy scale),
also behaves desirably in the specious-confined phase. In particular, the $\mathcal{C}$-function decreases monotonically from UV to IR in the specious-confined
phase as well. This is shown in Figure~\ref{CvslsmallBHcase2}, where a sharp drop in its magnitude (shown by vertical solid lines) is
explicitly evident. Interestingly, our holographic estimate for the length scale $\ell_c$ at which $\mathcal{C}$-function drop sharply is of the same order as was
observed in lattice QCD. For example, at vanishing $T$ and $\mu$ we find an estimate $\ell_c=0.931 \ GeV^{-1}$ whereas SU(3) lattice gauge setup suggested
$\ell_c \approx 4.3 \ GeV^{-1}$. Moreover, the temperature dependent behaviour of $\ell_c$ also qualitatively matches with lattice prediction.
The $(T, \ell_{c})$ holographic phase diagram is shown in Figure~\ref{lcritvsTMu0smallBHcase1}, and can be compared with the SU(2) lattice gauge theory
conjecture of \cite[Figure 8]{Buividovich:2008kq}.

We further like to emphasise that the above discussed richness in the structure of entanglement entropy in the specious-confined phase remains
true with finite $\mu$ as well. In particular, there are again novel connected/connected surface transitions with a sharp decrease in the magnitude
of $\mathcal{C}$-function at the critical point $\ell_c$. With finite $\mu$, the main difference appears in the magnitude of $\ell_c$, which attains
a higher value for higher $\mu$, and moreover approaches $\mu$ dependent constant value at low $T$.

\subsubsection{With small black hole background: two strip}
Since there is no connected to disconnected surface transition in the specious-confined phase, the corresponding phase diagram for two and higher
strips is much simpler than the standard confined phase. In particular, there will not be any disconnected phases like $S_C$ and $S_D$ (see
Figure~\ref{ES2equalstrips}) and only the connected  phases like $S_A$ and $S_B$ remain. The entanglement entropy expressions for $S_A$ and $S_B$
are again given by eq.~(\ref{eqnstrips}).

\begin{figure}[h!]
\begin{minipage}[b]{0.5\linewidth}
\centering
\includegraphics[width=2.8in,height=2.3in]{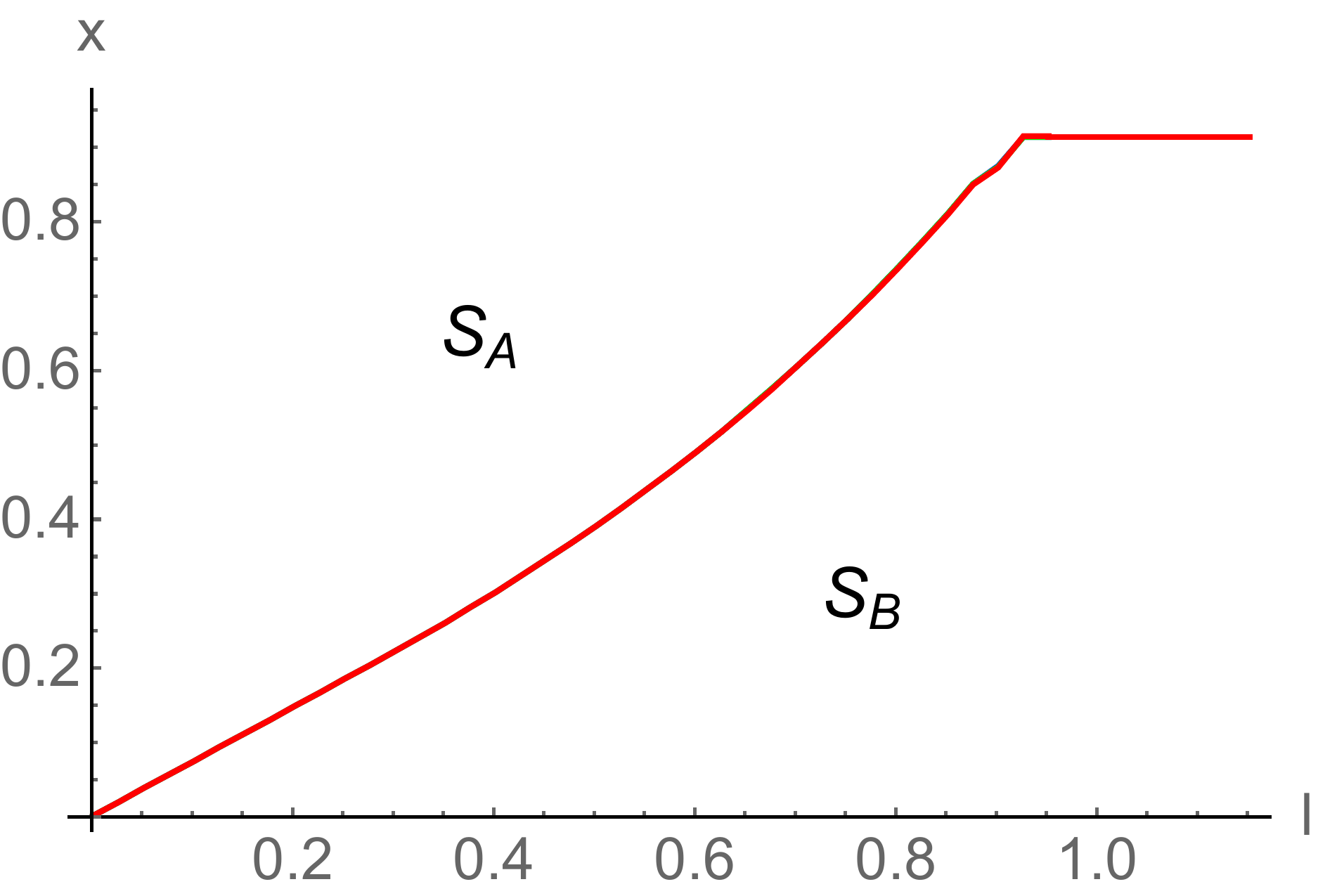}
\caption{ \small  Two strip phase diagram of the specious-confined phase for various values of $T$. The $S_A$ and $S_B$ phases
correspond to the two connected bulk surfaces of Figure~\ref{ES2equalstrips}. Here $\mu=0$ and red, green and blue curves correspond
to $T/T_c=0.9$, $0.8$ and $0.7$ respectively. In units \text{GeV}. }
\label{phaseDiag2stripsSmallBHMu0}
\end{minipage}
\hspace{0.4cm}
\begin{minipage}[b]{0.5\linewidth}
\centering
\includegraphics[width=2.8in,height=2.3in]{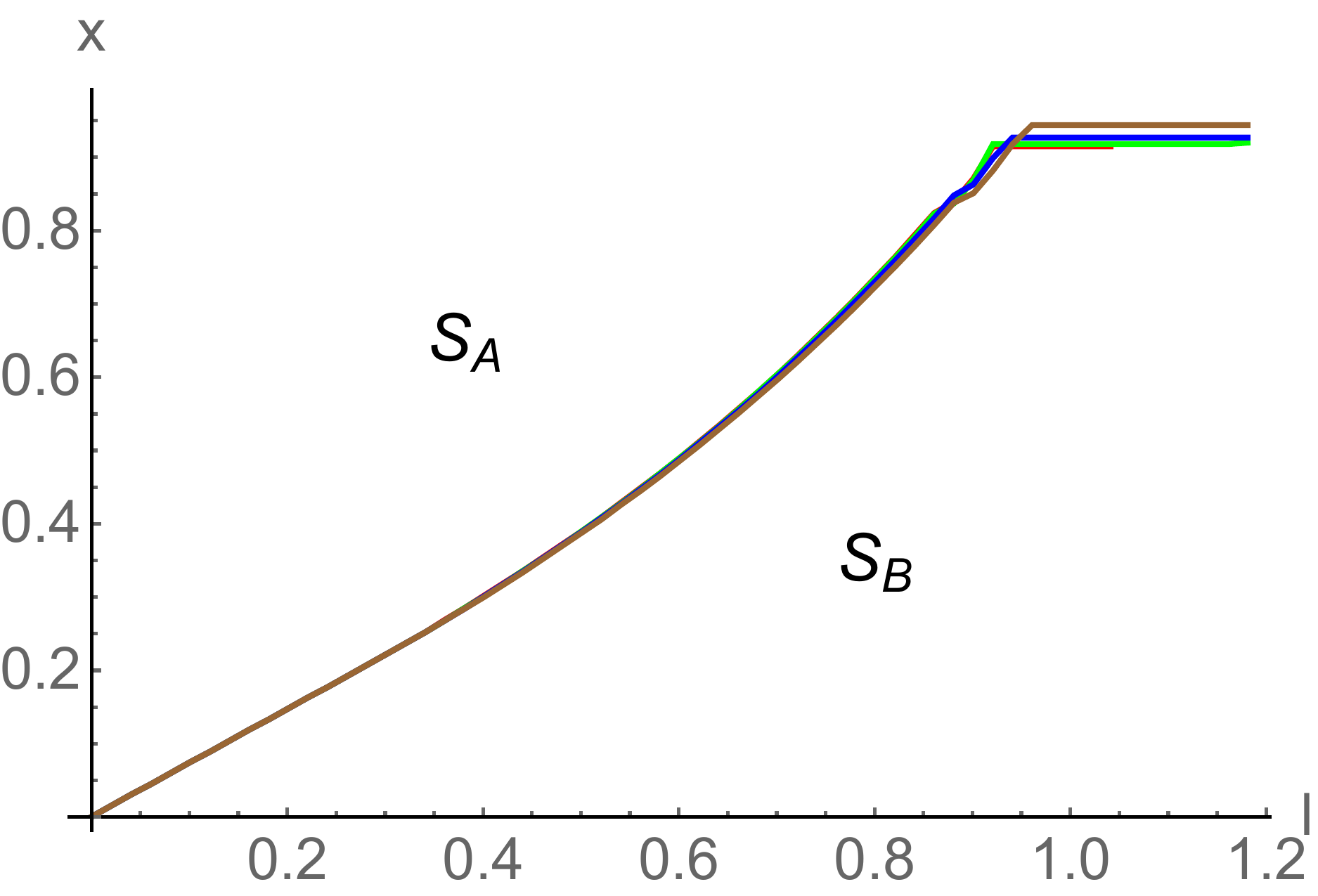}
\caption{\small Two strip phase diagram of the specious-confined phase for various values of $\mu$. The $S_A$ and $S_B$ phases
correspond to the two connected bulk surfaces of Figure~\ref{ES2equalstrips}. Here $T/T_c=0.9$ and red, green, blue and brown curves
correspond to $\mu=0$, $0.1$, $0.2$ and $0.3$ respectively. In units \text{GeV}. }
\label{phaseDiag2stripsSmallBHvsMuTPt9}
\end{minipage}
\end{figure}

The phase diagram with two strips at $\mu=0$ is shown in Figure~\ref{phaseDiag2stripsSmallBHMu0}. We again find a phase transition between
$S_A$ and $S_B$ phases. In particular, for a given $\ell$, $S_A$ phase has the lowest entanglement entropy for large $x$
whereas $S_B$ phase has the lowest entanglement entropy for small $x$. It is interesting to observe that this phase diagram is
quite similar to the two strip standard confined phase diagram if we remove the disconnected phases from the latter. In particular, the
transition line between $S_A/S_B$ is again almost constant for large $x, l > \ell_{c}$.  Moreover, the nature of the critical point $\ell_c$,
where non-analyticity in the entanglement entropy
appears, is also quite similar to the second tricritical point $\ell_{c2}$ of the standard confined phase. This once again emphasizes the closeness
of specious-confined phase with the standard confined phase. However here, as opposed to the standard confined phase, entanglement entropy order always
changes at the transition line. Our analysis further suggests only a mild dependence of the phase diagram on temperature.
In particular, $S_A$/$S_B$ transition line at different temperatures almost overlap with each other.

In Figure~\ref{phaseDiag2stripsSmallBHMu0}, two strip phase diagram with finite chemical potential at $T=0.9~T_c$ is shown.
We again find a similar type of $S_A/S_B$ transition as $x$ and $\ell$ are varied. The main difference arises in the magnitude of
$\ell_c$, which only gets enhanced with $\mu$. The higher $\mu$ therefore increases the parameter space of $S_B$. Although not presented
here for brevity, similar results occur at other temperatures as well.

\begin{figure}[h!]
\begin{minipage}[b]{0.5\linewidth}
\centering
\includegraphics[width=2.8in,height=2.3in]{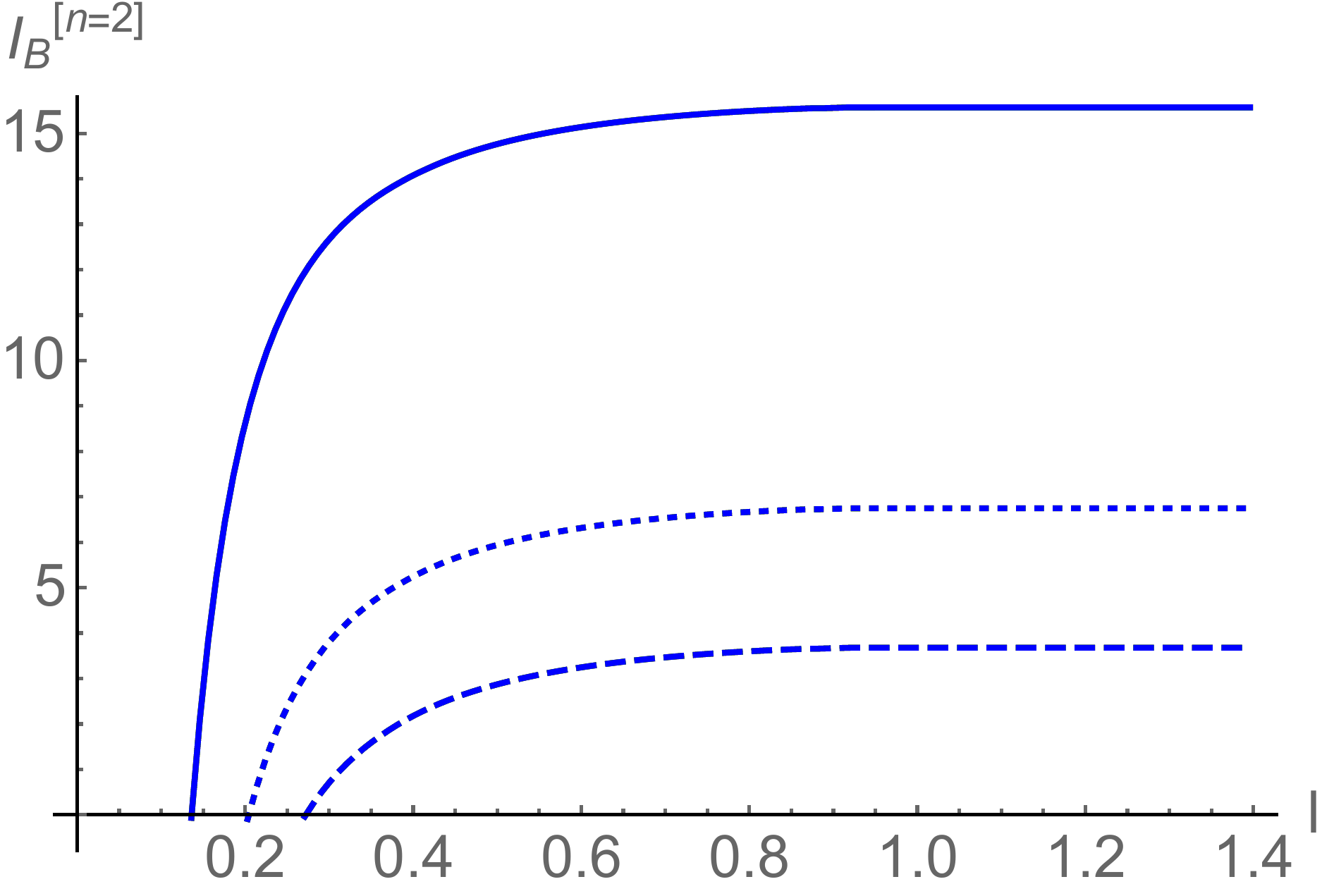}
\caption{ \small Mutual information $I_{B}^{[n=2]}$ as a function of $\ell$ for various values of $x$ and $T$. Here $\mu=0$ is fixed
and solid, dotted and dashed lines correspond to $x=0.1$, $0.15$ and $0.20$ respectively. The Red, green, blue and black curves correspond to
$T/T_c=0.9$, $0.8$, $0.7$ and $0$ respectively. In units \text{GeV}. }
\label{MutualinfoSmallBH2stripvsLvsXvsTMu0}
\end{minipage}
\hspace{0.4cm}
\begin{minipage}[b]{0.5\linewidth}
\centering
\includegraphics[width=2.8in,height=2.3in]{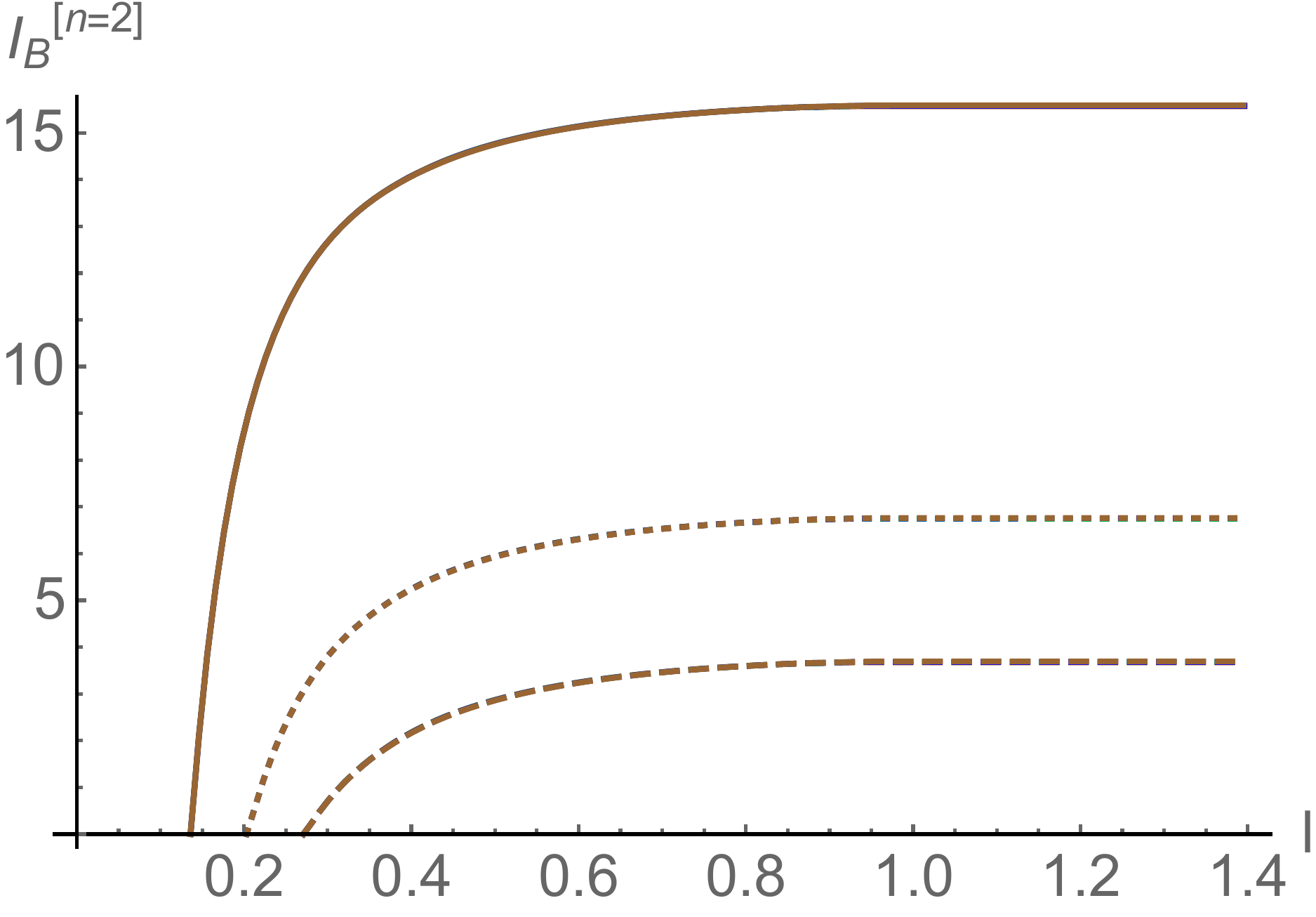}
\caption{\small Mutual information $I_{B}^{[n=2]}$ as a function of $\ell$ for various values of $x$ and $\mu$. Here $T/T_c=0.9$ is fixed and solid,
dotted and dashed lines correspond to $x=0.1$, $0.15$ and $0.20$ respectively. The Red, green, blue and brown curves correspond to $\mu=0$, $0.1$, $0.2$
and $0.3$ respectively. In units \text{GeV}.  }
\label{MutualinfoSmallBH2stripvsLvsXvsMuTPt9}
\end{minipage}
\end{figure}

It is also instructive to investigate the mutual information of the specious-confined phase and compared it with the standard confined phase.
The results for various temperatures are shown in Figure~\ref{MutualinfoSmallBH2stripvsLvsXvsTMu0}. $I_A^{[n=2]}$ is zero again, whereas $I_B^{[n=2]}$ always
satisfies the condition $I_{B}^{[n=2]} \geq 0$ and increases monotonically with $\ell$. Interestingly, $I_B$ approaches a temperature independent
constant value at large $\ell$. This behaviour should be contrasted from the deconfined phase where $I_B^{[n=2]}$ was instead found to approach a
temperature dependent constant value (see Figure~\ref{MutualinfoAdSBh2stripXpt1vsLvsT}). For completion, we have also included zero temperature $I_B^{[n=2]}$
behaviour.  We find that $I_B^{[n=2]}$ profile for various temperatures overlap with each other, both in small as well as in large $\ell$ regions, thereby
suggesting its non-thermal nature in the specious-confined phase. This interesting new result has not been discussed in lattice QCD community
yet, and it would be interesting to perform a similar temperature dependent analysis of the mutual information using lattice simulations and compare the
corresponding lattice results with the holographic prediction.

We further find that $I_B^{[n=2]}$ smoothly goes to zero as the $S_A$/$S_B$ transition line is approached.
Moreover, the order of mutual information also changes as we go from $S_A$ to $S_B$ and visa versa.  The phase transition in the
specious-confined phase is therefore always accompanied by a change in the order of mutual information, as opposed to the standard confined
phase where its order may or may not change depending on the nature of the transition line.

The effect of chemical potential on $I_{B}^{[n=2]}$ is shown in Figure~\ref{MutualinfoSmallBH2stripvsLvsXvsMuTPt9}. $I_B^{[n=2]}$ again exhibits the standard monotonic
behaviour with $\ell$ and asymptotes to a constant value at large $\ell$. Interestingly, different values of $\mu$ do not cause a significant variation
in $I_B^{[n=2]}$ and we find that $I_B$ curves for different $\mu$ actually overlap with each other. The $\mu$ independent
nature of $I_B^{[n=2]}$ is again an interesting and new result from holography, which might have an analogue realization in lattice QCD.

\subsubsection{With small black hole background: $n>2$ strips}
The entanglement phase diagram of $n>2$ strips is very similar to $n=2$ strips. Again, only $S_A$/$S_B$ type of entangling surface phase
transition appears. At a fixed $T$ and $\mu$, larger $n$ only increases the parameter space of $S_B$ in the small $\ell$ region whereas it remains almost
constant in the large $\ell$ region.

\begin{figure}[h!]
\begin{minipage}[b]{0.5\linewidth}
\centering
\includegraphics[width=2.8in,height=2.3in]{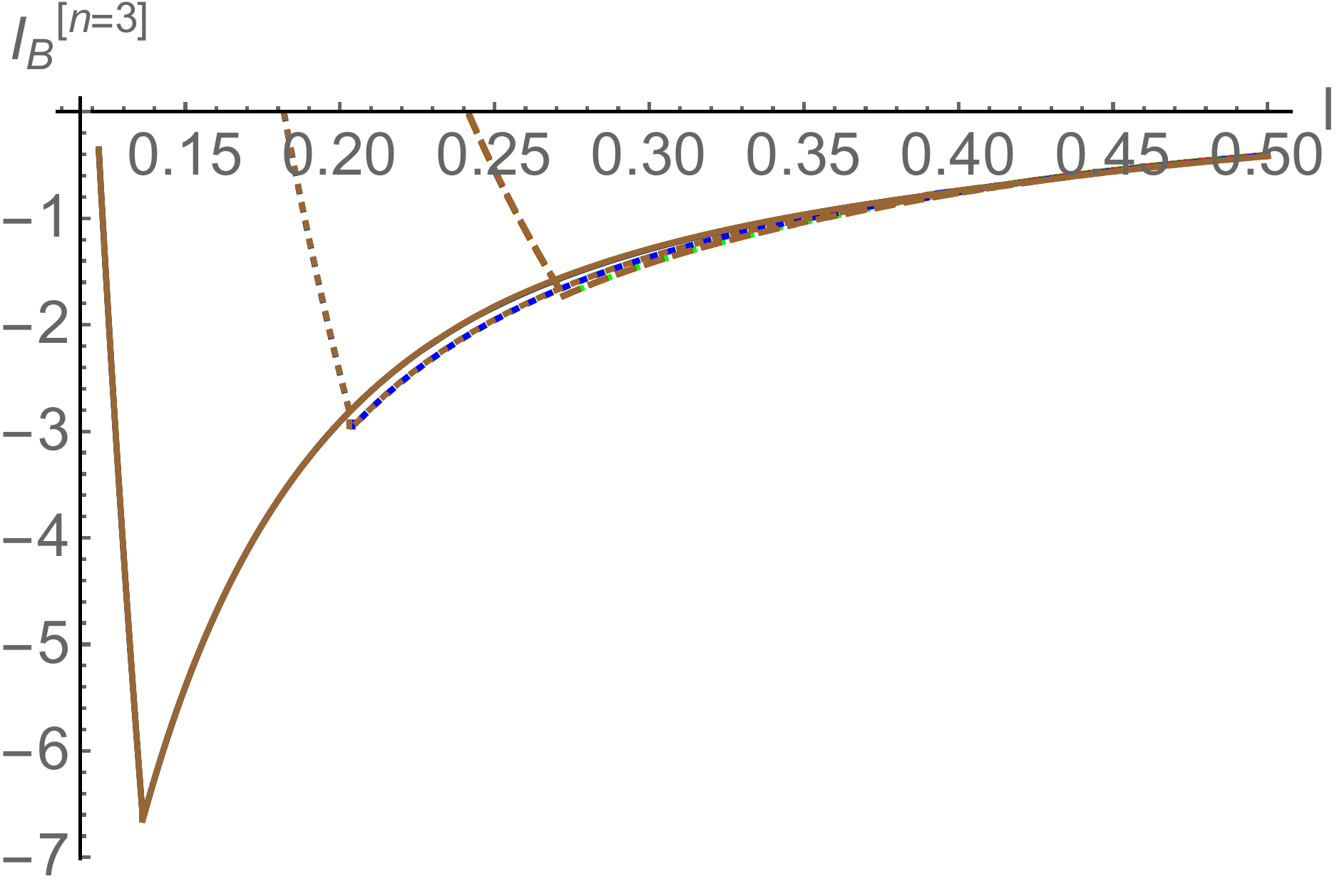}
\caption{ \small Tri-partite information $I_B^{[n=3]}$ as a function of $\ell$ for various values of chemical potential in the specious confined phase.
Here $T=0.9~T_c$ and solid, dotted and dashed lines correspond to $x$=$0.1$, $0.15$ and $0.2$ respectively. The red, green, blue and brown curves
correspond to $\mu=0$, $0.1$, $0.2$ and $0.3$ respectively. In units \text{GeV}.}
\label{I3vsLvsXVsMuTPt9}
\end{minipage}
\hspace{0.4cm}
\begin{minipage}[b]{0.5\linewidth}
\centering
\includegraphics[width=2.8in,height=2.3in]{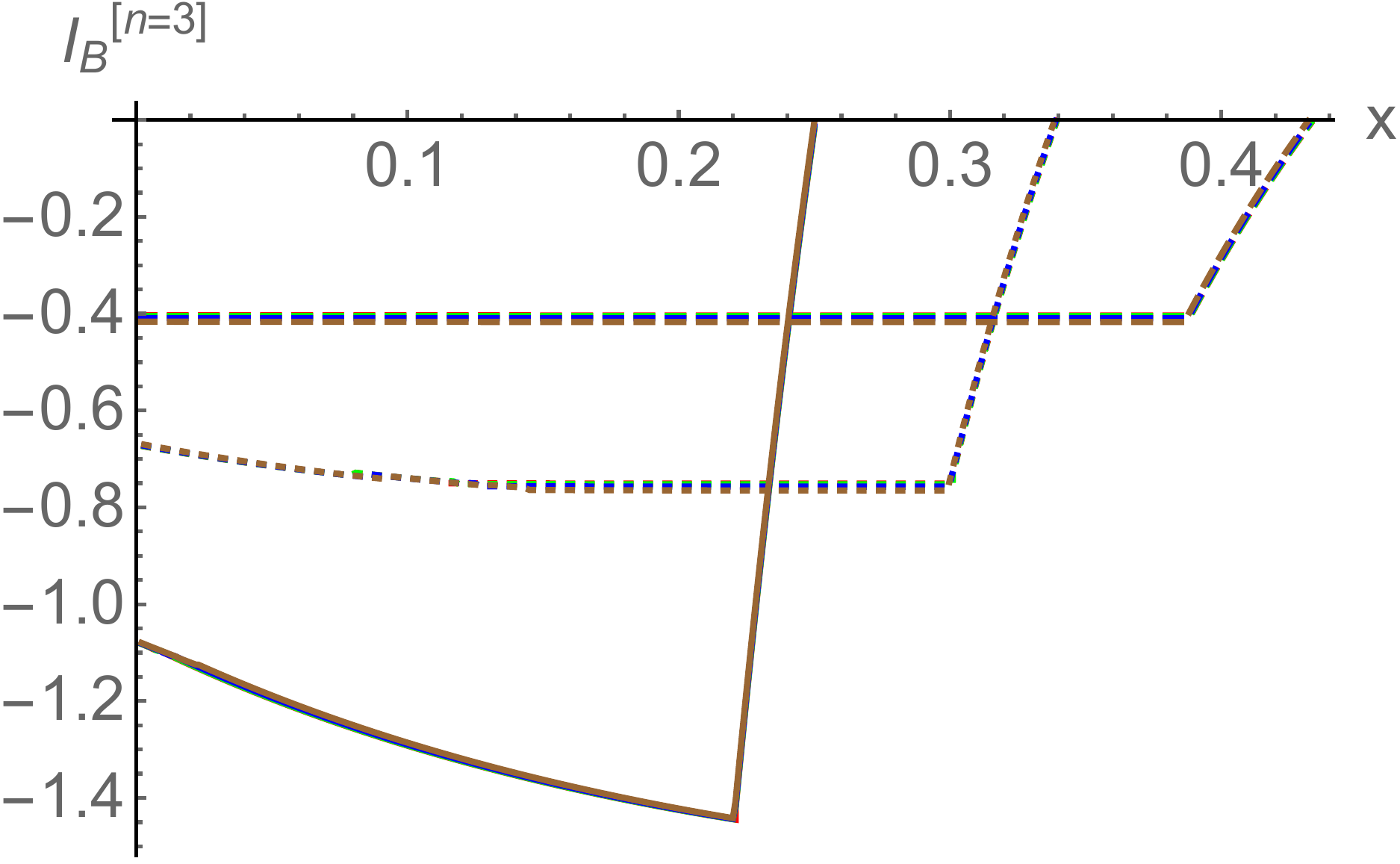}
\caption{\small Tri-partite information $I_B^{[n=3]}$ as a function of $x$ for various values of chemical potential in the specious confined phase.
Here $T=0.9~T_c$ and solid, dotted and dashed lines correspond to $\ell$=$0.3$, $0.4$ and $0.5$ respectively. The red, green, blue and brown curves
correspond to $\mu=0$, $0.1$, $0.2$ and $0.3$ respectively. In units \text{GeV}.}
\label{I3vsXvsLVsMuTPt9}
\end{minipage}
\end{figure}

The $3$-partite information as a function of $x$ and $\ell$ are shown in Figures~\ref{I3vsLvsXVsMuTPt9} and \ref{I3vsXvsLVsMuTPt9}. We find that $3$-partite information,
like mutual information, shows almost no dependence on temperature and chemical potential. In particular, various temperature and chemical
potential dependent profiles of the $3$-partite information overlap with each other, and even the length scale where non-analyticity in the $3$-partite information appears
does not change with temperature and chemical potential. The temperature independent behaviour of $3$-partite information in the specious confined phase
is therefore very different from the deconfined phase where $3$-partite information was instead found to vary with temperature
(see Figure~\ref{I3vsXvsLAdSBHMu0T1Pt2case1}).\\

Similarly, we find that even the $4$-partite information does not show any dependence on temperature and chemical potential. Although it is hard to explicitly
establish this result for a general $n$, however the structure of $n$-strip phase diagram and the general trend do suggest that the corresponding
$n$-partite information is independent of temperature and chemical potential as well. Again, this behaviour of $4$-partite information should be contrasted from the deconfined phase $4$-partite information where it does depend on temperature and chemical potential.

\subsubsection{With large black hole background: $n$ strips}
\begin{figure}[h!]
\begin{minipage}[b]{0.5\linewidth}
\centering
\includegraphics[width=2.8in,height=2.3in]{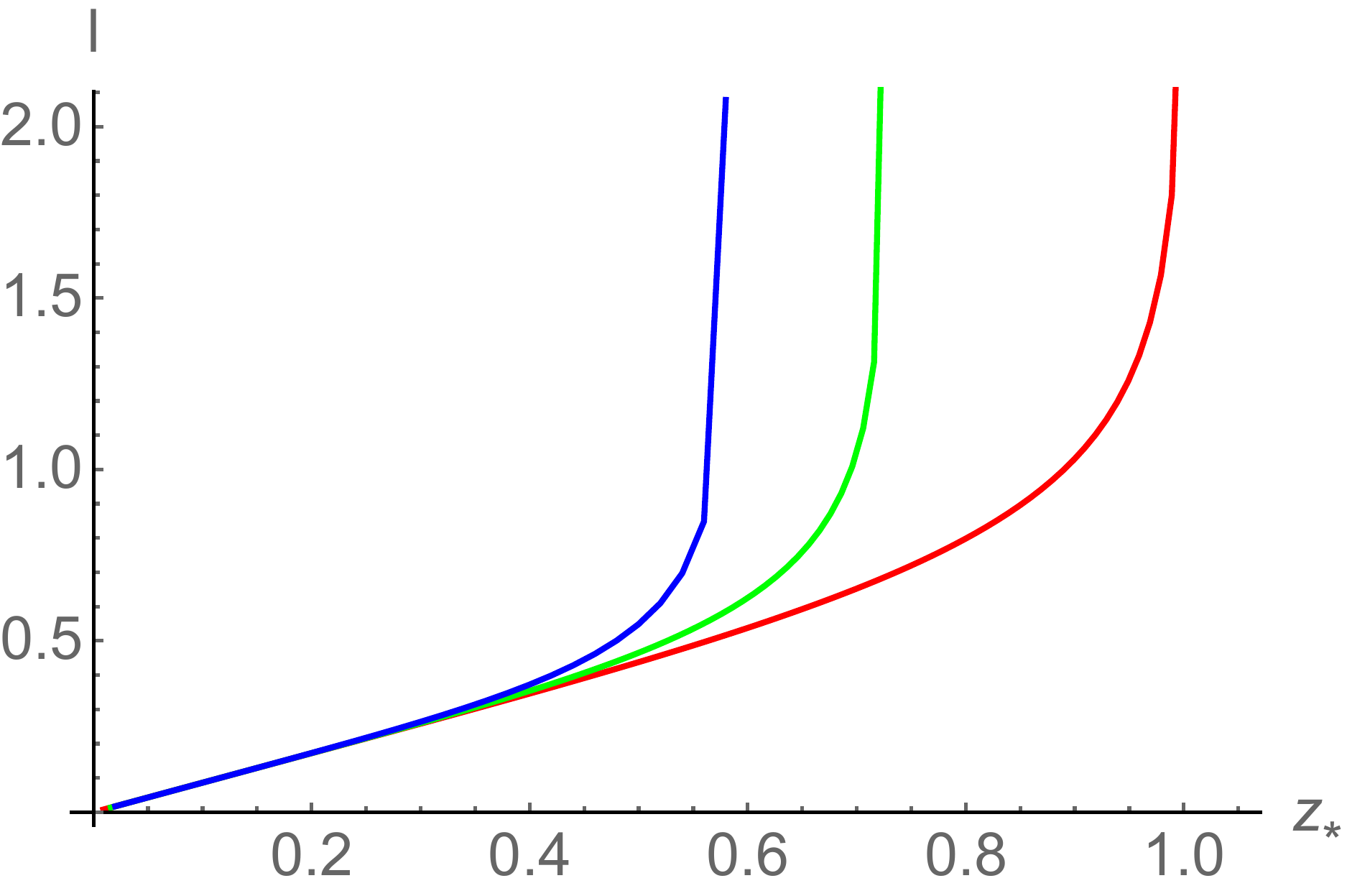}
\caption{ \small $\ell$ as a function of $z_*$ in the dual deconfinement phase of the large black hole. Here $\mu=0$ and red, green
and blue curves correspond to $T/T_{c}=1.2$, $1.6$ and $2.0$ respectively. In units \text{GeV}.}
\label{zsvslAdSlargeBHMu0case1}
\end{minipage}
\hspace{0.4cm}
\begin{minipage}[b]{0.5\linewidth}
\centering
\includegraphics[width=2.8in,height=2.3in]{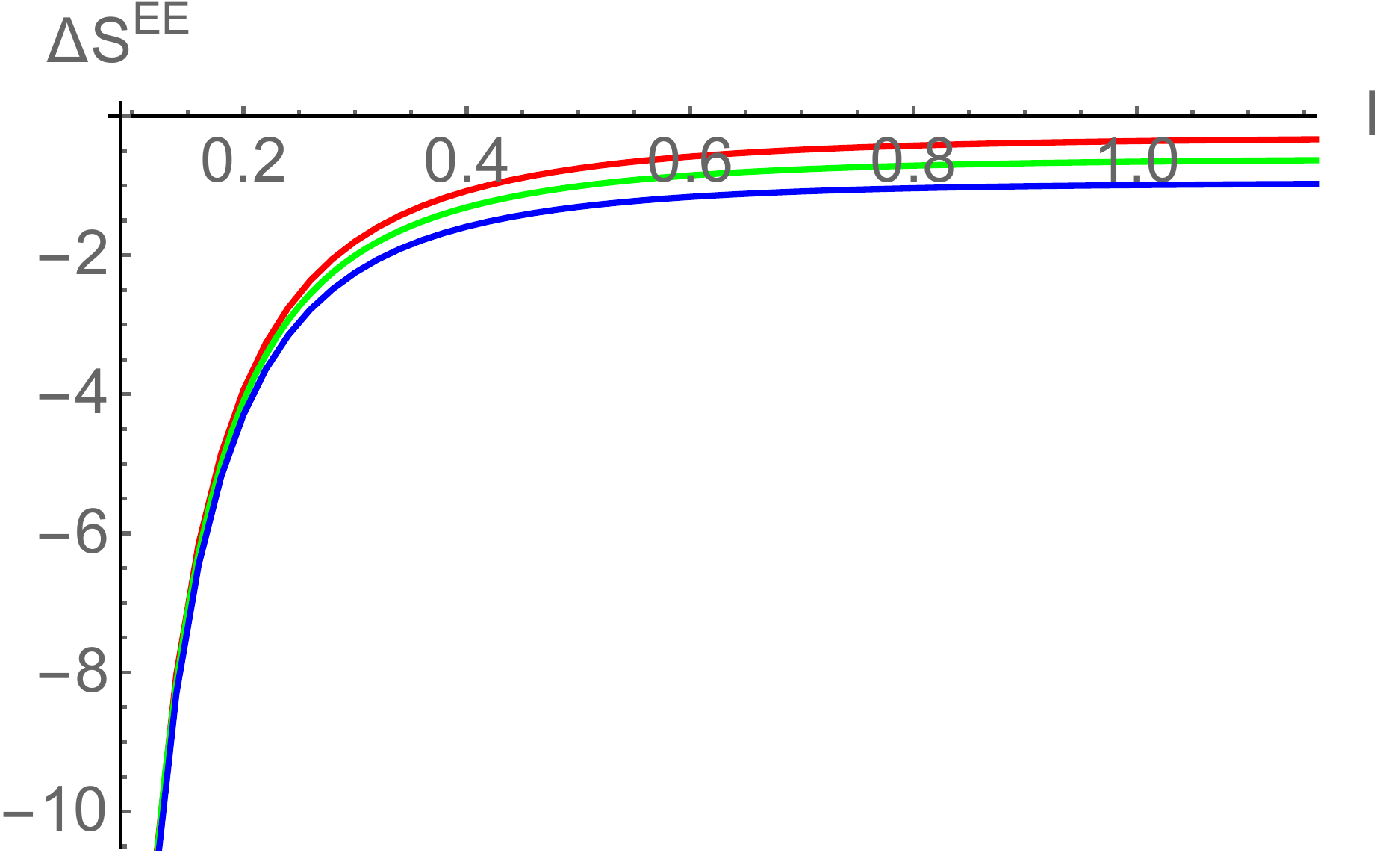}
\caption{\small $\Delta S^{EE}=S^{EE}_{con}-S^{EE}_{discon}$ as a function of $\ell$ in the dual deconfinement phase of the large black
hole. Here $\mu=0$ and red, green and blue curves correspond to $T/T_{c}=1.2$, $1.6$ and $2.0$ respectively. In units \text{GeV}.}
\label{lvsSEEAdSBhlargeMu0case1}
\end{minipage}
\end{figure}

Now, we will briefly mention the results for the dual deconfined phase which corresponds to the large black hole phase. The results for one strip
(or the entanglement entropy) are shown in Figures~\ref{zsvslAdSlargeBHMu0case1} and \ref{lvsSEEAdSBhlargeMu0case1}. There is again a one to one
relation between $\ell$ and $z_*$, with neither $\ell_{max}$ nor $\ell_{crit}$ exist. The connected entangling surface moves more and more towards
the horizon as the subsystem size increases, thereby probing deeper spacetime structure, and have lower entanglement entropy than the disconnected
surface at all $\ell$. Therefore, no connected/disconnected phase transition and $\ell_{crit}$ exist in the deconfined phase. These results are
similar to the deconfined phase results obtained in the previous section using the scale factor $P_1(z)$. In fact, we have checked by taking other forms
of the scale factor $P(z)$ as well that these results for the entanglement entropy remain the same in the dual deconfined phase. Moreover, even in inconsistent
models like soft and hard walls, similar results in the deconfined phases can be obtained. This suggests a universality in the entanglement structure
of the dual deconfined phase.\\

Similarly, for $n \geq 2$ strips, we do not find many differences from the deconfined phase results discussed in section 3.2. $S_A$ and $S_B$
phases and the corresponding phase diagram display the same features as were observed previously. In the phase diagram,
higher temperature again tries to enhance the parameter space of $S_A$ whereas higher chemical potential tries to enhance the parameter space
of $S_B$. The phase transition is again accompanied by a change in the order of mutual information in the deconfined phase. The mutual information also behaves desirably and display the same features as were observed previously
using $P_1(z)$. In particular, $I_{B}^{[n=2]}$ again asymptotically approaches to a temperature and chemical potential dependent constant value
(\textit{i.e.} its large $\ell$ asymptotic value gets enhanced with both temperature and chemical potential). The closeness of $P_2(z)$ deconfined phase with the
deconfined phase of $P_1(z)$ goes beyond the mutual information and we find that even the $3$ and $4$ partite information exhibit similar features in these
deconfined phases.

\subsubsection{Small/large hole phase transition and mutual information}
\begin{figure}[h!]
\centering
\includegraphics[width=2.8in,height=2.3in]{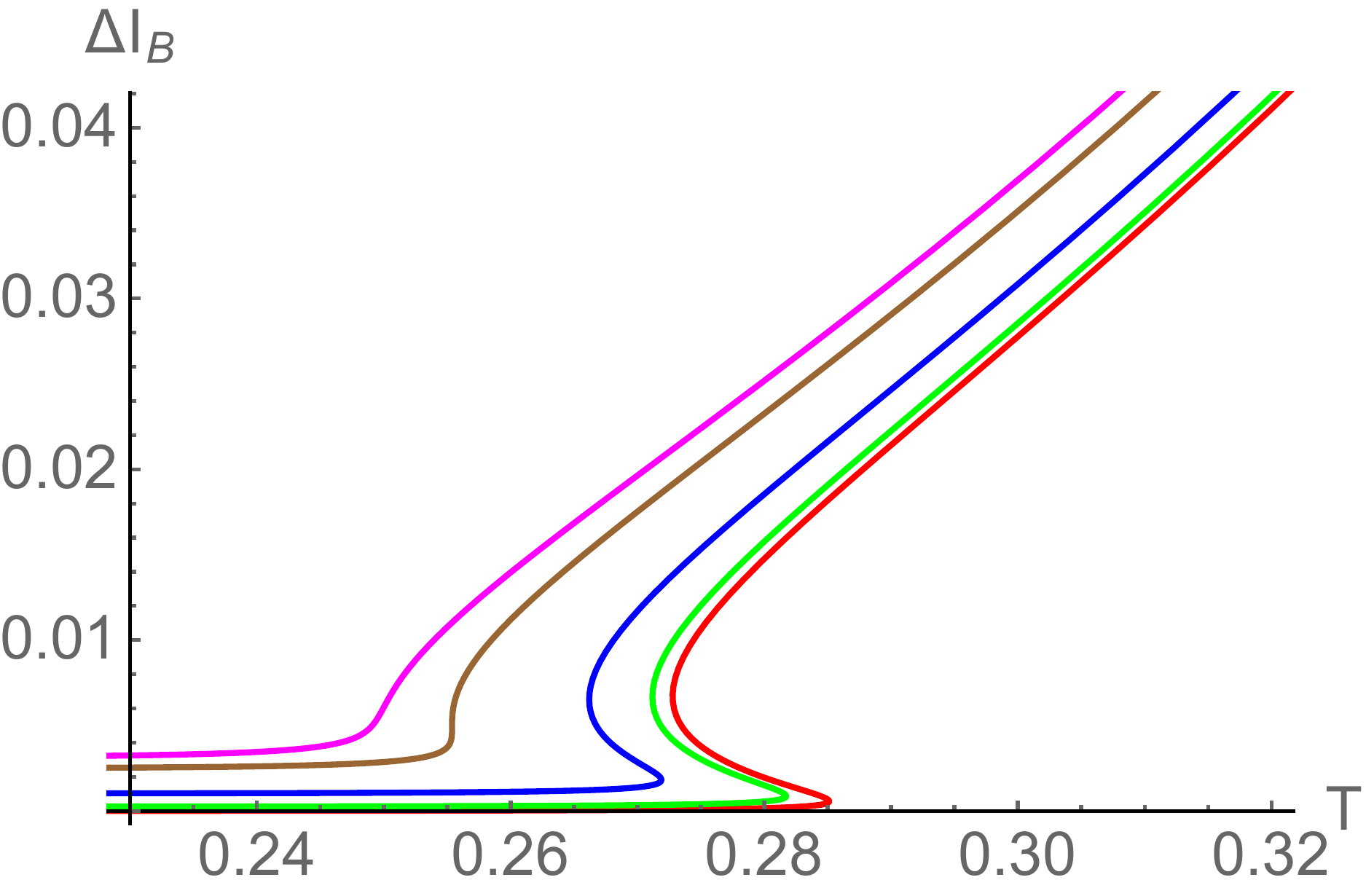}
\caption{ \small $\Delta I_B=I_{B}^{\text{Thermal-AdS}}-I_{B}^{\text{Black hole}}$ as a function of $T$ for various values $\mu$. Here $\ell=0.2$ and $x=0.1$
are used and red, green, blue, brown
and magenta curves correspond to $\mu=0$, $0.1$, $0.2$, $0.312$ and $0.35$ respectively. In units \text{GeV}.}
\label{IBvsTvsMu2stripsBHthermodynamicscase2}
\end{figure}
We close this section by analysis the mutual information in $T-\mu$ plane. The objective here is to see whether the $3$-partite information,
like the entanglement entropy, captures the small/large black hole (or the dual specious-confined/deconfined) phase transitions just as it did for
the thermal-AdS/black hole phase transition. Our results for the mutual information are shown in Figure~\ref{IBvsTvsMu2stripsBHthermodynamicscase2} for a
fixed $\ell=0.2 \ GeV^{-1}$ and $x=0.1 \ GeV^{-1}$, although the main results of our investigation remain unchanged even for other
values of $\ell$ and $x$. We again find that the structure of mutual information resembles remarkably well with the Bekenstein-Hawking entropy
(see Figure~\ref{TvsSBHvsMucase2}). In particular, the mutual information also exhibits three branches in $T-\mu$ plane. The branches
with positive slope are stable whereas the branch with a negative slope is unstable. These three branches exist only when $\mu<\mu_c$,
and for $\mu>\mu_c$ the negative slope branch ceases to exist and we have only one branch. Importantly, a critical feature like $\mu_c$
does not change when other values of $\ell$ and $x$ are considered. The mutual information therefore again experiences the deviations
in the spacetime geometry caused by the black hole phase transition. Moreover, we have checked that similar results also hold for $3$- and $4$-partite
information. Our analysis therefore once again suggests that the other information theoretic quantities can also be used to probe black hole (or
dual specious-confined/deconfined) phase transitions.

\section{Conclusions}
In this paper, using the holographic RT prescription of the entanglement entropy, we have investigated the mutual and $n$-partite information of a strongly coupled QCD
theory whose dual gravitational theory is described by a consistent phenomenological bottom-up Einstein-Maxwell-dilaton model. This is an extension of our previous work on the entanglement
entropy \cite{Dudal:2018ztm}, and the objective here was to further investigate how the mutual and $n$-partite information can shed new light
on the confinement mechanism. The excited dilaton field in this model allowed us to modify the nature of the gravity solutions by choosing
an appropriate form of the scale factor. With one form factor ($P_{1}(z)$), we found a thermal-AdS/black hole phase transition which on the dual boundary side
corresponds to the standard confined/deconfined phase transition. We then studied the entanglement entropy phase diagram by considering $n$ disjoint intervals. In the confining background, with $n\geq2$, we found a rich phase diagram consisting of four distinct connected and disconnected surfaces in the parameter space of
$\ell$ and $x$. With $n\geq2$, unlike $n=1$, the order of the entanglement entropy may or may not change as the transition point is crossed. We then
analyzed the mutual and $n$-partite information in the confining background and found that the mutual information is monogamous and that the $n$-partite information exhibits non-analyticity in its structure. In the deconfining background, on the other hand, only two connected phases appeared, leading to a much simpler phase diagram. We further found that higher temperature makes the parameter space of $S_B$ phase smaller whereas higher chemical potential makes the parameter space of $S_B$ larger, suggesting that one should look for the low temperature/high
chemical potential region of the QCD phase diagram to find a non-trivial profile of the mutual information. Moreover, the separation length at which non-analyticity in $3$-partite information appears is found to decrease with temperature whereas it enhances with chemical potential.

The second form factor ($P_{2}(z)$) instead leads to the small/large black hole phase transition which in the dual boundary side corresponds to the
specious-confined/deconfined phase transition. In this case, the entanglement structure of the deconfined phase is found to be similar to
the deconfined phase entanglement structure obtained using $P_{1}(z)$. However, the entanglement structure of the specious-confined phase displayed many
dissimilarities with the standard confined phase. In particular,
a novel connected/connected (instead of a connected/disconnected) surface phase transition appeared with $n=1$, which greatly modified its $n\geq2$
entanglement entropy phase diagram compared to the standard confined phase. The small black hole phase also allowed us to probe the effect of temperature and chemical potential on the entanglement phase diagram. We found that $n\geq2$ entanglement entropy phase diagram is almost independent of
temperature and chemical potential in small $\ell<\ell_{c}$ region whereas the phase space of $S_B$ gets slightly enhanced with chemical potential in large
$\ell>\ell_{c}$ region. Interestingly, the mutual information and $n$-partite information although behaving desirably in the specious confined phase, however, exhibit no
dependence on temperature and chemical potential. This is a new and interesting prediction from holography, but unfortunately, unlike the entanglement
entropy, we do not yet have any corresponding independent lattice result to compare our holographic result.

We further studied the Hawking/Page and small/large black hole phase transitions using the mutual and $n$-partite information and found that, just like the entanglement entropy,
these quantities also capture the essence of black hole phase transitions. Since the imprints of phase transition were also seen on the mutual and $n$-partite information, our analysis therefore extended the number of boundary observables that can be used to probe these phase transitions.

There are several directions in which the analysis of this paper can be expanded. In present work, we have considered the oversimplified case of equal strip
and separation lengths. This makes the phase space effectively two dimensional. However, for the case of $n$ unequal strips and separations lengths the phase space
would be $2n-1$ dimensional.
Analysis of this multi-dimensional phase space structure would although be bit tedious, however it might shed new
light on the QCD phases, especially in the confining background. Another direction to extend our work is to include a background magnetic field and use
the entanglement structure to investigate (inverse) magnetic catalysis. We hope to come back to these issues soon.

\section*{Acknowledgments}
I am grateful to D. Dudal for useful discussions and for giving valuable comments. I would like to thank D. Dudal and P. Roy for careful reading of the manuscript and pointing out the necessary corrections. This work is supported by the Department of Science and Technology, Government of India under the Grant Agreement number IFA17-PH207 (INSPIRE Faculty Award).


\begin{thebibliography}{99}


 \bibitem{Maldacena:1997re}
  J.~M.~Maldacena,
  ``The Large $N$ limit of superconformal field theories and supergravity,''
  Int.\ J.\ Theor.\ Phys.\  {\bf 38}, 1113 (1999)
  [Adv.\ Theor.\ Math.\ Phys.\  {\bf 2}, 231 (1998)]
  [hep-th/9711200].

 \bibitem{Gubser:1998bc}
  S.~S.~Gubser, I.~R.~Klebanov and A.~M.~Polyakov,
  ``Gauge theory correlators from noncritical string theory,''
  Phys.\ Lett.\ B {\bf 428}, 105 (1998)
  [hep-th/9802109].

  \bibitem{Witten:1998qj}
  E.~Witten,
  ``Anti-de Sitter space and holography,''
  Adv.\ Theor.\ Math.\ Phys.\  {\bf 2}, 253 (1998)
  [hep-th/9802150].

 \bibitem{Ryu:2006bv}
  S.~Ryu and T.~Takayanagi,
  ``Holographic derivation of entanglement entropy from AdS/CFT,''
  Phys.\ Rev.\ Lett.\  {\bf 96}, 181602 (2006)
  [hep-th/0603001].

  \bibitem{Ryu:2006ef}
  S.~Ryu and T.~Takayanagi,
  ``Aspects of Holographic Entanglement Entropy,''
  JHEP {\bf 0608}, 045 (2006)
  [hep-th/0605073].

\bibitem{Lewkowycz:2013nqa}
  A.~Lewkowycz and J.~Maldacena,
  ``Generalized gravitational entropy,''
  JHEP {\bf 1308}, 090 (2013)
  [arXiv:1304.4926 [hep-th]].


  \bibitem{Headrick:2014cta}
  M.~Headrick, V.~E.~Hubeny, A.~Lawrence and M.~Rangamani,
  ``Causality \& holographic entanglement entropy,''
  JHEP {\bf 1412}, 162 (2014)
  doi:10.1007/JHEP12(2014)162
  [arXiv:1408.6300 [hep-th]].

\bibitem{VanRaamsdonk:2010pw}
  M.~Van Raamsdonk,
  ``Building up spacetime with quantum entanglement,''
  Gen.\ Rel.\ Grav.\  {\bf 42}, 2323 (2010)
  [Int.\ J.\ Mod.\ Phys.\ D {\bf 19}, 2429 (2010)]
  [arXiv:1005.3035 [hep-th]].

  \bibitem{Balasubramanian:2013lsa}
  V.~Balasubramanian, B.~D.~Chowdhury, B.~Czech, J.~de Boer and M.~P.~Heller,
  ``Bulk curves from boundary data in holography,''
  Phys.\ Rev.\ D {\bf 89}, no. 8, 086004 (2014)
  [arXiv:1310.4204 [hep-th]].

\bibitem{Hayden:2011ag}
  P.~Hayden, M.~Headrick and A.~Maloney,
  ``Holographic Mutual Information is Monogamous,''
  Phys.\ Rev.\ D {\bf 87}, no. 4, 046003 (2013)
  [arXiv:1107.2940 [hep-th]].

\bibitem{Headrick:2010zt}
  M.~Headrick,
  ``Entanglement Renyi entropies in holographic theories,''
  Phys.\ Rev.\ D {\bf 82}, 126010 (2010)
  [arXiv:1006.0047 [hep-th]].

\bibitem{Allais:2011ys}
  A.~Allais and E.~Tonni,
  ``Holographic evolution of the mutual information,''
  JHEP {\bf 1201}, 102 (2012)
  [arXiv:1110.1607 [hep-th]].


\bibitem{Witten9803}
  E.~Witten,
  ``Anti-de Sitter space, thermal phase transition, and confinement in gauge theories,''
  Adv.\ Theor.\ Math.\ Phys.\  {\bf 2} (1998) 505
  [hep-th/9803131].

\bibitem{Sakai0412}
 T.~Sakai and S.~Sugimoto,
  ``Low energy hadron physics in holographic QCD,''
  Prog.\ Theor.\ Phys.\  {\bf 113} (2005) 843
  [hep-th/0412141].

\bibitem{Sakai0507}
 T.~Sakai and S.~Sugimoto,
  ``More on a holographic dual of QCD,''
  Prog.\ Theor.\ Phys.\  {\bf 114} (2005) 1083
  [hep-th/0507073].

   \bibitem{Polchinski0003}
  J.~Polchinski and M.~J.~Strassler,
  ``The String dual of a confining four-dimensional gauge theory,''
  hep-th/0003136.


\bibitem{Kruczenski0311}
  M.~Kruczenski, D.~Mateos, R.~C.~Myers and D.~J.~Winters,
  ``Towards a holographic dual of large $N_c$ QCD,''
  JHEP {\bf 0405} (2004) 041
  [hep-th/0311270].


\bibitem{Klebanov0007}
  I.~R.~Klebanov and M.~J.~Strassler,
  ``Supergravity and a confining gauge theory: Duality cascades and chi SB resolution of naked singularities,''
  JHEP {\bf 0008} (2000) 052
  [hep-th/0007191].

\bibitem{Karch0205}
A.~Karch and E.~Katz,
  ``Adding flavor to AdS/CFT,''
  JHEP {\bf 0206} (2002) 043
  [hep-th/0205236].

\bibitem{Erlich}
  J.~Erlich, E.~Katz, D.~T.~Son and M.~A.~Stephanov,
  ``QCD and a holographic model of hadrons,''
  Phys.\ Rev.\ Lett.\  {\bf 95} (2005) 261602
  [hep-ph/0501128].

  \bibitem{Herzog0608}
  C.~P.~Herzog,
  ``A Holographic Prediction of the Deconfinement Temperature,''
  Phys.\ Rev.\ Lett.\  {\bf 98} (2007) 091601
  [hep-th/0608151].


\bibitem{Karch0602}
  A.~Karch, E.~Katz, D.~T.~Son and M.~A.~Stephanov,
  ``Linear confinement and AdS/QCD,''
  Phys.\ Rev.\ D {\bf 74} (2006) 015005
  [hep-ph/0602229].


\bibitem{Gubser:2008yx}
  S.~S.~Gubser, A.~Nellore, S.~S.~Pufu and F.~D.~Rocha,
  ``Thermodynamics and bulk viscosity of approximate black hole duals to finite temperature quantum chromodynamics,''
  Phys.\ Rev.\ Lett.\  {\bf 101}, 131601 (2008)
  [arXiv:0804.1950 [hep-th]].

  \bibitem{Gubser:2008ny}
  S.~S.~Gubser and A.~Nellore,
  ``Mimicking the QCD equation of state with a dual black hole,''
  Phys.\ Rev.\ D {\bf 78}, 086007 (2008)
  [arXiv:0804.0434 [hep-th]].

  \bibitem{DeWolfe:2011ts}
  O.~DeWolfe, S.~S.~Gubser and C.~Rosen,
  ``Dynamic critical phenomena at a holographic critical point,''
  Phys.\ Rev.\ D {\bf 84}, 126014 (2011)
  [arXiv:1108.2029 [hep-th]].


  \bibitem{Gursoy:2010fj}
  U.~G\"ursoy, E.~Kiritsis, L.~Mazzanti, G.~Michalogiorgakis and F.~Nitti,
  ``Improved Holographic QCD,''
  Lect.\ Notes Phys.\  {\bf 828}, 79 (2011)
  [arXiv:1006.5461 [hep-th]].

\bibitem{Gursoy}
 U.~G\"ursoy, E.~Kiritsis, L.~Mazzanti and F.~Nitti,
  ``Deconfinement and Gluon Plasma Dynamics in Improved Holographic QCD,''
  Phys.\ Rev.\ Lett.\  {\bf 101} (2008) 181601
  [arXiv:0804.0899 [hep-th]].

\bibitem{Jarvinen:2015ofa}
  M.~J\"arvinen,
  ``Massive holographic QCD in the Veneziano limit,''
  JHEP {\bf 1507}, 033 (2015)
  [arXiv:1501.07272 [hep-ph]].


  \bibitem{Callebaut:2011ab}
  N.~Callebaut, D.~Dudal and H.~Verschelde,
  ``Holographic rho mesons in an external magnetic field,''
  JHEP {\bf 1303}, 033 (2013)
  [arXiv:1105.2217 [hep-th]].

  \bibitem{Dudal:2015wfn}
  D.~Dudal, D.~R.~Granado and T.~G.~Mertens,
  ``No inverse magnetic catalysis in the QCD hard and soft wall models,''
  Phys.\ Rev.\ D {\bf 93} (2016) no.12,  125004
  [arXiv:1511.04042 [hep-th]].

\bibitem{Callebaut:2013ria}
  N.~Callebaut and D.~Dudal,
  ``Transition temperature(s) of magnetized two-flavor holographic QCD,''
  Phys.\ Rev.\ D {\bf 87}, no. 10, 106002 (2013)
  [arXiv:1303.5674 [hep-th]].

  \bibitem{Dudal:2014jfa}
  D.~Dudal and T.~G.~Mertens,
  ``Melting of charmonium in a magnetic field from an effective AdS/QCD model,''
  Phys.\ Rev.\ D {\bf 91}, 086002 (2015)
  [arXiv:1410.3297 [hep-th]].

\bibitem{Dudal:2018rki}
  D.~Dudal
   and T.~G.~Mertens,
  ``Holographic estimate of heavy quark diffusion in a magnetic field,''
  Phys.\ Rev.\ D {\bf 97}, no. 5, 054035 (2018)
  [arXiv:1802.02805 [hep-th]].


\bibitem{Fang:2015ytf}
  Z.~Fang, S.~He and D.~Li,
  ``Chiral and Deconfining Phase Transitions from Holographic QCD Study,''
  Nucl.\ Phys.\ B {\bf 907}, 187 (2016)
  [arXiv:1512.04062 [hep-ph]].


\bibitem{Giataganas:2017koz}
  D.~Giataganas, U.~G\"ursoy and J.~F.~Pedraza,
  ``Strongly-coupled anisotropic gauge theories and holography,''
  arXiv:1708.05691 [hep-th].


 \bibitem{Panero:2009tv}
  M.~Panero,
  ``Thermodynamics of the QCD plasma and the large-$N$ limit,''
  Phys.\ Rev.\ Lett.\  {\bf 103}, 232001 (2009)
  [arXiv:0907.3719 [hep-lat]].

   \bibitem{Paula}
 W.~de Paula, T.~Frederico, H.~Forkel and M.~Beyer,
  ``Dynamical AdS/QCD with area-law confinement and linear Regge trajectories,''
  Phys.\ Rev.\ D {\bf 79} (2009) 075019
  [arXiv:0806.3830 [hep-ph]].


  \bibitem{Noronha:2010hb}
  J.~Noronha,
  ``The Heavy Quark Free Energy in QCD and in Gauge Theories with Gravity Duals,''
  Phys.\ Rev.\ D {\bf 82}, 065016 (2010)
  [arXiv:1003.0914 [hep-th]].


  \bibitem{He:2013qq}
  S.~He, S.~Y.~Wu, Y.~Yang and P.~H.~Yuan,
  ``Phase Structure in a Dynamical Soft-Wall Holographic QCD Model,''
  JHEP {\bf 1304}, 093 (2013)
  [arXiv:1301.0385 [hep-th]].

 \bibitem{Yang:2015aia}
  Y.~Yang and P.~H.~Yuan, ``Confinement-deconfinement phase transition for heavy quarks in a soft wall holographic QCD model,''
  JHEP {\bf 1512}, 161 (2015)
  [arXiv:1506.05930 [hep-th]].

   \bibitem{Cai:2012xh}
  R.~G.~Cai, S.~He and D.~Li,
  ``A hQCD model and its phase diagram in Einstein-Maxwell-Dilaton system,''
  JHEP {\bf 1203}, 033 (2012)
  [arXiv:1201.0820 [hep-th]].


\bibitem{Knaute:2017opk}
  J.~Knaute, R.~Yaresko and B.~K\"ampfer,
 ``Holographic QCD phase diagram with critical point from Einstein-Maxwell-dilaton dynamics,''
  Phys.\ Lett.\ B {\bf 778}, 419 (2018)
  [arXiv:1702.06731 [hep-ph]].


\bibitem{Arefeva:2018hyo}
  I.~Aref'eva and K.~Rannu,
  ``Holographic Anisotropic Background with Confinement-Deconfinement Phase Transition,''
  JHEP {\bf 1805}, 206 (2018)
  [arXiv:1802.05652 [hep-th]].


\bibitem{Buividovich:2008kq}
  P.~V.~Buividovich and M.~I.~Polikarpov,
  ``Numerical study of entanglement entropy in SU(2) lattice gauge theory,''
  Nucl.\ Phys.\ B {\bf 802} (2008) 458
  [arXiv:0802.4247 [hep-lat]].


\bibitem{Buividovich:2008gq}
  P.~V.~Buividovich and M.~I.~Polikarpov,
  ``Entanglement entropy in gauge theories and the holographic principle for electric strings,''
  Phys.\ Lett.\ B {\bf 670} (2008) 141
  [arXiv:0806.3376 [hep-th]].

\bibitem{Itou:2015cyu}
  E.~Itou, K.~Nagata, Y.~Nakagawa, A.~Nakamura and V.~I.~Zakharov,
  ``Entanglement in Four-Dimensional SU(3) Gauge Theory,''
  PTEP {\bf 2016} (2016) no.6,  061B01
  [arXiv:1512.01334 [hep-th]].

\bibitem{Rabenstein:2018bri}
  A.~Rabenstein, N.~Bodendorfer, P.~Buividovich and A.~Schäfer,
  ``Lattice study of R\'enyi entanglement entropy in $SU(N_c)$ lattice Yang-Mills theory with $N_c = 2, 3, 4$,''
  arXiv:1812.04279 [hep-lat].


\bibitem{Klebanov0709}
I.~R.~Klebanov, D.~Kutasov and A.~Murugan,
  ``Entanglement as a probe of confinement,''
  Nucl.\ Phys.\ B {\bf 796} (2008) 274
  [arXiv:0709.2140 [hep-th]].



\bibitem{Kola1403}
 U.~Kol, C.~Nunez, D.~Schofield, J.~Sonnenschein and M.~Warschawski,
  ``Confinement, Phase Transitions and non-Locality in the Entanglement Entropy,''
  JHEP {\bf 1406} (2014) 005
  [arXiv:1403.2721 [hep-th]].

\bibitem{Fujita0806}
M.~Fujita, T.~Nishioka and T.~Takayanagi,
``Geometric Entropy and Hagedorn/Deconfinement Transition,''
  JHEP {\bf 0809} (2008) 016
  [arXiv:0806.3118 [hep-th]].

\bibitem{Lewkowycz}
  A.~Lewkowycz,
  ``Holographic Entanglement Entropy and Confinement,''
  JHEP {\bf 1205} (2012) 032
  [arXiv:1204.0588 [hep-th]].

\bibitem{Kim}
N.~Kim,
  ``Holographic entanglement entropy of confining gauge theories with flavor,''
  Phys.\ Lett.\ B {\bf 720} (2013) 232.

\bibitem{Ghodrati}
M.~Ghodrati,
  ``Schwinger Effect and Entanglement Entropy in Confining Geometries,''
  Phys.\ Rev.\ D {\bf 92} (2015) no.6,  065015
  [arXiv:1506.08557 [hep-th]].

\bibitem{Ali-Akbari:2017vtb}
  M.~Ali-Akbari and M.~Lezgi,
  ``Holographic QCD, entanglement entropy, and critical temperature,''
  Phys.\ Rev.\ D {\bf 96}, no. 8, 086014 (2017)
  [arXiv:1706.04335 [hep-th]].

\bibitem{Knaute:2017lll}
  J.~Knaute and B.~K\"ampfer,
  ``Holographic Entanglement Entropy in the QCD Phase Diagram with a Critical Point,''
  Phys.\ Rev.\ D {\bf 96}, no. 10, 106003 (2017)
  [arXiv:1706.02647 [hep-ph]].


\bibitem{Anber:2018ohz}
  M.~M.~Anber and B.~J.~Kolligs,
  ``Entanglement entropy, dualities, and deconfinement in gauge theories,''
  arXiv:1804.01956 [hep-th].


\bibitem{Dudal:2016joz}
  D.~Dudal and S.~Mahapatra,
  ``Confining gauge theories and holographic entanglement entropy with a magnetic field,''
  JHEP {\bf 1704}, 031 (2017)
  [arXiv:1612.06248 [hep-th]].

\bibitem{Dudal:2018ztm}
  D.~Dudal and S.~Mahapatra,
  ``Interplay between the holographic QCD phase diagram and entanglement entropy,''
  JHEP {\bf 1807}, 120 (2018)
  [arXiv:1805.02938 [hep-th]].


\bibitem{Gursoy:2018ydr}
  U.~Gürsoy, M.~Järvinen, G.~Nijs and J.~F.~Pedraza,
  ``Inverse Anisotropic Catalysis in Holographic QCD,''
  arXiv:1811.11724 [hep-th].

\bibitem{Calabrese:2009ez}
  P.~Calabrese, J.~Cardy and E.~Tonni,
  ``Entanglement entropy of two disjoint intervals in conformal field theory,''
  J.\ Stat.\ Mech.\  {\bf 0911}, P11001 (2009)
  [arXiv:0905.2069 [hep-th]].

\bibitem{Calabrese:2010he}
  P.~Calabrese, J.~Cardy and E.~Tonni,
  ``Entanglement entropy of two disjoint intervals in conformal field theory II,''
  J.\ Stat.\ Mech.\  {\bf 1101}, P01021 (2011)
  [arXiv:1011.5482 [hep-th]].

\bibitem{Balasubramanian:2011at}
  V.~Balasubramanian, A.~Bernamonti, N.~Copland, B.~Craps and F.~Galli,
  ``Thermalization of mutual and tripartite information in strongly coupled two dimensional conformal field theories,''
  Phys.\ Rev.\ D {\bf 84}, 105017 (2011)
  [arXiv:1110.0488 [hep-th]].

\bibitem{Fischler:2012uv}
  W.~Fischler, A.~Kundu and S.~Kundu,
  ``Holographic Mutual Information at Finite Temperature,''
  Phys.\ Rev.\ D {\bf 87}, no. 12, 126012 (2013)
  [arXiv:1212.4764 [hep-th]].
  
  \bibitem{Kundu:2016dyk}
  S.~Kundu and J.~F.~Pedraza,
  ``Aspects of Holographic Entanglement at Finite Temperature and Chemical Potential,''
  JHEP {\bf 1608}, 177 (2016)
  [arXiv:1602.07353 [hep-th]].
  

\bibitem{Casini:2015woa}
  H.~Casini, M.~Huerta, R.~C.~Myers and A.~Yale,
  ``Mutual information and the F-theorem,''
  JHEP {\bf 1510}, 003 (2015)
  [arXiv:1506.06195 [hep-th]].

\bibitem{Morrison:2012iz}
  I.~A.~Morrison and M.~M.~Roberts,
  ``Mutual information between thermo-field doubles and disconnected holographic boundaries,''
  JHEP {\bf 1307}, 081 (2013)
  [arXiv:1211.2887 [hep-th]].

\bibitem{Fonda:2014cca}
  P.~Fonda, L.~Giomi, A.~Salvio and E.~Tonni,
  ``On shape dependence of holographic mutual information in AdS$_{4}$,''
  JHEP {\bf 1502}, 005 (2015)
  [arXiv:1411.3608 [hep-th]].

  \bibitem{Alishahiha:2014jxa}
  M.~Alishahiha, M.~R.~Mohammadi Mozaffar and M.~R.~Tanhayi,
  ``On the Time Evolution of Holographic n-partite Information,''
  JHEP {\bf 1509}, 165 (2015)
  [arXiv:1406.7677 [hep-th]].

  \bibitem{MolinaVilaplana:2011xt}
  J.~Molina-Vilaplana and P.~Sodano,
  ``Holographic View on Quantum Correlations and Mutual Information between Disjoint Blocks of a Quantum Critical System,''
  JHEP {\bf 1110}, 011 (2011)
  [arXiv:1108.1277 [quant-ph]].

  \bibitem{Asplund:2013zba}
  C.~T.~Asplund and A.~Bernamonti,
  ``Mutual information after a local quench in conformal field theory,''
  Phys.\ Rev.\ D {\bf 89}, no. 6, 066015 (2014)
  [arXiv:1311.4173 [hep-th]].

\bibitem{Agon:2015ftl}
  C.~Agón and T.~Faulkner,
  ``Quantum Corrections to Holographic Mutual Information,''
  JHEP {\bf 1608}, 118 (2016)
  [arXiv:1511.07462 [hep-th]].

\bibitem{Mozaffar:2015xue}
  M.~R.~Mohammadi Mozaffar, A.~Mollabashi and F.~Omidi,
  ``Holographic Mutual Information for Singular Surfaces,''
  JHEP {\bf 1512}, 082 (2015)
  [arXiv:1511.00244 [hep-th]].

\bibitem{Tanhayi:2015cax}
  M.~R.~Tanhayi,
  ``Thermalization of Mutual Information in Hyperscaling Violating Backgrounds,''
  JHEP {\bf 1603}, 202 (2016)
  [arXiv:1512.04104 [hep-th]].

  \bibitem{Hosseini:2015gua}
  S.~M.~Hosseini and A.~Veliz-Osorio,
  ``Entanglement and mutual information in two-dimensional nonrelativistic field theories,''
  Phys.\ Rev.\ D {\bf 93}, no. 2, 026010 (2016)
  [Phys.\ Rev.\ D {\bf 93}, 026010 (2016)]
  [arXiv:1510.03876 [hep-th]].

\bibitem{Mirabi:2016elb}
  S.~Mirabi, M.~R.~Tanhayi and R.~Vazirian,
  ``On the Monogamy of Holographic $n$-partite Information,''
  Phys.\ Rev.\ D {\bf 93}, no. 10, 104049 (2016)
  [arXiv:1603.00184 [hep-th]].

\bibitem{Hartnoll:2014ppa}
  S.~A.~Hartnoll and R.~Mahajan,
  ``Holographic mutual information and distinguishability of Wilson loop and defect operators,''
  JHEP {\bf 1502}, 100 (2015)
  [arXiv:1407.8191 [hep-th]].

\bibitem{Cardy:2013nua}
  J.~Cardy,
  ``Some results on the mutual information of disjoint regions in higher dimensions,''
  J.\ Phys.\ A {\bf 46}, 285402 (2013)
  doi:10.1088/1751-8113/46/28/285402
  [arXiv:1304.7985 [hep-th]].

\bibitem{Larkoski:2014pca}
  A.~J.~Larkoski, J.~Thaler and W.~J.~Waalewijn,
  ``Gaining (Mutual) Information about Quark/Gluon Discrimination,''
  JHEP {\bf 1411}, 129 (2014)
  doi:10.1007/JHEP11(2014)129
  [arXiv:1408.3122 [hep-ph]].



\bibitem{Ben-Ami:2014gsa}
  O.~Ben-Ami, D.~Carmi and J.~Sonnenschein,
  ``Holographic Entanglement Entropy of Multiple Strips,''
  JHEP {\bf 1411}, 144 (2014)
  [arXiv:1409.6305 [hep-th]].


\bibitem{Dudal:2017max}
  D.~Dudal and S.~Mahapatra,
  ``Thermal entropy of a quark-antiquark pair above and below deconfinement from a dynamical holographic QCD model,''
  Phys.\ Rev.\ D {\bf 96}, no. 12, 126010 (2017)
  [arXiv:1708.06995 [hep-th]].

\bibitem{Mahapatra:2018gig}
  S.~Mahapatra and P.~Roy,
  ``On the time dependence of holographic complexity in a dynamical Einstein-dilaton model,''
  JHEP {\bf 1811}, 138 (2018)
  [arXiv:1808.09917 [hep-th]].

\bibitem{Zhang:2018qnt}
  S.~J.~Zhang,
  ``Subregion complexity and confinement–deconfinement transition in a holographic QCD model,''
  Nucl.\ Phys.\ B {\bf 938}, 154 (2019)
  [arXiv:1808.08719 [hep-th]].


\bibitem{Lucini:2003zr}
  B.~Lucini, M.~Teper and U.~Wenger,
  ``The High temperature phase transition in SU($N$) gauge theories,''
  JHEP {\bf 0401}, 061 (2004)
  [hep-lat/0307017].

  \bibitem{Ratti:2018ksb}
  C.~Ratti,
  ``Lattice QCD and heavy ion collisions: a review of recent progress,''
  Rept.\ Prog.\ Phys.\  {\bf 81}, no. 8, 084301 (2018)
  [arXiv:1804.07810 [hep-lat]].

\bibitem{deForcrand:2002hgr}
  P.~de Forcrand and O.~Philipsen,
  ``The QCD phase diagram for small densities from imaginary chemical potential,''
  Nucl.\ Phys.\ B {\bf 642}, 290 (2002)
  [hep-lat/0205016].


\bibitem{Brewer:2018abr}
  J.~Brewer, S.~Mukherjee, K.~Rajagopal and Y.~Yin,
  ``Searching for the QCD critical point via the rapidity dependence of cumulants,''
  Phys.\ Rev.\ C {\bf 98}, no. 6, 061901 (2018)
  [arXiv:1804.10215 [hep-ph]].

\bibitem{Nishioka:2006gr}
  T.~Nishioka and T.~Takayanagi,
  ``AdS Bubbles, Entropy and Closed String Tachyons,''
  JHEP {\bf 0701}, 090 (2007)
  [hep-th/0611035].


\bibitem{Johnson:2013dka}
  C.~V.~Johnson,
  ``Large $N$ Phase Transitions, Finite Volume, and Entanglement Entropy,''
  JHEP {\bf 1403}, 047 (2014)
  doi:10.1007/JHEP03(2014)047
  [arXiv:1306.4955 [hep-th]].

   \bibitem{Dey:2015ytd}
  A.~Dey, S.~Mahapatra and T.~Sarkar,
  ``Thermodynamics and Entanglement Entropy with Weyl Corrections,''
  Phys.\ Rev.\ D {\bf 94}, no. 2, 026006 (2016)
  [arXiv:1512.07117 [hep-th]].

\bibitem{Caceres:2015vsa}
  E.~Caceres, P.~H.~Nguyen and J.~F.~Pedraza,
  ``Holographic entanglement entropy and the extended phase structure of STU black holes,''
  JHEP {\bf 1509}, 184 (2015)
  [arXiv:1507.06069 [hep-th]].

\bibitem{Zeng:2015wtt}
  X.~X.~Zeng and L.~F.~Li,
  ``Van der Waals phase transition in the framework of holography,''
  Phys.\ Lett.\ B {\bf 764}, 100 (2017)
  [arXiv:1512.08855 [hep-th]].


\bibitem{Gubser:2000nd}
  S.~S.~Gubser,
  ``Curvature singularities: The Good, the bad, and the naked,''
  Adv.\ Theor.\ Math.\ Phys.\  {\bf 4}, 679 (2000)
  [hep-th/0002160].

  \bibitem{Chamblin:1999tk}
  A.~Chamblin, R.~Emparan, C.~V.~Johnson and R.~C.~Myers,
  ``Charged AdS black holes and catastrophic holography,''
  Phys.\ Rev.\ D {\bf 60}, 064018 (1999)
  [hep-th/9902170].

\bibitem{Chamblin:1999hg}
  A.~Chamblin, R.~Emparan, C.~V.~Johnson and R.~C.~Myers,
  ``Holography, thermodynamics and fluctuations of charged AdS black holes,''
  Phys.\ Rev.\ D {\bf 60}, 104026 (1999)
  [hep-th/9904197].

\bibitem{Mahapatra:2016dae}
  S.~Mahapatra,
  ``Thermodynamics, Phase Transition and Quasinormal modes with Weyl corrections,''
  JHEP {\bf 1604}, 142 (2016)
  [arXiv:1602.03007 [hep-th]].

\bibitem{Caldarelli:1999xj}
  M.~M.~Caldarelli, G.~Cognola and D.~Klemm,
  ``Thermodynamics of Kerr-Newman-AdS black holes and conformal field theories,''
  Class.\ Quant.\ Grav.\  {\bf 17}, 399 (2000)
  [hep-th/9908022].


\end{thebibliography}
\end{document}